\documentclass[11pt]{article}
\pdfoutput=1

\usepackage[usenames,dvipsnames,svgnames,table]{xcolor} % use color
\usepackage[obeyspaces,hyphens,spaces]{url}
\usepackage{jcapmod}
\usepackage[normalem]{ulem}

\usepackage{shorthand}
\usepackage{mathtools}
\usepackage{booktabs}
\usepackage[english]{babel}
\usepackage{amsmath,amssymb,amsbsy,amstext, amsthm, simplewick, amsfonts,braket}
\usepackage{hyperref}
\usepackage{graphicx}
\usepackage[small]{caption}
\usepackage{slashed}
\usepackage[separate-uncertainty=true,multi-part-units=single]{siunitx}
\usepackage{upgreek}
\usepackage{framed}
\usepackage{wrapfig}
\usepackage{multirow}
\usepackage{bbm}
\usepackage{enumerate}

\usepackage[utf8x]{inputenc}
\usepackage{selinput}
\usepackage{bm}
\usepackage{float}
\usepackage{geometry}
\usepackage{yfonts}
\usepackage{caption}
\usepackage{subcaption}
\usepackage{sidecap}
\usepackage{longtable}
\usepackage{anyfontsize}
\usepackage{dsfont}
\usepackage{tikz}
\usepackage{relsize}
\usepackage{tcolorbox}

\usepackage{tikz}
\usetikzlibrary{calc}
\newcommand*\circled[1]{\tikz[baseline=(char.base)]{
    \node[shape=circle, draw, inner sep=1pt, 
        minimum height={\f@size*1.6},] (char) {\vphantom{WAH1g}#1};}}
\makeatother
\newcommand{\slab}[1]{{\textsc{#1}}}

\newcommand{\minus}{{\scalebox {0.75}[1.0]{$-$}}}

\newcommand{\floq}[1]{{\scalebox{0.65}{$(#1)$}}}

\usepackage{colortbl}
\definecolor{lightgreen}{cmyk}{0.2, 0, 0.2, 0.2}
\definecolor{lightgray2}{cmyk}{0.1,0.1,0,0.1}
\definecolor{Red2}{RGB}{214, 39, 40}
\definecolor{Blue2}{RGB} {31, 119, 180}
\definecolor{Orange2}{RGB}{255, 127, 14}
\definecolor{Green2}{RGB}{44, 160, 44}
\definecolor{greyish2}{rgb}{.96,.96,.96}

\makeatletter
\newlength{\apb@width}
\newcommand{\autoparbox}[2][c]{\settowidth{\apb@width}{#2}\parbox[#1]{\apb@width}{#2}}

\makeatother


\allowdisplaybreaks[1]
\newcommand{\ped}[1]{\textormath{\textsubscript{#1}}{_{\mathrm{#1}}}}
\newcommand{\ap}[1]{\textormath{\textsuperscript{#1}}{^{\mathrm{#1}}}}

\DeclarePairedDelimiter{\abs}{\lvert}{\rvert}

\renewcommand{\vec}[1]{\boldsymbol{\mathbf{#1}}}

\newcommand{\dd}{\mathop{\mathrm{d}\!}{}}

\renewcommand{\Im}{\operatorname{Im}}
\renewcommand{\Re}{\operatorname{Re}}

\definecolor{pyBlue}{RGB}{31, 119, 180}
\definecolor{pyRed}{RGB}{214, 39, 40}
\definecolor{pyGreen}{RGB}{44, 160, 44}
\definecolor{pyBlue2}{RGB}{0, 111, 237}
\definecolor{pyRed2}{RGB}{224, 52, 36}
\definecolor{Mathematica1}{rgb}{0.368417, 0.506779, 0.709798}
\definecolor{Mathematica2}{rgb}{0.880722, 0.611041, 0.142051}
\definecolor{Mathematica3}{rgb}{0.560181, 0.691569, 0.194885}
\definecolor{Mathematica4}{rgb}{0.922526, 0.385626, 0.209179}

\def\beq{\begin{equation}}
\def\eeq{\end{equation}}

\begin{document}

\pagenumbering{roman}
\begin{titlepage}
\baselineskip=15.5pt \thispagestyle{empty}

\phantom{h}
\vspace{1cm}
\begin{center}
{\fontsize{22}{0}\selectfont  \bfseries  Resonant history of gravitational atoms\\\vskip 12pt in black hole binaries}\\
\end{center}

\vspace{0.5cm}
\begin{center}
{\fontsize{12}{18}\selectfont Giovanni Maria Tomaselli,$^{1}$ Thomas F.M. Spieksma$^{2}$ and Gianfranco Bertone$^{1}$}

\end{center}

\vspace{-0.1cm}
\begin{center}
\vskip 6pt
\textit{$^1$ Gravitation Astroparticle Physics Amsterdam (GRAPPA),
University of Amsterdam, \\ Science Park 904, 1098 XH Amsterdam, The Netherlands}

\vskip 6pt
\textit{$^2$ Niels Bohr International Academy, Niels Bohr Institute, \\
Blegdamsvej 17, 2100 Copenhagen, Denmark}
\end{center}

\vspace{1.2cm}
\hrule \vspace{0.3cm}
\noindent {\bf Abstract}\\[0.1cm]
Rotating black holes can produce superradiant clouds of ultralight bosons. When the black hole is part of a binary system, its cloud can undergo resonances and ionization. These processes leave a distinct signature on the gravitational waveform that depends on the cloud's properties. To determine the state of the cloud by the time the system enters the band of future millihertz detectors, we study the chronological sequence of resonances encountered during the inspiral. For the first time, we consistently take into account the nonlinearities induced by the orbital backreaction and we allow the orbit to have generic eccentricity and inclination. We find that the resonance phenomenology exhibits striking new features. Resonances can ``start'' or ``break'' above critical thresholds of the parameters, which we compute analytically, and induce dramatic changes in eccentricity and inclination. Applying these results to realistic systems, we find two possible outcomes. If the binary and the cloud are sufficiently close to counter-rotating, the cloud survives in its original state until the system enters in band; otherwise, the cloud is destroyed during a resonance at large separations, but leaves an imprint on the eccentricity and inclination. In both scenarios, we characterize the observational signatures, with particular focus on future gravitational wave detectors.
\vskip10pt
\hrule
\vskip10pt

\end{titlepage}

\thispagestyle{empty}
\setcounter{page}{2}

\tableofcontents

\newpage
\pagenumbering{arabic}
\setcounter{page}{1}

\clearpage
\section{Introduction}

The advent of gravitational-wave (GW) astronomy has opened up new opportunities to test fundamental physics. One promising avenue focuses on the environments surrounding black holes (BHs). When a binary inspirals inside a sufficiently dense medium, the orbital motion is modified, and the resulting GW signal is affected accordingly \cite{Barausse:2014tra}. Intermediate to extreme mass ratio inspirals form the perfect testbed for this phenomenon, as (1) the perturbation from the companion is not strong enough to completely disrupt the environment and (2) the binary spends many orbital cycles \emph{in band} allowing environmental effects to build up throughout the inspiral.\footnote{Although the impact of the environment is less strong, one can also look for environmental effects in the equal mass ratio case using the already available LIGO-Virgo data \cite{CanevaSantoro:2023aol}.} Future GW detectors such as LISA or the Einstein Telescope will be able to probe these types of binaries \cite{LISA:2017pwj,Baker:2019nia,Maggiore:2019uih}, and potentially infer properties of the environment from the waveform with great accuracy \cite{Barausse:2007dy,Speri:2022upm,Eda:2013gg,Kavanagh:2020cfn,Cole:2022yzw,Becker:2022wlo,Karydas:2024fcn,Kavanagh:2024lgq,Garg:2024oeu,Bamber:2022pbs,Aurrekoetxea:2023jwk}. This area of research has recently attracted conspicuous interest.

\vskip 4pt
One intriguing type of environment is comprised of ultralight bosons. Such particles are predicted from high energy physics as a possible solution to the strong CP problem (i.e., the QCD axion \cite{Weinberg:1977ma,Wilczek:1977pj,Peccei:1977hh}), but they can also arise from string compactifications \cite{Arvanitaki:2009fg,Svrcek:2006yi} and have been proposed as possible dark matter candidates \cite{Bergstrom:2009ib,Marsh:2015xka,Hui:2016ltb,Ferreira:2020fam}. As it turns out, rotating BHs are unstable against perturbations of the boson field. In particular, when the Compton wavelength of the bosonic field is comparable to the size of the BH, an efficient energy and angular momentum extraction from the BH is possible. Consequently,  a ``cloud'' is formed around the BH, often referred to as a ``gravitational atom.'' This process is termed \emph{superradiance} and has been studied extensively in the literature \cite{ZelDovich1971,ZelDovich1972,Starobinsky:1973aij,Brito:2015oca}. While superradiance occurs for bosons of any spin, in this work we focus on scalars, due to their simplicity and stronger theoretical motivation. As no couplings to other standard model particles are required, GWs form the ideal tool to study gravitational atoms, especially since they have been shown to exhibit a rich phenomenology  in binary systems~\cite{Zhang:2018kib,Baumann:2018vus,Zhang:2019eid,Baumann:2019ztm,Baumann:2021fkf,Baumann:2022pkl,Tomaselli:2023ysb,Brito:2023pyl,Duque:2023cac}.

\vskip 4pt
In particular, in previous works \cite{Baumann:2018vus,Baumann:2019ztm}, it was found that during a binary inspiral the cloud induces not only secular effects, such as dynamical friction or accretion, but also resonant behavior. At certain \emph{resonance frequencies}, where the orbital frequency of the binary matches the energy difference between two eigenstates of the cloud, the gravitational perturbation from the companion is resonantly enhanced and a full transfer from one state to another can be induced. The accompanied change in the angular momentum of the cloud must then be compensated for by the binary. Depending on the nature of the resonance, this leads to \emph{floating} or \emph{sinking} orbits, where the cloud releases or absorbs angular momentum from the binary, and hence the inspiral is either stalled or sped-up. Consequently, a potentially detectable dephasing is left in the GW signal. The orbital frequency at which these resonances happen can be predicted, and thus they serve as a direct and unique probe of the properties of the cloud. In addition to their floating or sinking nature, resonances can be divided into three different types, depending on the energy difference between the eigenstates involved. These are called hyperfine, fine, or Bohr resonances, which occur in this chronological order. The former two happen ``early'' in the inspiral (at radii far larger than the cloud's radius) and are all of the floating type. These will thus only affect the GW signal \emph{indirectly}: the binary frequency is too low for GW detectors to pick up the signal, but the resonance can change the late-time evolution of the system. Conversely, Bohr resonances happen ``late'' in the inspiral (at radii comparable with the cloud's radius) and can be either floating or sinking. As these happen while the signal is in band, they can affect the GW signal \emph{directly}.

\vskip 4pt
Due to the exciting observational signatures, various works have studied these resonances. However, to date, they have all made simplifying assumptions, which turn out to crucially alter the behavior of the system. This includes ignoring the backreaction from the resonance \cite{Takahashi:2021eso}, assuming a quasi-circular and equatorial orbit, or including just the strongest multipole of the gravitational perturbation \cite{Zhang:2018kib,Ding:2020bnl,Tong:2021whq,Du:2022trq,Takahashi:2023flk}. A nonzero eccentricity was only considered in \cite{Berti:2019wnn}, yet at a time when the behavior of the resonances had still not been fully understood. In this work, we relax all of the aforementioned assumptions to study resonances in gravitational atoms in full generality and use this to determine the evolution of the cloud-binary system throughout the inspiral. Crucially, by studying the nonlinear system, we find that there exists a critical threshold above which an \emph{adiabatic} floating resonance is initiated. Below the threshold, there is a negligible transfer, or a \emph{non-adiabatic} resonance. Furthermore, we discover mechanisms that induce a \emph{resonance breaking}, shutting down the transition before it is complete: these are due to the decay of the state excited by the resonance, or to a slow time variation of the parameters. Additionally, we allow for generically inclined and eccentric orbits and study how resonances impact the orbital parameters: while the eccentricity is forced towards fixed points, the inclination angle is always tilted towards a co-rotating configuration. Finally, we take into account the impact of ionization \cite{Baumann:2021fkf,Baumann:2022pkl,Tomaselli:2023ysb} in the evolution of the system and include all relevant multipoles.

\begin{figure} 
\centering
\includegraphics{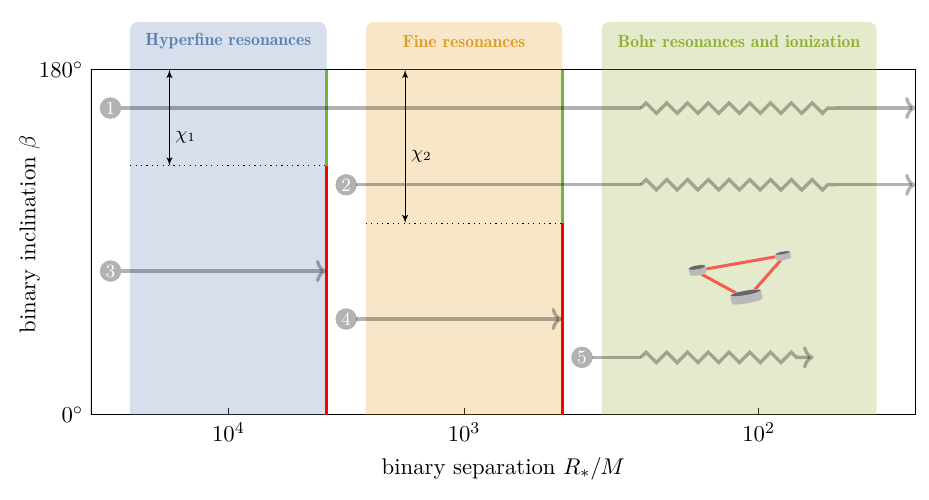}
\caption{Illustration of the possible outcomes of the resonant history of the cloud-binary system. The inspiral starts with the cloud in its initial state, either $\ket{211}$ or $\ket{322}$. Only systems {\normalsize \textcircled{\footnotesize 1}}--{\normalsize \textcircled{\footnotesize 2}} whose inclination angle is within intervals $\chi_1$ and $\chi_2$ from the counter-rotating ($\beta=\SI{180}{\degree}$) configuration are able to move past the hyperfine and fine resonances with the cloud still intact (green vertical lines). These later give rise to observational signatures in the form of ionization and Bohr resonances. Others {\normalsize \textcircled{\footnotesize 3}}--{\normalsize \textcircled{\footnotesize 4}} are destroyed by the hyperfine or fine resonances (red vertical lines). Binary systems that form at small enough separations may be able to skip early resonances {\normalsize \textcircled{\footnotesize 5}}.}
\label{fig:history_211}
\end{figure}

\vskip 4pt
Using these results, we lay out, for the first time, a systematic study of resonances for realistic parameters, focusing on intermediate and extreme mass ratios. Starting from states commonly populated by superradiance, we evolve the binary taking into account energy losses from both GWs and ionization. By doing so, we determine the resonant evolution of the cloud and the impact of the resonances on the binary's orbit. We find that, for generic orbital configurations, the cloud is often destroyed early in the inspiral. This is due to floating resonances that transfer the cloud to states that decay much quicker than the duration of the resonance. However, when the orbital inclination is within a certain interval centered around a counter-rotating configuration, all hyperfine and fine floating resonances are either non-adiabatic or break prematurely, allowing the cloud to survive until it enters the Bohr regime. Conversely, we find that all sinking resonances for typical parameters have a negligible impact on the cloud. A schematic illustration of these conclusions is given in Figure~\ref{fig:history_211}.

\vskip 4pt
To accurately model a binary inspiral in a non-vacuum spacetime, one needs to combine secular effects, such as dynamical friction, with resonant behavior. In a recent work \cite{Tomaselli:2023ysb}, we studied the former, whose associated energy losses can be much larger than those emitted by GWs. In this work, we combine these results with the resonant behavior, to obtain a self-consistent treatment of binary inspirals in a gravitational atom. We hope these two works serve as a guideline towards future studies in a fully relativistic setup, which should be able to describe correctly even the final stages of the inspiral. Recent work on this \cite{Brito:2023pyl} has already shown encouraging progress and new insights in the rich phenomenology of the problem.

\vskip 4pt
The new observational prospects discovered in this paper are further elaborated in the companion Letter \cite{Tomaselli:2024dbw}.

\paragraph{Outline} The outline of the paper is as follows. In Section~\ref{sec:setup}, we briefly review our setup and introduce necessary definitions and equations. Then, in Section~\ref{sec:resonance-pheno}, we study resonances at the nonlinear level, determining when they are adiabatic and when they break, and extending the framework to eccentric and inclined orbits. In Section~\ref{sec:types-of-resonances}, we discuss the different types of resonances. Then, in Section~\ref{sec:history}, we turn to a realistic setting and unveil the full history of the cloud and binary as function of the parameters. In Section~\ref{sec:observational-signatures}, we discuss the observational signatures this leads to. We conclude in Section~\ref{sec:conclusions}. Appendices \ref{sec:hyperfine-angular-momenutm}, \ref{sec:breaKING}, \ref{sec:ion-at-resonance}, \ref{sec:211_200}, and \ref{sec:variables} contain technical details.

\paragraph{Notation and conventions} We work in natural units, $G=\hbar=c=1$. The larger BH has a mass $M$ and dimensionless spin $\tilde a$, with $0\le\tilde a<1$. The mass and radial distance of the smaller object are denoted by $M_*\equiv qM$ and $R_*$, where $q$ is the mass ratio, while the orbital frequency is $\Omega$ and the mass of the cloud is $M\ped{c}$. The gravitational fine structure constant is $\alpha=\mu M$, where $\mu$ is the mass of the scalar field. We write most of our results in an explicit scaling form, with respect to the following set of benchmark parameters: $M=10^4M_\odot$, $M\ped{c}=0.01M$, $q=10^{-3}$ and $\alpha=0.2$. The eigenstate of the cloud before encountering a resonance will be denoted by $\ket{n_a\ell_am_a}$, while any other eigenstate involved in the resonance by $\ket{n_b\ell_bm_b}$. The cloud's wavefunction will then be expanded in as a linear superposition of energy eigenstates as $\ket{\psi}=c_a\ket{a}+\sum_bc_b\ket{b}$.

\paragraph{Code availability} The code used in this work and in \cite{Tomaselli:2023ysb} is publicly available on 
\href{https://github.com/thomasspieksma/GrAB/}{GitHub}.

\section{Setup}
\label{sec:setup}

\begin{figure} 
\centering
\includegraphics[width=0.8\textwidth]{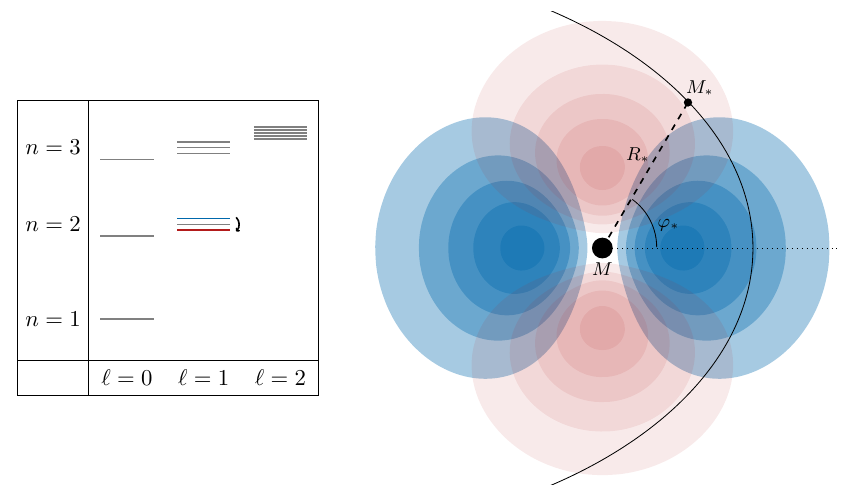}
\caption{Schematic illustration of the cloud-binary system. The primary object of mass $M$ is surrounded by a superradiantly grown scalar cloud of mass $M\ped{c}$. The secondary object of mass $M_*$ perturbs it through its gravitational potential, causing a mixing between different states of the cloud. The blue and red regions are a faithful representation of the mass densities of the states $\ket{211}$ and $\ket{21\,\minus1}$ on the equatorial plane, but the BH size is not to scale. In the box, we show the bound state spectrum of the gravitational atom for the first few values of $n$.}
\label{fig:SchematicIllustration}
\end{figure}

The goal of this section is to lay down our framework. We closely follow a previous work \cite{Tomaselli:2023ysb}, and thus we refer the reader there for more details.

\vskip 4pt
Via BH superradiance, bosonic fields can extract energy and angular momentum from rotating BHs, analogous to the Penrose process. The key condition for this process to occur is that the boson's frequency $\omega_\slab{B}$ is smaller than the angular velocity of the event horizon $\Omega_{H}$, i.e., $\omega_\slab{B} < m\Omega_{H}$, where $m$ is the azimuthal quantum number in the BH frame. If the boson is massive, then the superradiantly amplified waves can get trapped around the BH, allowing their number to grow exponentially. In particular, when the gravitational radius of the BH ($r_{g}$) and the Compton wavelength of the field ($\lambda_{c}$) are of the same order, i.e.,
\beq
\alpha \equiv \frac{r_{g}}{\lambda_{c}} = \mu M \lesssim 1\,,
\eeq
this growth becomes efficient on astrophysical timescales, and a cloud of particles can build up around the BH. The system is then often referred to as a gravitational atom. 

\vskip 4pt
In this work, we will stay in the nonrelativistic limit, where the Klein-Gordon equation reduces to a Schrödinger equation, and thus its solutions are the hydrogenic eigenfunctions, defined as
\beq
\psi_{n \ell m}(t, \vec{r})=R_{n \ell}(r) Y_{\ell m}(\theta, \phi) e^{-i\left(\omega_{n \ell m}-\mu\right) t}\,,
\label{eqn:eigenstates}
\eeq
where $n, \ell$, and $m$ are the principal, angular momentum, and azimuthal quantum numbers, respectively. These satisfy the usual relations, $n > \ell $, $\ell \geq 0$, and $\ell \geq |m|$, while $Y_{\ell m}$ are the spherical harmonics and $R_{n \ell}$ the hydrogenic radial functions.

\vskip 4pt
Due to the dissipative nature of the BH horizon, the eigenstates of the cloud, written as $\ket{n\ell m}$, are only quasi-stationary and have complex eigenfrequencies, i.e., $\omega_{n\ell m} = (\omega_{n \ell m})\ped{R} + i(\omega_{n \ell m})\ped{I}$. A certain eigenstate or \textit{mode} is superradiant when $(\omega_{n \ell m})\ped{I} > 0$, which can only happen if $m > 0$. The fastest growing mode is $\ket{n\ell m} = \ket{211}$ and the maximum mass the cloud can obtain is approximately $M_{\rm c} = 0.1M$  \cite{Hui:2022sri,Herdeiro:2021znw,East:2017ovw}.

\vskip 4pt
The energy of different modes are defined through the real part of the eigenfrequency, which in the limit $\alpha\ll1$ reads \cite{Baumann:2019eav}
\beq
\epsilon_{n\ell m}\equiv(\omega_{n\ell m})\ped{R}=\mu\biggl(1-\frac{\alpha^{2}}{2 n^{2}}-\frac{\alpha^{4}}{8 n^{4}}-\frac{(3 n-2 \ell-1) \alpha^{4}}{n^{4}(\ell+1 / 2)}+\frac{2 \tilde{a} m \alpha^{5}}{n^{3} \ell(\ell+1 / 2)(\ell+1)}+\mathcal{O}(\alpha^{6})\biggr)\,.
\label{eq:eigenenergy}
\eeq
As we will see later, it is useful to define different types of energy splittings: we say that two states have a \emph{Bohr} ($\Delta n\ne0$), \emph{fine} ($\Delta n=0$, $\Delta\ell\ne0$) or \emph{hyperfine} ($\Delta n=0$, $\Delta\ell=0$, $\Delta m\ne0$) splitting. A schematic illustration of the bound state spectrum of the gravitational atom is shown in Figure~\ref{fig:SchematicIllustration}. 

\vskip 4pt
We work in the reference frame of the central BH with mass $M$, where $\vec{r} = \{r, \theta, \phi\}$ and the coordinates of the companion with mass $M_{*}=qM$ are $\mathbf{R}_{*} = \{R_*, \theta_*, \varphi_*\}$. In our conventions, the gravitational perturbation from the companion in the Newtonian approximation is defined as
\beq
\label{eqn:V_star}
V_*(t,\vec r)=-\sum_{\ell_*=0}^\infty\sum_{m_*=-\ell_*}^{\ell_*}\frac{4\pi q\alpha}{2\ell_*+1}Y_{\ell_*m_*}(\theta_*,\varphi_*)Y_{\ell_*m_*}^*(\theta,\phi)\,F(r)\,,
\eeq
where
\beq
F(r)=
\begin{cases}
\dfrac{r^{\ell_*}}{R_*^{\ell_*+1}}\Theta(R_*-r)+\dfrac{R_*^{\ell_*}}{r^{\ell_*+1}}\Theta(r-R_*)&\text{for }\ell_*\ne1\,,\\[12pt]
\biggl(\dfrac{R_*}{r^2}-\dfrac{r}{R_*^2}\biggr)\Theta(r-R_*)&\text{for }\ell_*=1\,.
\end{cases}
\label{eqn:F(r)}
\eeq
On a generic orbit, the perturbation \eqref{eqn:V_star} induces a mixing between the cloud's bound state $\ket{n_b \ell_b m_b}$ and another state $\ket{n_a\ell_am_a}$ through the matrix element \cite{Baumann:2018vus,Baumann:2019ztm} 
\beq
\braket{n_a\ell_a m_a|V_*(t,\vec r)|n_b\ell_bm_b}=-\sum_{\ell_*,m_*}\frac{4\pi\alpha q}{2\ell_*+1}Y_{\ell_*m_*}(\theta_*,\varphi_*)\, I_r(t) \, I_{\Omega}(t)\,,
\label{eqn:MatrixElement}
\eeq
where $I_r$ and $I_\Omega$ are integrals over radial and angular variables, respectively. The following \emph{selection rules} need to be satisfied in order for $I_{\Omega}$ to be nonzero
\begin{align}
(\text{S1})\qquad & m_*=m_b - m_a\,,\\
(\text{S2})\qquad & \ell_a+\ell_*+\ell_b=2 p,\qquad \text{for }p \in \mathbb{Z}\,, \label{eqn:S2}\\
(\text{S3})\qquad & \abs{\ell_b-\ell_a} \leq \ell_* \leq \ell_b+\ell_a\,. \label{eqn:S3}
\end{align}
Furthermore, we will often expand the spherical harmonic $Y_{\ell_*m_*}$ appearing in \eqref{eqn:MatrixElement} in terms of the form $Y_{\ell_* g}(\pi/2, 0)$ (where $g$ is a summation index), which is zero whenever $\ell_*$ and $g$ have opposite parity.

\section{Resonance phenomenology}
\label{sec:resonance-pheno}

As first shown in \cite{Baumann:2019ztm}, while the companion perturbs the cloud at a slowly increasing frequency, transitions between modes are induced, analogous to the ones described in quantum mechanics by Landau and Zener \cite{zener1932non,landau1932theorie}. This process can exert a strong backreaction on the orbit, giving rise to ``floating'' and ``sinking'' orbits. In this section, we study these transitions for generic orbits and at the nonlinear level, by including the backreaction in the frequency evolution self-consistently. We do not, however, worry about the astrophysically relevant range of parameters just yet, nor about whether the phenomena we discover here can actually occur after well-motivated initial conditions. Such ``realistic'' cases, to which we often refer to, will only be defined and studied in Section~\ref{sec:history}, where the general results found here will turn out to be crucial in determining the evolution of the cloud-binary system.

\vskip 4pt
In Section~\ref{sec:two-state-resonances} we review the setup and the well-known results, which we extend to inclined and eccentric orbits in Section~\ref{sec:eccentric-inclined-resonances}. Then, in Section~\ref{sec:backreaction} we include the backreaction, thus coupling the resonating states to the evolution of the binary parameters. The phenomenology of the resulting nonlinear system is explored in Section~\ref{sec:floating} and Section~\ref{sec:sinking}, for the floating and sinking cases, respectively.

\subsection{Two-state resonances}
\label{sec:two-state-resonances}

The matrix element \eqref{eqn:MatrixElement} of the gravitational perturbation $V_*$ between two states $\ket a=\ket{n_a\ell_am_a}$ and $\ket b=\ket{n_b\ell_bm_b}$ is an oscillatory function of the true anomaly of the orbit $\varphi_*$,
\beq
\braket{a|V_{*}(t)|b} = \sum_{g\in \mathbb{Z}} \eta^\floq{g} e^{ig \varphi_*}\,.
\label{eqn:eta-circular}
\eeq
On equatorial co-rotating quasi-circular orbits, the only nonzero term is $g=m_b-m_a\equiv\Delta m$ (on counter-rotating orbits, $g$ has opposite sign), and the coefficients $\eta^\floq{g}$ only depend on time through $\Omega(t)$. Restricting our attention to the two-state system, the Hamiltonian is thus
\beq
\mathcal H=\begin{pmatrix}-\Delta\epsilon/2 & \eta^\floq{g}e^{ig\varphi_*}\\ \eta^\floq{g}e^{-ig\varphi_*} & \Delta\epsilon/2\end{pmatrix}\,,
\label{eqn:schrodinger-hamiltonian}
\eeq
where $\Delta\epsilon=\epsilon_b-\epsilon_a$ is the energy difference between $\ket{b}$ and $\ket{a}$. As in \cite{Baumann:2019ztm}, it is useful to rewrite (\ref{eqn:schrodinger-hamiltonian}) in a \emph{dressed} frame, where the fast oscillatory terms $e^{ig\varphi_*}$ are traded for a slow evolution of the energies. This is done by means of a unitary transformation,
\beq
\begin{pmatrix}c_a\\ c_b\end{pmatrix}=\begin{pmatrix}e^{ig\varphi_*/2} & 0\\ 0 & e^{-ig\varphi_*/2}\end{pmatrix}\begin{pmatrix}\tilde c_a\\ \tilde c_b\end{pmatrix}\,,
\eeq
where $c_j=\braket{j|\psi}$ (with $j=a,b$) are the Schrödinger frame coefficients, while $\tilde c_a$ and $\tilde c_b$ are the dressed frame coefficients. Because $\abs{c_i}^2=\abs{\tilde c_i}^2$, we will drop the tildes in the following discussion. In the dressed frame, the Schrödinger equation reads
\beq
\frac\dd{\dd t}\!\begin{pmatrix}c_a\\ c_b\end{pmatrix}=-i\mathcal H\ped{D}\begin{pmatrix}c_a\\ c_b\end{pmatrix},\qquad\mathcal H\ped{D}=\begin{pmatrix}-(\Delta\epsilon-g\Omega)/2 & \eta^\floq{g}\\ \eta^\floq{g} & (\Delta\epsilon-g\Omega)/2\end{pmatrix}\,.
\label{eqn:dressed-hamiltonian}
\eeq
When $\Omega(t)\equiv\dot\varphi_*$ is specified, (\ref{eqn:dressed-hamiltonian}) determines the evolution of the population of the two states.

\vskip 4pt
Without including the backreaction of the resonance on the orbit, $\Omega(t)$ is exclusively determined by external factors, such as the energy losses due to GW emission or cloud ionization, which induce a frequency chirp. These effects typically have a nontrivial dependence on $\Omega$ itself, widely varying in strength at different points of the inspiral. However, the resonances described by (\ref{eqn:dressed-hamiltonian}) are restricted to a bandwidth $\Delta\Omega\sim\eta^\floq{g}$. This is typically narrow enough to allow us to approximate the external energy losses, as well as any other $\Omega$-dependent function, with their value at the resonance frequency,
\beq
\Omega_0=\frac{\Delta\epsilon}{g}\,.
\eeq
Around $\Omega_0$, we can linearize the frequency chirp and write $\Omega=\gamma t$. For concreteness, in this section we will assume that external energy losses are only due to GW emission, in which case
\beq
\gamma=\frac{96}5\frac{qM^{5/3}\Omega_0^{11/3}}{(1+q)^{1/3}}\,.
\label{eqn:gamma_gws}
\eeq
It is particularly convenient to rewrite the Schrödinger equation in terms of dimensionless variables and parameters:
\beq
\label{eqn:dimensionless-dressed-schrodinger}
\frac\dd{\dd\tau}\!\begin{pmatrix}c_a\\ c_b\end{pmatrix}=-i\begin{pmatrix}\omega/2 & \sqrt{Z}\\ \sqrt{Z} & -\omega/2\end{pmatrix}\begin{pmatrix}c_a\\ c_b\end{pmatrix}\,,
\eeq
where the frequency chirp now reads $\omega=\tau$, and we defined
\beq
\label{eqn:coefficients}
\tau=\sqrt{\abs{g}\gamma}\,t\,,\qquad\omega=\frac{\Omega-\Omega_0}{\sqrt{\gamma/\abs{g}}}\,,\qquad Z=\frac{(\eta^\floq{g})^2}{\abs{g}\gamma}\,.
\eeq
The initial conditions at $\tau\to-\infty$ we are interested in are those where only one state is populated, say $c_a=1$ and $c_b=0$. The only dimensionless parameter of \eqref{eqn:dimensionless-dressed-schrodinger} is the so-called ``Landau-Zener parameter'' $Z$, which determines uniquely the evolution of the system and its state at $\tau\to+\infty$. In fact, the populations at $\tau\to+\infty$ can be derived analytically and are given by the Landau-Zener formula:
\beq
\abs{c_a}^2=e^{-2\pi Z}\,,\qquad \abs{c_b}^2=1-e^{-2\pi Z}\,.
\label{eqn:lz}
\eeq
For $2\pi Z\gg1$ the transition can be classified as \emph{adiabatic}, meaning that the process is so slow that the cloud is entirely transferred from $\ket{a}$ to $\ket{b}$. Conversely, for $2\pi Z\ll1$, the transition is \emph{non-adiabatic}, with a partial or negligible transfer occurring.\footnote{The adiabaticity of a resonance is not related to the adiabaticity of the orbital evolution, which is always assumed to hold throughout our work.}

\vskip 4pt
There is one final remark to be made before we proceed. Dealing with two-state transitions is a good approximation as long as the frequency width of the resonance, $\Delta\Omega\sim\eta^\floq{g}$, is much narrower than the distance (in frequency) from the closest resonance. The latter becomes extremely small for hyperfine resonances, especially on generic orbits, where $g$ can take values different from $\Delta m$. In some cases formula \eqref{eq:eigenenergy} can indeed return an exact degeneracy of two resonances, up to $\mathcal O(\alpha^5)$. We have thoroughly checked, by numerical computation of the eigenfrequencies up to $\mathcal O(\alpha^6)$, that in all realistic cases the resonances are indeed narrow enough for the two-state approximation to hold.

\subsection{Resonances on eccentric and inclined orbits}
\label{sec:eccentric-inclined-resonances}

We now extend the treatment of Section~\ref{sec:two-state-resonances} to orbits with nonzero eccentricity or inclination, explaining what changes for the resonant frequencies and the overlap coefficients $\eta^\floq{g}$.

\vskip 4pt
Let us start with eccentric co-rotating orbits. In the quasi-circular case, equation (\ref{eqn:eta-circular}) manifestly separates a ``fast'' and a ``slow'' motion: the former originates from $\varphi_*$ varying over the course of an orbit, while the latter is due to the dependence of the coefficients $\eta^\floq{g}$ on $\Omega(t)$ (and can be safely neglected). It will be helpful to work with a variable that performs the same trick on eccentric orbits: the \emph{mean anomaly}
\beq
\tilde\varphi_*(t)=\int^t\Omega(t')\dd t'\,.
\eeq
Because $\varphi_*$ itself is an oscillating function of $\tilde\varphi_*$, we can write
\beq
\braket{a|V_{*}(t)|b} = \sum_{g\in \mathbb{Z}} \tilde\eta^\floq{g} e^{ig\tilde\varphi_*}\,,
\label{eqn:eta-tilde}
\eeq
where the coefficients $\tilde\eta^\floq{g}$ only depend on time through $\Omega(t)$. For simplicity, in the following discussion we will drop the tildes, with the different definition of $\eta^\floq{g}$ for nonzero eccentricity left understood.

\vskip 4pt
For a given eccentricity $\varepsilon\ne0$, multiple terms of (\ref{eqn:eta-tilde}), each corresponding to a different value of $g$, can be nonzero. As a consequence, a resonance between two given states can be triggered at different points of the inspiral, at the frequencies $\Omega_0^\floq{g}=\Delta\epsilon/g$, for any integer $g$ (provided that it has the same sign as $\Delta\epsilon$). The numerical evaluation of the coefficients $\eta^\floq{g}$ requires to Fourier expand $V_*$ in the time domain, at the orbital frequency $\Omega=\Omega_0^\floq{g}$. This can be done with techniques similar to \cite{Tomaselli:2023ysb}, where the same matrix element was evaluated between a bound and an unbound state. The coefficient $\eta^\floq{\Delta m}$ is special because it is the only one with a finite, nonzero limit for $\varepsilon\to0$, where it reduces to its circular-orbit counterpart. For all other values of $g$, instead, $\eta^\floq{g}$ vanishes for $\varepsilon\to0$. Even at moderately large $\varepsilon$, the coefficient $\eta^\floq{\Delta m}$ remains significantly larger than all the others

\vskip 4pt
Let us now look at circular but inclined orbits. Here, the Fourier coefficients $\eta^\floq{g}$ acquire a dependence on the inclination angle $\beta$, where $\beta=0$ and $\beta=\pi$ correspond to the co-rotating and counter-rotating scenarios. The functional dependence can be readily extracted by evaluating the perturbation \eqref{eqn:V_star} using the identity \cite{wigner}
\beq
Y_{\ell_*m_*}(\theta_*,\varphi_*)=\sum_{g=-\ell_*}^{\ell_*}d^\floq{\ell_*}_{m_*,g}(\beta)Y_{\ell_*g}\biggl(\frac\pi2,0\biggr)e^{ig\Omega t}\,.
\label{eqn:Y-decomposed}
\eeq
Here, $d^\floq{\ell_*}_{m_*,g}(\beta)$ is a Wigner small $d$-matrix and is responsible for the angular dependence of the coupling, $\eta^\floq{g}\propto d^\floq{\ell_*}_{m_*,g}(\beta)$. Its functional form takes on a simple expression in many of the physically interesting cases, as we will discuss explicitly in Section~\ref{sec:types-of-resonances}. We thus see that inclined orbits also trigger resonances at $\Omega=\Omega_0^\floq{g}=\Delta\epsilon/g$, but this time $g$ can only assume a finite number of different values. Similar to the eccentric case, $g=\Delta m$ is special, because it is the only case where $d^\floq{\ell_*}_{m_*,g}(\beta)$ does not vanish for $\beta\to0$, as the resonance survives in the equatorial co-rotating limit. Similarly, in the counter-rotating case $\beta\to\pi$, the only surviving value is $g=-\Delta m$.

\vskip 4pt
Similar techniques can be applied in the eccentric \emph{and} inclined case, where the overlap can be expanded in two sums, each with its own index, say $g_\varepsilon$ and $g_\beta$. We do not explicitly compute $\eta^\floq{g}$ in the general case, as the understanding developed so far is sufficient to move forward and characterize the phenomenology in realistic cases.

\subsection{Backreaction on the orbit}
\label{sec:backreaction}

We now include the backreaction on the orbit, allowing for generic nonzero eccentricity \emph{and} inclination. During a resonance, the energy and angular momentum contained in the cloud change over time: this variation must be compensated by an evolution of the binary parameters, the (dimensionless) frequency $\omega$, eccentricity $\varepsilon$ and inclination $\beta$. In turn, this backreaction impacts the Schrödinger equation \eqref{eqn:dimensionless-dressed-schrodinger}, which directly depends on $\omega$. The result is a coupled nonlinear system of ordinary differential equations, describing the co-evolution of the cloud and the binary, which we derive in this section.

\vskip 4pt
To describe the evolution of $\omega$, $\varepsilon$, and $\beta$ we need three equations. These are the conservation of energy and of two components of the angular momentum: the projection along the BH spin and the projection on the equatorial plane. The conservation of energy reads
\beq
\frac\dd{\dd t}\bigl(E+E\ped{c}\bigr)=-\gamma f(\varepsilon)\,\frac{qM^{5/3}}{3(1+q)^{1/3}\Omega_0^{1/3}}\,,
\label{eqn:E-balance}
\eeq
where $\gamma$ was defined in \eqref{eqn:gamma_gws} and the binary's and cloud's energies are
\beq
E=-\frac{qM^{5/3}\Omega^{2/3}}{2(1+q)^{1/3}}\,,\qquad E\ped{c}=\frac{M\ped{c}}\mu(\epsilon_a\abs{c_a}^2+\epsilon_b\abs{c_b}^2)\,,
\eeq
while the function
\beq
f(\varepsilon)=\frac{1+\frac{73}{24}\varepsilon^2+\frac{37}{96}\varepsilon^4}{(1-\varepsilon^2)^{7/2}}
\eeq
quantifies the dependence of GW energy losses on the eccentricity \cite{Peters:1963ux,Peters:1964zz}. Similarly, the conservation of the angular momentum components requires
\begin{align}
\label{eqn:Lz-balance}
\frac\dd{\dd t}\bigl(L\cos\beta+S\ped{c}\bigr)=-h(\varepsilon)\gamma\,\frac{qM^{5/3}}{3(1+q)^{1/3}\Omega_0^{4/3}}\cos\beta\,,\\
\label{eqn:Lx-balance}
\frac\dd{\dd t}\bigl(L\sin\beta\bigr)=-h(\varepsilon)\gamma\,\frac{qM^{5/3}}{3(1+q)^{1/3}\Omega_0^{4/3}}\sin\beta\,.
\end{align}
where
\beq
L=\frac{qM^{5/3}}{(1+q)^{1/3}}\frac{\sqrt{1-\varepsilon^2}}{\Omega^{1/3}},\qquad S\ped{c}=\frac{M\ped{c}}\mu(m_a\abs{c_a}^2+m_b\abs{c_b}^2)\,,
\eeq
and
\beq
h(\varepsilon)=\frac{1+\frac78\varepsilon^2}{(1-\varepsilon^2)^2}\,.
\eeq
Before proceeding, there are two issues the reader might worry about. First, depending on the resonance, the spin of the cloud during the transition might also have equatorial components, and should thus appear in \eqref{eqn:Lx-balance}. Second, the BH spin breaks spherical symmetry, therefore the equatorial projection of the angular momentum should not be conserved. Clearly, in the Newtonian limit this is not a problem, but one might still question whether it is consistent to treat within this framework hyperfine resonances, whose very existence is due to a nonzero BH spin in the first place. We address both these issues in Appendix~\ref{sec:hyperfine-angular-momenutm}, where we justify our assumptions, and proceed here to study the dynamics of the previous equations.

\vskip 4pt
Equations \eqref{eqn:E-balance}, \eqref{eqn:Lz-balance} and \eqref{eqn:Lx-balance} can be put in a dimensionless form as follows:
\begin{align}
\label{eqn:dimensionless-omega-evolution}
\frac{\dd\omega}{\dd\tau}&=f(\varepsilon)-B\frac{\dd\abs{c_b}^2}{\dd\tau}\,,\\
\label{eqn:dimensionless-eccentricity-evolution}
C\frac\dd{\dd\tau}\sqrt{1-\varepsilon^2}&=\sqrt{1-\varepsilon^2}\biggl(f(\varepsilon)-B\frac{\dd\abs{c_b}^2}{\dd\tau}\biggr)+B\frac{\Delta m}g\frac{\dd\abs{c_b}^2}{\dd\tau}\cos\beta-h(\varepsilon)\,,\\
\label{eqn:dimensionless-inclination-evolution}
C\sqrt{1-\varepsilon^2}\frac{\dd\beta}{\dd\tau}&=-B\frac{\Delta m}g\frac{\dd\abs{c_b}^2}{\dd\tau}\sin\beta\,,
\end{align}
where we defined the dimensionless parameters
\beq
B=\frac{3M\ped{c}}M\frac{\Omega_0^{4/3}((1+q)M)^{1/3}}{q\alpha\sqrt{\gamma/\abs{g}}}(-g)\,,\qquad C=\frac{3\Omega_0}{\sqrt{\gamma/\abs{g}}}\,.
\label{eqn:BC}
\eeq
The Schrödinger equation \eqref{eqn:dimensionless-dressed-schrodinger} remains unchanged, but it should be kept in mind that $Z$ now depends on $\varepsilon$ and $\beta$ through $\eta^\floq{g}$ (instead, the dependence on $\omega$ can still be neglected if the resonance is narrow enough).

\vskip 4pt
The parameter $B$ controls the strength of the backreaction. As can be seen from \eqref{eqn:dimensionless-omega-evolution}, a positive $B>0$ (i.e., $g<0$ and $\Delta\epsilon<0$) will slow down the frequency chirp, giving rise to a \emph{floating} orbit and generally making the resonance more adiabatic. Conversely, $B<0$ (i.e., $g>0$ and $\Delta\epsilon>0$) induces \emph{sinking} orbits and makes resonances less adiabatic. By extension, we will refer to floating resonances and sinking resonances to denote the type of backreaction they induce. A summary of the main variables used to describe the resonances and their backreaction is given in Appendix~\ref{sec:variables}.

\subsection{Floating orbits}
\label{sec:floating}

Backreaction of the floating type ($B>0$) turns out to be the most relevant case for realistic applications, so we make a detailed study of its phenomenology here. When the backreaction is strong, the evolution of the system exhibits a very well-defined phase of floating orbit. We are then concerned with three aspects.
\begin{enumerate}
\item Under what conditions is a floating resonance initiated? We answer this question with a simple analytical prescription, which is found and discussed in Section~\ref{sec:floating-adiabatic}.
\item How does the system evolve during the float? This is addressed in Section~\ref{sec:floating-evolution-e-beta}, where we study the evolution of the eccentricity and inclination.
\item When does a floating resonance end? In Section~\ref{sec:resonance-breaking} we show that several phenomena can \emph{break} (and end) the resonance before the transition from $\ket{a}$ to $\ket{b}$ is complete, and compute accurately the conditions under which this phenomenon happens.
\end{enumerate}

\subsubsection{Adiabatic or non-adiabatic}
\label{sec:floating-adiabatic}

From Section~\ref{sec:two-state-resonances}, we know that if $B=0$, then a fraction $1-e^{-2\pi Z}$ of the cloud is transferred during the resonances. For $2\pi Z\gg1$, this value is already very close to 1. Adding the backreaction does not change this conclusion: the resonance stays adiabatic and a complete transfer from $\ket{a}$ to $\ket{b}$ is observed. Assuming, for simplicity, quasi-circular orbits ($\varepsilon=0$), the duration of the floating orbit can be easily read off~(\ref{eqn:dimensionless-omega-evolution}):
\beq
\Delta t\ped{float}=\frac{B}{\sqrt{\abs{g}\gamma}}=\frac{3M\ped{c}}M\frac{\Omega_0^{4/3}((1+q)M)^{1/3}}{q\alpha\gamma}(-g)\,.
\label{eqn:t-float}
\eeq
This is independent of the strength of the perturbation $\eta^\floq{g}$, and corresponds to the time it takes for the external energy losses to dissipate the energy of the two-state system. For nonzero eccentricity instead, one must integrate $f(\varepsilon)$ over time to determine the duration of the float.

\begin{figure} 
\centering
\includegraphics{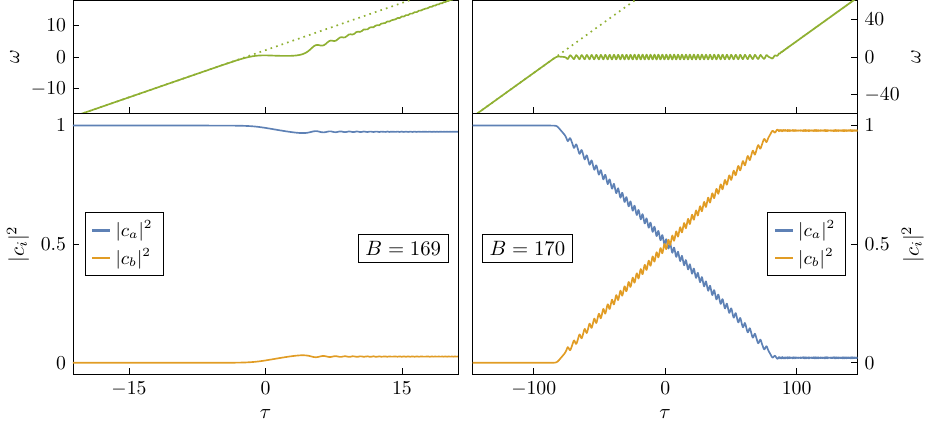}
\caption{Numerical solution of the nonlinear system (\ref{eqn:dimensionless-dressed-schrodinger})-(\ref{eqn:dimensionless-omega-evolution}). In both panels we set $Z=0.001$, whereas we choose the values of $B$ to be $169$ (\emph{left panel}) and $170$ (\emph{right panel}), slightly below or above the adiabaticity threshold (the limit value of $ZB$ differs slightly from the one given in (\ref{eqn:1/2piadiabatic}), due to finite-$Z$ corrections). In the left panel, a non-adiabatic transition is observed. Conversely, in the right panel, we find an adiabatic transition and the consequent formation of a floating orbit, whose duration matches the predicted $\Delta t\ped{float}=B/\sqrt{\abs{g}\gamma}$. The dotted lines represent the evolution of $\omega$ in absence of backreaction.}
\label{fig:floating-backreacted-resonance}
\end{figure}

\vskip 4pt
The situation for $2\pi Z\ll1$ is, in principle, much less clear: with $B=0$ the resonance would be non-adiabatic, but backreaction tends to make it more adiabatic. Let us once again restrict to quasi-circular orbits for simplicity. By careful numerical study of equations (\ref{eqn:dimensionless-dressed-schrodinger}) and (\ref{eqn:dimensionless-omega-evolution}), we find that the long-time behavior of the system is predicted by the parameter $ZB$ alone. Depending on its value, two qualitatively different outcomes are possible:
\beq
\text{if}\quad2\pi Z\ll1\quad \mathrm{and} \quad
\begin{cases}
ZB<0.1686\ldots\quad\longrightarrow\quad\text{very non-adiabatic,}\\[12pt]
ZB>0.1686\ldots\quad\longrightarrow\quad\text{very adiabatic.}
\end{cases}
\label{eqn:1/2piadiabatic}
\eeq
In the upper case, a negligible fraction of the cloud is transferred and the time evolution of $\omega$ is almost exactly linear. Conversely, in the bottom case, the cloud is entirely transferred from $\ket{a}$ to $\ket{b}$ and $\omega$ is stalled for an amount of time $\Delta t\ped{float}$ given by (\ref{eqn:t-float}), during which it oscillates around zero. Intermediate behaviors are not possible, unless the value of $ZB$ is extremely fine-tuned. Numerical solutions of (\ref{eqn:dimensionless-dressed-schrodinger}) and (\ref{eqn:dimensionless-omega-evolution}) are shown in Figure~\ref{fig:floating-backreacted-resonance}, choosing parameters in such a way to illustrate the two cases in \eqref{eqn:1/2piadiabatic}.

\vskip 4pt
We can give an approximate derivation of the previous result as follows. As long as $\abs{c_b}^2$ is small enough, the backreaction term in (\ref{eqn:dimensionless-omega-evolution}) is negligible, hence $\omega$ evolves linearly and the final populations approximate the Landau-Zener result (\ref{eqn:lz}), giving $\abs{c_b}^2\approx2\pi Z$. As the unbackreacted transition happens in the time window $\abs{\tau}\lesssim1$, we see from (\ref{eqn:dimensionless-omega-evolution}) that the backreaction becomes significant when $1\lesssim B\cdot2\pi Z\implies ZB\gtrsim1/(2\pi)\approx0.159\ldots$ Given the minimal numerical difference between this coefficient and the one given in (\ref{eqn:1/2piadiabatic}), for simplicity we will often write the relevant condition for an adiabatic resonance simply as $2\pi ZB\gtrless1$. The slow-down effect on the evolution of $\omega(\tau)$ enjoys a positive-feedback mechanism: the slower $\omega$ evolves, the more the transition is adiabatic, meaning that $\abs{c_b}^2$ is larger, which further slows down $\omega(\tau)$, and so on. This explains why no intermediate behaviors are observed: once the backreaction goes over a certain critical threshold, the process becomes self-sustaining.

\vskip 4pt
The picture outlined so far changes slightly when the eccentricity is nonzero. First, if the binary had a constant eccentricity $\varepsilon_0$, we could simply replace $\gamma\to\gamma f(\varepsilon_0)$ to conclude that the critical threshold for adiabaticness becomes
\beq
2\pi ZB\gtrless f(\varepsilon_0)^{3/2}\,.
\label{eqn:2piZB-epsilon0}
\eeq
When the eccentricity is allowed to vary starting from the initial value $\varepsilon_0$, equation \eqref{eqn:2piZB-epsilon0} still correctly predicts whether the system enters a floating orbit phase. However, the transfer might no longer be complete, as the resonance might \emph{break}. This aspect will be discussed in Section~\ref{sec:resonance-breaking}.

\subsubsection{Evolution of eccentricity and inclination}
\label{sec:floating-evolution-e-beta}

\begin{figure} 
\centering
\includegraphics{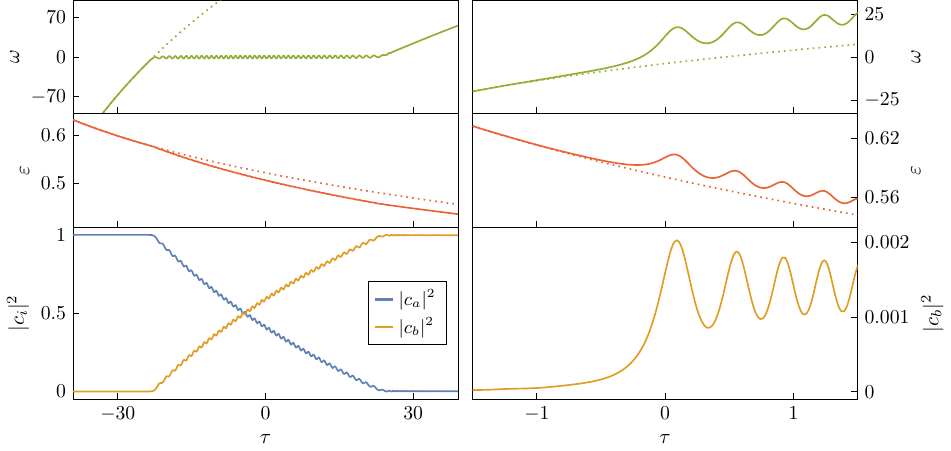}
\caption{Floating (\emph{left panel}) and sinking (\emph{right panel}) resonances on eccentric orbits, with $\Delta m/g=1$. We display the value of the frequency $\omega$, the eccentricity $\varepsilon$, and the populations $\abs{c_a}^2$ and $\abs{c_b}^2$ as function of $\tau$, obtained by solving equations (\ref{eqn:dimensionless-dressed-schrodinger}), (\ref{eqn:dimensionless-omega-evolution}), and (\ref{eqn:dimensionless-eccentricity-evolution}) with $\beta=0$ numerically. The parameters used for the floating case are $Z=0.03$, $B=250$, $C=1000$, while for the sinking case we used $Z=0.01$, $B=-10000$, $C=100$. The dotted lines represent the evolution of $\omega$ and $\varepsilon$ in absence of backreaction. Even though the impact of the resonance on the eccentricity might look mild, the effect is actually dramatic when seen as a function of $\omega$, as shown in Figure~\ref{fig:omega-eccentricity}.}
\label{fig:eccentric-backreacted-resonance}
\end{figure}

\begin{figure} 
\centering
\includegraphics{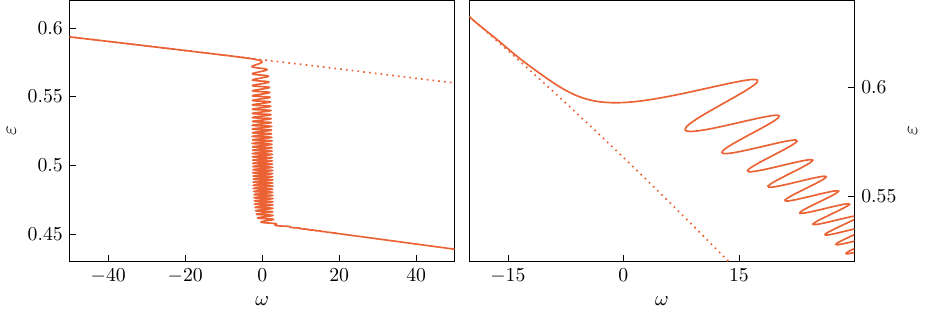}
\caption{Same resonances as in Figure~\ref{fig:eccentric-backreacted-resonance}, but now the evolution of eccentricity is shown as a function of the frequency, for floating (\emph{left panel}) and sinking (\emph{right panel}) orbits. The dashed lines represent the vacuum evolution.}
\label{fig:omega-eccentricity}
\end{figure}

The left panel of Figure~\ref{fig:eccentric-backreacted-resonance} shows a numerical solution of the coupled nonlinear equations \eqref{eqn:dimensionless-dressed-schrodinger}, \eqref{eqn:dimensionless-omega-evolution}, \eqref{eqn:dimensionless-eccentricity-evolution} and \eqref{eqn:dimensionless-inclination-evolution}, for an equatorial co-rotating ($\beta=0$) but eccentric ($\varepsilon\ne0$) system, undergoing a floating orbit with $g=\Delta m$. The state dynamics is largely similar to what we described in Section~\ref{sec:floating-adiabatic}. The most interesting new effect concerns the evolution of the eccentricity, which can be seen to decrease during the float, at a rate faster than the circularization provided by GW emission. The same numerical solution is shown as function of frequency in Figure~\ref{fig:omega-eccentricity}. 

\vskip 4pt
The evolution of the eccentricity during a float can be studied analytically by plugging $\dd\omega/\dd\tau\approx0$ into equations \eqref{eqn:dimensionless-eccentricity-evolution} and \eqref{eqn:dimensionless-inclination-evolution}, which become
\begin{align}
\label{eqn:eccentricity-evolution-floating}
C\frac\dd{\dd\tau}\sqrt{1-\varepsilon^2}&=\frac{\Delta m}gf(\varepsilon)\cos\beta-h(\varepsilon)\,,\\
\label{eqn:inclination-evolution-floating}
C\sqrt{1-\varepsilon^2}\frac{\dd\beta}{\dd\tau}&=-\frac{\Delta m}gf(\varepsilon)\sin\beta\,.
\end{align}
For resonances with $\beta=0$ and $g=\Delta m$, such as the one shown in Figures~\ref{fig:eccentric-backreacted-resonance} and~\ref{fig:omega-eccentricity}, a small-$\varepsilon$ expansion leads to the following solution:
\beq
\varepsilon(t)\approx\varepsilon_0\,e^{-\frac{22}{18}\gamma t/\Omega_0}\,.
\label{eqn:circulatization-floating}
\eeq
This result should be compared to the GW-induced circularization in absence of backreaction,
\beq
\varepsilon(t)\approx\varepsilon_0\,e^{-\frac{19}{18}\gamma t/\Omega_0}\,.
\label{eqn:circulatization-GW}
\eeq
Therefore, not only is the orbit stalled at $\Omega(t)\approx\Omega_0$ for a potentially long time, given in (\ref{eqn:t-float}), during which the eccentricity keeps reducing; but it also goes down at a faster rate than in the vacuum, as can be seen comparing (\ref{eqn:circulatization-floating}) with (\ref{eqn:circulatization-GW}). The longer the resonance, the more the binary is circularized.

\vskip 4pt
This result holds for co-rotating resonances with $g=\Delta m$, which are the only ones surviving in the small-$\varepsilon$ limit and usually have the largest coupling $\eta^\floq{g}$ even at moderately large eccentricities. The dynamics are different in other cases. Remaining in the equatorial co-rotating case ($\beta=0$), eccentric binaries can also undergo (usually weaker) resonances where $g\ne\Delta m$. In this case, \eqref{eqn:eccentricity-evolution-floating} has a different behavior: if $\abs{\Delta m/g}<1$, then there is a fixed point $\bar\varepsilon>0$ such that if $\varepsilon<\bar\varepsilon$ then $\varepsilon$ increases, while if $\varepsilon>\bar\varepsilon$ then $\varepsilon$ decreases. For example, for $\Delta m/g=1/2$, we have $\bar\varepsilon\approx0.46$ and the eccentricity approaches the fixed point according to
\beq
\varepsilon(t)\approx0.46+(\varepsilon_0-0.46)e^{-3.49\gamma t/\Omega_0}\,.
\label{eqn:eccentrification-floating-DeltamOverg=0.5}
\eeq
Floating resonances with $\abs{\Delta m/g}>1$ will instead circularize the binary even quicker than \eqref{eqn:circulatization-floating}. As $\varepsilon$ decreases, however, so does $Z$: eventually, the perturbation becomes too weak and the resonance stops, generically leaving the cloud in a mixed state as the inspiral resumes. This aspect will be discussed in Section~\ref{sec:resonance-breaking}.

\begin{figure} 
\centering
\includegraphics{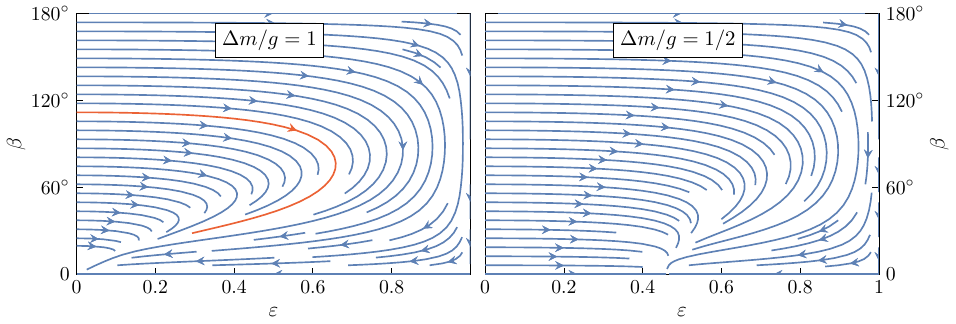}
\caption{Flow in the eccentricity-inclination plane $(\varepsilon,\beta)$ determined by equations \eqref{eqn:dimensionless-eccentricity-evolution} and \eqref{eqn:dimensionless-inclination-evolution} under the assumption that the system is on a floating orbit, i.e., $\dd\abs{c_b}^2/\dd\tau=f(\varepsilon)/B$, for two different values of $\Delta m/g$. The highlighted arrow [{\color{Mathematica4}red}] roughly depicts the trajectory followed by the system in Figure~\ref{fig:eccentric-inclined-backreacted-resonance}.}
\label{fig:streamplot_inclination-eccentricity}
\end{figure}

\begin{figure} 
\centering
\includegraphics{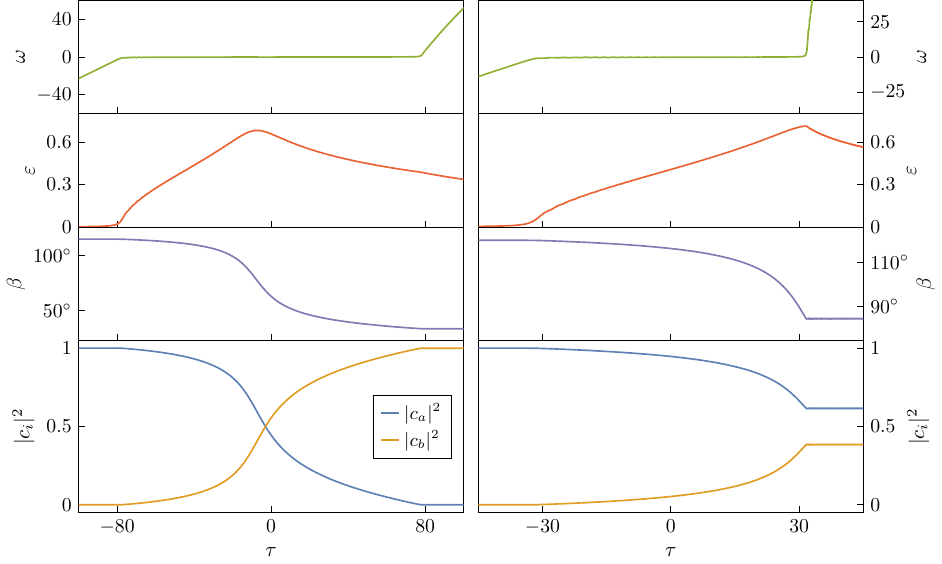}
\caption{Numerical solution of equations \eqref{eqn:dimensionless-dressed-schrodinger}, \eqref{eqn:dimensionless-omega-evolution}, \eqref{eqn:dimensionless-eccentricity-evolution} and \eqref{eqn:dimensionless-inclination-evolution} with parameters $Z=0.001$, $B=1000$, $D=4/3$ and $g=\Delta m$. For simplicity we ignore that in realistic cases $Z$ depends on the eccentricity, and we keep it constant instead. The system is initialized with eccentricity $\varepsilon_0=0$. A complete transition is achieved when the initial inclination is $\beta_0=\SI{115}{\degree}$ (\emph{left panel}), while a ``broken resonance'' is observed when $\beta_0=\SI{120}{\degree}$ (\emph{right panel}), with the float abruptly ending when \eqref{eqn:epsilon-breaking} is satisfied. In both cases, the system follows the trajectories indicated in Figure~\ref{fig:streamplot_inclination-eccentricity} until the resonance ends or breaks.}
\label{fig:eccentric-inclined-backreacted-resonance}
\end{figure}

\vskip 4pt
The possibilities described so far are a particular case of the general dynamics, which includes the evolution of the inclination $\beta$. The flow induced by equations \eqref{eqn:eccentricity-evolution-floating} and \eqref{eqn:inclination-evolution-floating} in the $(\varepsilon,\beta)$ plane is shown in Figure~\ref{fig:streamplot_inclination-eccentricity}, where the dynamics on the $x$ axis are described by equations \eqref{eqn:circulatization-floating} (\emph{left panel}) and \eqref{eqn:eccentrification-floating-DeltamOverg=0.5} (\emph{right panel}). Perhaps the most striking feature of Figure~\ref{fig:streamplot_inclination-eccentricity} is the fact that the system is violently pulled away from inclined circular orbits ($y$ axis). In fact, $\dd\varepsilon/\dd\tau$ diverges for $\varepsilon\to0$ and finite $\beta$, meaning that the validity of equations \eqref{eqn:eccentricity-evolution-floating} and \eqref{eqn:inclination-evolution-floating} must somehow break down in that limit. The explanation for this behavior is that it is inconsistent to assume that the system undergoes an adiabatic floating resonance on inclined circular orbits: eccentricity \emph{must} increase before the onset of the resonance. This is precisely the behavior observed in Figure~\ref{fig:eccentric-inclined-backreacted-resonance}, where equations \eqref{eqn:dimensionless-dressed-schrodinger}, \eqref{eqn:dimensionless-omega-evolution}, \eqref{eqn:dimensionless-eccentricity-evolution} and \eqref{eqn:dimensionless-inclination-evolution} are solved numerically starting from $\varepsilon_0=0$ and $\beta_0\ne0$. If $2\pi ZB>f(\varepsilon_0)^{3/2}$, then the system enters the floating orbit and starts to follow the trajectories shown in Figure~\ref{fig:streamplot_inclination-eccentricity}. The total ``distance'' in the $(\beta,\varepsilon)$ plane traveled by the system by the time the transition completes depends on a single dimensionless ``distance parameter,''
\beq
D\equiv\frac{B}C=\frac{\gamma\Delta t\ped{float}}{\Omega_0}\,.
\label{eqn:D}
\eeq
However, it is also possible that the transition stops before fully completing, as shown in Figure~\ref{fig:eccentric-inclined-backreacted-resonance} (\emph{right panel}). This is the subject of Section~\ref{sec:resonance-breaking}.

\subsubsection{Resonance breaking}
\label{sec:resonance-breaking}

When some parameters are allowed to vary with time, the floating orbit dynamics described in Section~\ref{sec:floating-adiabatic} and \ref{sec:floating-evolution-e-beta} feature a new phenomenon, which we call \emph{resonance breaking}, and has been shown already in Figure~\ref{fig:eccentric-inclined-backreacted-resonance} (\emph{right panel}). The goal of this section is to determine analytically under which conditions a floating resonance breaks. Three different cases of parameter variation are encountered in realistic scenarios.
\begin{enumerate}
\item The binary eccentricity $\varepsilon$ changes with time, as seen in Figure~\ref{fig:eccentric-inclined-backreacted-resonance}. The eccentricity is the only binary parameter that appears explicitly in \eqref{eqn:dimensionless-dressed-schrodinger} and \eqref{eqn:dimensionless-omega-evolution}, while a change in $\beta$ only acts through a variation of $Z$.
\item As a consequence of changing $\varepsilon$ and $\beta$, the strength of the perturbation $\eta^\floq{g}$, and thus the Landau-Zener parameter $Z$, changes as well.
\item The total mass of the cloud changes with time if state $\ket{b}$ has $(\omega_{n\ell m})\ped{I}\ne0$: as a consequence, the Schrödinger equation \eqref{eqn:dimensionless-dressed-schrodinger} is modified to
\beq
\frac\dd{\dd\tau}\!\begin{pmatrix}c_a\\ c_b\end{pmatrix}=-i\begin{pmatrix}\omega/2 & \sqrt{Z}\\ \sqrt{Z} & -\omega/2-i\Gamma\end{pmatrix}\begin{pmatrix}c_a\\ c_b\end{pmatrix}\,,
\label{eqn:schrodinger-Gamma}
\eeq
where $\Gamma \equiv (\omega_{n\ell m})\ped{I}/\sqrt{\gamma\abs{g}}$, and care must be paid in the definition of $B$.
\end{enumerate}
All three effects come with two possible signs, one of which ``weakens'' the resonance and potentially breaks it, while the other ``reinforces'' it: in the first category we have the increase of eccentricity, the decrease of $Z$ and the cloud decay ($\Gamma<0$).\footnote{The superradiant amplification of state $\ket{b}$, that is, $\Gamma>0$, is never encountered for floating resonances anyway.}

\vskip 4pt
To understand under what conditions a resonance breaks, it is insightful to study the evolution of $\omega$ during the float. To zeroth order, $\omega$ is identically zero, but Figures~\ref{fig:floating-backreacted-resonance}, \ref{fig:eccentric-backreacted-resonance}, and \ref{fig:eccentric-inclined-backreacted-resonance} hint towards a nontrivial dynamics to higher order, with small oscillatory features of varying frequency. Let us try to find an equation of motion for the sole $\omega$, in the vanilla case with $\dd\varepsilon/\dd\tau=\dd Z/\dd\tau=\Gamma=0$, where no resonance break is expected. By taking the derivative of \eqref{eqn:dimensionless-omega-evolution} and repeatedly using Schrödinger's equation, we find
\beq
\frac{\dd^2\omega}{\dd\tau^2}=-B\frac{\dd\abs{c_b}^2}{\dd\tau^2}=-2ZB(1-2\abs{c_b}^2)+\sqrt ZB(c_a^*c_b+c_ac_b^*)\omega\,.
\label{eqn:d2omegadt2}
\eeq
Remarkably, the equation of motion obeyed by $\omega$ closely resembles a harmonic oscillator whose (squared) frequency is $-\sqrt ZB(c_a^*c_b+c_ac_b^*)$. It is thus natural to study this quantity: by directly applying Schrödinger's equation, we find
\beq
\sqrt Z\frac\dd{\dd\tau}(c_a^*c_b+c_ac_b^*)=\omega\frac{\dd\abs{c_b}^2}{\dd\tau}\,.
\label{eqn:dcacbcbcadt}
\eeq
We notice that equations \eqref{eqn:d2omegadt2} and \eqref{eqn:dcacbcbcadt} form a closed system of ordinary differential equations (because in the vanilla case $\abs{c_b}^2=(\tau-\omega)/B$), through which it is possible to prove mathematically a number of interesting properties of the system, such as the fact that at small $Z$ the evolution is entirely determined by $ZB$, as thoroughly described in Section~\ref{sec:floating-adiabatic}.

\vskip 4pt
For the scope of this section it is, however, sufficient to assume that the quantity $c_a^*c_b+c_ac_b^*$ evolves slowly during a float, with a timescale of $\Delta t\ped{float}$, similar to $\abs{c_b}^2$. Equation \eqref{eqn:d2omegadt2} can then be solved in a WKB approximation as
\beq
\omega\approx\frac{2\sqrt Z(1-2\abs{c_b}^2)}{c_a^*c_b+c_ac_b^*}+\frac{A\,Z^{-1/8}B^{-1/4}}{(-c_a^*c_b-c_ac_b^*)^{1/4}}\cos\biggl(Z^{1/4}B^{1/2}\int_0^\tau\sqrt{-c_a^*c_b-c_ac_b^*}\dd\tau'+\delta\biggr)\,,
\label{eqn:omega-solution-oscillator}
\eeq
where $A$ is a constant and $\delta$ is a phase. As the fast oscillations average out, we can plug the first, non-oscillatory, term of \eqref{eqn:omega-solution-oscillator} into \eqref{eqn:dcacbcbcadt} and integrate to find $c_a^*c_b+c_ac_b^*\approx-\sqrt{1-(1-2\abs{c_b}^2)^2}$. The resulting solution for $\omega$,
\beq
\omega\approx-\frac{2\sqrt Z(1-2\abs{c_b}^2)}{\sqrt{1-(1-2\abs{c_b}^2)^2}}+\text{oscillatory terms}\,,
\eeq
is well behaved for the entire duration of the float, only diverging before ($\abs{c_b}^2=0$) or after ($\abs{c_b}^2=1$) the resonance.

\vskip 4pt
The same analytical approach can be applied to the cases mentioned above, with varying $\varepsilon$ or $Z$, or $\Gamma\ne0$. A ``master equation'', where all three effects are turned on at the same time, is derived and shown in Appendix~\ref{sec:breaKING}. Here, we find it more illuminating to study them one at a time. The outcome in realistic cases may then be approximated by only retaining the strongest of the three effects.

\vskip 4pt
When the eccentricity is not a constant, the time derivative of \eqref{eqn:dimensionless-omega-evolution} contains the additional term $\dd f(\varepsilon)/\dd\tau$. As a result, the equation of motion for $\omega$ and the expression of $c_a^*c_b+c_ac_b^*$ are both modified. The final result, which has been thoroughly checked against numerical solutions of the full system \eqref{eqn:dimensionless-dressed-schrodinger}--\eqref{eqn:dimensionless-omega-evolution}--\eqref{eqn:dimensionless-eccentricity-evolution}--\eqref{eqn:dimensionless-inclination-evolution}, is
\beq
\omega\approx\frac{\frac{\dd f(\varepsilon)}{\dd\tau}-2ZB(1-2\abs{c_b}^2)}{\sqrt{ZB^2(1-(1-2\abs{c_b}^2)^2)-(f(\varepsilon)^2-f(\varepsilon_0)^2)}}+\text{oscillatory terms}\,.
\eeq
If $\varepsilon$ increases from its initial value $\varepsilon_0$, then the denominator can hit zero before the transition is complete, and the resonance breaks. The population remaining in state $\ket{a}$ and the binary eccentricity at resonance breaking satisfy
\beq
4ZB^2(\abs{c_a}^2-\abs{c_a}^4)=f(\varepsilon)^2-f(\varepsilon_0)^2\,,
\label{eqn:epsilon-breaking}
\eeq
which can be compared with the numerical solution in Figure~\ref{fig:eccentric-inclined-backreacted-resonance} (\emph{right panel}). Despite the simplicity of \eqref{eqn:epsilon-breaking}, a numerical integration is still needed, in principle, to determine $\varepsilon$ as function of $\abs{c_a}^2$, and so whether a resonance will break. We can, however, make a simple conservative estimate by noting that the left-hand side can be at maximum $ZB^2$. If the system follows a trajectory in the $(\varepsilon,\beta)$ plane (cf.~Figure~\ref{fig:streamplot_inclination-eccentricity}) that significantly increases its eccentricity, such that
\beq
f(\varepsilon)>\sqrt ZB
\label{eqn:epsilon-breaking-simple}
\eeq
at some point, then the resonance must necessarily break.

\begin{figure} 
\centering
\includegraphics{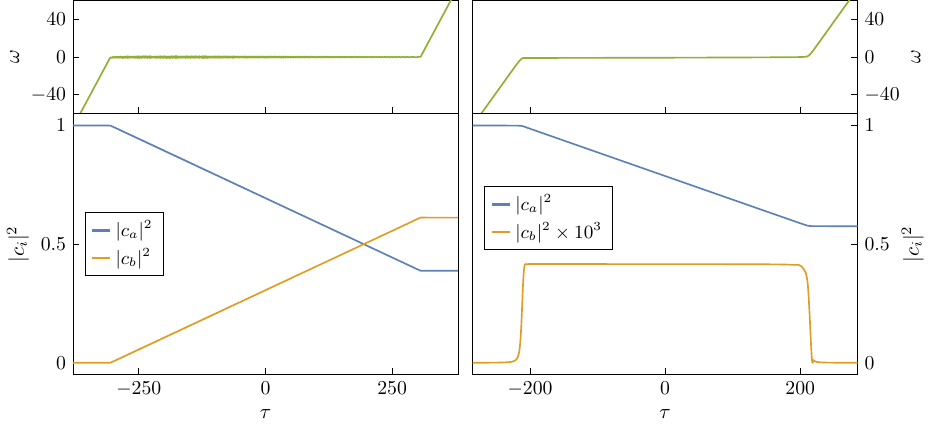}
\caption{Numerical solution of equations \eqref{eqn:dimensionless-dressed-schrodinger} and \eqref{eqn:dimensionless-omega-evolution} with (initial) parameters $Z=0.001$ and $B=1000$. A $Z$-breaking occurs when $Z$ is slowly reduced over time, with the resonance ending when \eqref{eqn:Z-breaking} is satisfied (\emph{left panel}). A $\Gamma$-breaking is observed when $Z$ is kept fixed but state $\ket{b}$ is given a nonzero decay width $\Gamma=1.2$, with the resonance ending when \eqref{eqn:gamma-breaking} is satisfied (\emph{right panel}).}
\label{fig:broken-resonance}
\end{figure}

\vskip 4pt
If, instead, $Z$ is allowed to vary while $\varepsilon$ is kept constant, then new terms appear when taking the time derivative of the Schrödinger equation (used in the second equality of \eqref{eqn:d2omegadt2}), and \eqref{eqn:omega-solution-oscillator} becomes a damped harmonic oscillator. Similar to the previous case, the resonance breaks when $c_a^*c_b+c_ac_b^*=0$, which is equivalent to
\beq
4ZB^2(\abs{c_a}^2-\abs{c_a}^4)=f(\varepsilon)^2\biggl(1-\frac{Z}{Z_0}\biggr)\,.
\label{eqn:Z-breaking}
\eeq
We illustrate this phenomenon in Figure~\ref{fig:broken-resonance} (\emph{left panel}), by solving numerically \eqref{eqn:dimensionless-dressed-schrodinger} and \eqref{eqn:dimensionless-omega-evolution} while $Z$ slowly reduces over time. Analogous considerations as before can be applied to extract from \eqref{eqn:Z-breaking} the approximate point of resonance breaking without performing a numerical integration.

\vskip 4pt
Taking into account a nonzero decay width $\Gamma$, while keeping $\varepsilon$ and $Z$ constant, requires more care. Because $\abs{c_a}^2+\abs{c_b}^2$ is no longer a constant, equation \eqref{eqn:dimensionless-omega-evolution} is now written as
\beq
\frac{\dd\omega}{\dd\tau}=f(\varepsilon)-\frac{B}{\Delta\epsilon}\biggl(\epsilon_a\frac{\dd\abs{c_a}^2}{\dd\tau}+\epsilon_b\frac{\dd\abs{c_b}^2}{\dd\tau}+2\Gamma\epsilon_b\abs{c_b}^2\biggr)=f(\varepsilon)+B\frac{\dd\abs{c_a}^2}{\dd\tau}\,,
\label{eqn:dimensionless-omega-evolution-Gamma}
\eeq
where the constant parameter $B$ is computed according to \eqref{eqn:BC}, using the value of the mass of the cloud before the start of the resonance. Furthermore, due to the modified Schrödinger equation, formula \eqref{eqn:dcacbcbcadt} becomes
\beq
\sqrt Z\frac\dd{\dd\tau}(c_a^*c_b+c_ac_b^*)=-\omega\frac{\dd\abs{c_a}^2}{\dd\tau}-\Gamma\sqrt Z(c_a^*c_b+c_ac_b^*)\,.
\label{eqn:dcacbcbcadt_Gamma}
\eeq
As we will show later (cf.~Figure~\ref{fig:tfloat_tdecay}), in almost all realistic cases state $\ket{b}$ decays much faster than the duration of the resonance, i.e., $\tau\ped{decay}\equiv(2\Gamma)^{-1}\ll B$. As a consequence, its population $\abs{c_b}^2$ during a floating orbit stays approximately constant, at a value $\abs{c_b}^2=f(\varepsilon)/(2\Gamma B)$, where the state decay is balanced by the transitions from $\ket{a}$ to $\ket{b}$. As this saturation value is typically very small, we will neglect it. Under this assumption, we can solve \eqref{eqn:dcacbcbcadt_Gamma} by plugging in the expression for $\omega$, given in \eqref{eqn:breaKING-harmonic-oscillator-solution}, obtaining
\beq
c_a^*c_b+c_ac_b^*\approx\sqrt{\frac{f(\varepsilon)(2ZB\abs{c_a}^2-f(\varepsilon)\Gamma)}{\Gamma ZB^2}}\,,
\eeq
and conclude that the resonance breaks when the remaining population in the initial state is
\beq
\abs{c_a}^2\approx\frac{f(\varepsilon)\Gamma}{2ZB}\,.
\label{eqn:gamma-breaking}
\eeq
This result is confirmed by a numerical solution of \eqref{eqn:dimensionless-dressed-schrodinger} and \eqref{eqn:dimensionless-omega-evolution} with nonzero $\Gamma$, as shown in Figure~\ref{fig:broken-resonance} (\emph{right panel}). Resonances where this quantity is larger than 1 do not exhibit a floating orbit at all, showing an ``immediate'' breaking.

\vskip 4pt
We refer to the three types of resonance breaking as $\varepsilon$-breaking, $Z$-breaking and $\Gamma$-breaking. A summary of the respective conditions is given below.

\begin{center}
\begin{tabular}{c|c|c} 
$\varepsilon$-breaking & $Z$-breaking & $\Gamma$-breaking \\ 
\hline
$f(\varepsilon)\gtrsim\sqrt ZB$ & $Z/Z_0\lesssim1-ZB^2/f(\varepsilon)^2$ & $\abs{c_a}^2\lesssim f(\varepsilon)\Gamma/(2ZB)$ \\ 
\end{tabular}
\end{center}

\subsection{Sinking orbits}
\label{sec:sinking}

\vskip 4pt
Let us now turn our attention to sinking orbits, corresponding to $B<0$, where backreaction tends to make the resonance less adiabatic. This case turns out to not be as dramatically relevant as floating orbits for the resonant history of the system. However, it is important for direct GW signatures. For this reason, we will only study the aspects of it with observational consequences.

\vskip 4pt
All the observable sinking resonances have $2\pi Z\ll1$. In this case, the final population in state $\ket{b}$, as predicted by \eqref{eqn:lz}, is very small, and this quantity is further reduced by the backreaction. In the regime where this correction is dominant, we can find a rough approximation for the total population transferred by only keeping the backreaction term in (\ref{eqn:dimensionless-omega-evolution}). Further assuming $\abs{c_a}^2\approx1$ and $\dot c_b\approx0$, we can substitute in the second component of (\ref{eqn:dimensionless-dressed-schrodinger}) and obtain $\abs{c_b}^2\approx(Z/B^2)^{1/3}$, where we assumed, for simplicity, quasi-circular orbits.\footnote{The validity of the assumption will become clear in Section \ref{sec:history}.} This result is confirmed by numerical tests, modulo a multiplicative factor: we find
\beq
B\ll -\frac1Z,\qquad\abs{c_b}^2\approx3.7\,\biggl(\frac{Z}{B^2}\biggr)^{1/3}\,.
\label{eqn:sinkingpopulation}
\eeq
This formula is accurate for $2\pi Z\ll1$ and provides a slight underestimate of the final population for moderately large $Z$.

\vskip 4pt
Sinking orbits backreact on the orbit by increasing both the orbital frequency and the binary eccentricity, as shown in Figure~\ref{fig:eccentric-backreacted-resonance} and \ref{fig:omega-eccentricity} (\emph{right panels}). At the same, time both $\Omega$ and $\varepsilon$ feature long-lived oscillations after the resonance. These oscillations slowly die out, so that a ``jump'' in the $\Omega$ and $\varepsilon$ is the only mark left after a long time. The non-monotonic behavior of $\Omega$ was already observed in \cite{Baumann:2019ztm}, where it was also speculated that sinking orbits could yield large eccentricities (becoming ``kicked orbits''). Our results confirm that the oscillations are not an artifact of having considered quasi-circular orbits and further show that the increase of the eccentricity is also not monotonic. However, for the realistic cases analyzed in Section~\ref{sec:history}, the increase in eccentricity due to sinking orbits turns out to be negligible.

\section{Three types of resonances}
\label{sec:types-of-resonances}

Resonances can be divided in three distinct categories, depending on the energy splitting between the two states, as computed from (\ref{eq:eigenenergy}). \emph{Hyperfine} resonances occur between states with same $n$ and $\ell$ but different $m$; they have the smallest energy splitting and thus occur the earliest in the inspiral, as the corresponding resonant orbital frequency is smallest. Then, \emph{fine} (same $n$, different $\ell$) and \emph{Bohr} resonances (different $n$) follow, the latter having the largest splittings. The tools developed in Section~\ref{sec:resonance-pheno} apply to all of them: the character of a resonance is only determined by the parameters $2\pi Z$ and $B$; its impact on eccentricity and inclination is quantified by $D$, and its duration (in case it is a floating resonance) is $\Delta t\ped{float}$. In principle, the recipe to determine the co-evolution of the binary and the cloud is clear: (1) pick the earliest resonance, (2) determine its character and backreaction by computing $Z$, $B$, $D$, and $\Delta t\ped{float}$, (3) update the state of binary and cloud accordingly, and (4) move to the next resonance and repeat. We will indeed execute this algorithm in Section~\ref{sec:history}. To be as generic as possible and explore a wide parameter space, it will prove useful to find the scalings of the relevant quantities with $M$, $M\ped{c}$, $q$, $\alpha$, and $\tilde a$. Different types of resonances have different scalings, so we analyze them here systematically.

\subsection{Hyperfine resonances}
\label{sec:hyperfine}

Let us start with hyperfine resonances. From (\ref{eq:eigenenergy}), we see that the energy splitting (and thus the resonant frequency) scales as $\Omega_0\propto M^{-1}\alpha^6\tilde a$. The corresponding orbital separation is $R_0\propto M\alpha^{-4}\tilde a^{-2/3}$. This strong $\alpha$-dependence places hyperfine resonances at distances parametrically much larger than the cloud's size. At such large orbital separations, the cloud's ionization is very inefficient, and thus the only significant mechanism for energy loss is the GW emission. As this too is a very weak effect, other phenomena might potentially be relevant, including astrophysical interactions connected to the binary formation mechanism. We will postpone the discussion of these complications to Section~\ref{sec:history} and assume for now that formula (\ref{eqn:gamma_gws}) applies, giving a chirp rate of $\gamma\propto qM^{-2}\alpha^{22}\tilde a^{11/3}$. This information is already enough to determine the scaling of three key quantities:
\beq
B\propto\frac{M\ped{c}}Mq^{-3/2}\alpha^{-4}\tilde a^{-1/2}\,,\qquad \Delta t\ped{float}\propto M\ped{c}q^{-2}\alpha^{-15}\tilde a^{-7/3}\,,\qquad D\propto \frac{M\ped{c}}Mq^{-1}\alpha\tilde a^{1/3}\,.
\label{eqn:hyperfine-B-tfloat-exponent}
\eeq

\vskip 4pt
The scaling of the Landau-Zener parameter $Z$ depends instead on the overlap coefficient $\eta^\floq{g}$. Given the hierarchy of length scales, $R_0\gg r\ped{c}$, the ``inner'' term in (\ref{eqn:F(r)}) dominates the radial integral $I_r$. At fixed $\ell_*\ne1$, we thus have
\beq
\begin{split}
\eta^\floq{g}&\propto q\alpha\,d^\floq{\ell_*}_{\Delta m,g}(\beta)\, I_r=q\alpha\,d^\floq{\ell_*}_{\Delta m,g}(\beta)\int_0^\infty \frac{r^{\ell_*}}{R_0^{\ell_*+1}}R_{n\ell}(r)^2r^2\dd r\\
&\propto M^{-1}q\alpha^{2\ell_*+5}\tilde a^{2(\ell_*+1)/3}d^\floq{\ell_*}_{\Delta m,g}(\beta)\,,
\end{split}
\eeq
and so
\beq
Z\propto q\alpha^{4\ell_*-12}\tilde a^{(4\ell_*-7)/3}\bigl(d^\floq{\ell_*}_{\Delta m,g}(\beta)\bigr)^2\,.
\eeq
The dipole $\ell_*=1$ is an exception for two reasons: (a) its inner term in (\ref{eqn:F(r)}) vanishes, (b) its ``outer'' term is not simply $r^{\ell_*}/R_*^{\ell_*+1}$. However, hyperfine resonances connect states with same $\ell$: from the selection rule \eqref{eqn:S2}, only even values of $\ell_*$ contribute. We can thus safely ignore the dipole. The rest of the multipole expansion can be seen as a power series in the small parameter $r\ped{c}/R_0$, the smallest $\ell_*$ giving the strongest contribution. Because selection rules require $\ell_*\ge \abs{g}=-g$,\footnote{Strictly speaking, this constraint only applies on circular orbits. In general, the same inequality applies to $g_\beta$ instead.} a resonance with a given value of $g$ will be dominated by $\ell_*=-g$. The only two cases we will encounter in Section~\ref{sec:history} are
\begin{align}
\label{eqn:hyperfine-g=-2-Z}
g=-2&\qquad Z\propto q\alpha^{-4}\tilde a^{1/3}\bigl(d^\floq{2}_{\Delta m,g}(\beta)\bigr)^2\,,\\
\label{eqn:hyperfine-g=-4-Z}
g=-4&\qquad Z\propto q\alpha^4\tilde a^{3}\bigl(d^\floq{4}_{\Delta m,g}(\beta)\bigr)^2\,.
\end{align}
Furthermore, the assumption $\ell_*=-g$ allows us to write the explicit expression for the angular dependence of $Z$ as
\beq
d^\floq{-g}_{\Delta m,g}(\beta)\propto\sin^{\Delta m-g}(\beta/2)\cos^{-\Delta m-g}(\beta/2)\,.
\label{eqn:dgg}
\eeq

\subsection{Fine resonances}

Most of the assumptions made for hyperfine resonances work in the fine case too. The resonant frequency now scales as $\Omega_0\propto M^{-1}\alpha^5$ and, similar to before, we arrive to
\beq
B\propto\frac{M\ped{c}}Mq^{-3/2}\alpha^{-7/2}\,,\qquad \Delta t\ped{float}\propto M\ped{c}q^{-2}\alpha^{-38/3}\,,\qquad D\propto \frac{M\ped{c}}Mq^{-1}\alpha^{2/3}\,.
\eeq
The scaling of the overlap coefficient reads $\eta^\floq{g}\propto qM^{-1}\alpha^{(4\ell_*+13)/3}$, and we get
\beq
Z\propto q\alpha^{(8\ell_*-29)/3}\bigl(d^\floq{\ell_*}_{\Delta m,g}(\beta)\bigr)^2\,.
\eeq
The main difference with the previous case resides in the possible values of $\ell_*$. Fine resonances connect states with different values of $\ell$, and most of the cases we will study in Section~\ref{sec:history} will have odd values of $\ell_*$. For $g=-3$, all the previous arguments apply and the octupole $\ell_*=3$ is the dominant contribution. For $g=-1$, the extreme weakness of the dipole at large distances again leaves the octupole as the most important term; because now $\ell_*\ne-g$, however, the angular dependence will have a form different from (\ref{eqn:dgg}), which we will describe on a case-by-case basis in Section~\ref{sec:history}. There is one further exception to this: if $\ell_a+\ell_b=1$, then the selection rule \eqref{eqn:S3} forbids all $\ell_*\ge2$. Only in this case (corresponding to the $\ket{211}\to\ket{200}$ resonance) the dipole is entirely responsible for the coupling between the two states. Its anomalous expression (\ref{eqn:F(r)}) endows $\eta^\floq{g}$ (and thus $Z$) with a non-power-law dependence on $\alpha$: given the peculiarity of this case, we will treat it explicitly in Appendix~\ref{sec:211_200}.

\subsection{Bohr resonances}
\label{sec:Bohr_res}

\begin{figure} 
\centering
\includegraphics{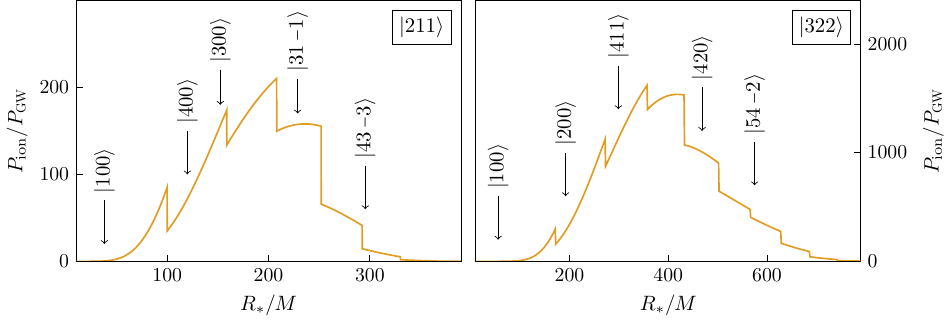}
\caption{Position of a few selected Bohr resonances, compared to $P\ped{ion}/P_\slab{gw}$, i.e., the ratio of the ionization power to the power emitted in GWs, shown here for a circular counter-rotating orbit and for a cloud in the $\ket{211}$ (\emph{left panel}) or $\ket{322}$ (\emph{right panel}) state. We assumed $M\ped{c}/M=0.01$ and $\alpha=0.2$, but the relative position of the resonances and the shape of the curve do not depend on the parameters.}
\label{fig:ionization-bohr-resonances}
\end{figure}

Bohr resonances are a different story. States with different principal quantum number $n$ have different energies to leading order, meaning that the resonant orbits are placed at distances comparable to the cloud's size. There is no parametric separation between the two, as now $R_0\propto M\alpha^{-2}\propto r\ped{c}$. At these orbital distances, the cloud's ionization is generally a more effective mechanism for energy loss than GWs. We prove this point in Figure~\ref{fig:ionization-bohr-resonances}, where the position of several Bohr resonances is shown on top of the ionization-to-GWs power ratio, computed as in \cite{Baumann:2021fkf,Baumann:2022pkl,Tomaselli:2023ysb} on circular orbits. This latter quantity scales as
\beq\label{eqn:P_ion}
\frac{P\ped{ion}}{P_\slab{gw}}\bigg|_{R_*=R_0}\propto\frac{M\ped{c}}M\alpha^{-5}\,.
\eeq
With the possible exception of transitions to $\ket{100}$, as they happen extremely late in the inspiral, Bohr resonances and ionization thus happen \emph{at the same time}. This observation raises two points.
\begin{enumerate}
\item Formula (\ref{eqn:gamma_gws}) for the chirp rate $\gamma$ is no longer accurate, as ionization must now be included.
\item The derivation of the expression for $P\ped{ion}$ laid down in \cite{Baumann:2021fkf} assumes that the system is away from bound-to-bound state resonances.
\end{enumerate}
In Appendix~\ref{sec:ion-at-resonance} we extend the framework of \cite{Baumann:2021fkf} to describe the ionization of a system actively in resonance. Although this requires the addition of new terms, their effect is generally negligible for realistic parameters. It is thus a good approximation to simply adjust the value of $\gamma$ by a factor $1+P\ped{ion}/P_\slab{gw}\approx P\ped{ion}/P_\slab{gw}$, where $P\ped{ion}$ is computed as in \cite{Baumann:2021fkf,Baumann:2022pkl,Tomaselli:2023ysb}. The last approximation holds whenever $P\ped{ion}\gg P_\slab{gw}$ and is always satisfied, unless the resonance involves $\ket{100}$ or the value of $\alpha$ is exceptionally large.

\vskip 4pt
Under these assumptions, we arrive to
\beq\label{eqn:B_bohr}
B\propto\sqrt{\frac{M\ped{c}}M}q^{-3/2}\,,\qquad \Delta t\ped{float}\propto Mq^{-2}\alpha^{-3}\,,\qquad D\propto \frac{M\ped{c}}Mq^{-1}\,.
\eeq
These quantities now also have a $\beta$-dependence, due to $P\ped{ion}$ having different values for different inclinations. However, we will see in Section~\ref{sec:history} that this detail is not relevant, so we neglect it here. As for the overlap $\eta^\floq{g}$, there is now no clear hierarchy of multipoles. Luckily, $R_0$ has the same $\alpha$-scaling as the argument of the hydrogenic wavefunctions $R_{n\ell}$: with an appropriate change of variable, we can show that
\beq
\eta^\floq{g}\propto M^{-1}q\alpha^3d^\floq{\ell_*}_{\Delta m,g}(\beta)\,.
\label{eqn:eta-bohr}
\eeq
The $\beta$-dependence in (\ref{eqn:eta-bohr}) can be written in terms of a Wigner small $d$-matrix only when there is a single value of $\ell_*$ that contributes. As this is the case for many of the Bohr resonances we will encounter in Section~\ref{sec:history}, we keep that factor explicit here. Finally, the Landau-Zener parameter scales as
\beq\label{eqn:Z_bohr}
Z\propto \frac{M}{M\ped{c}}q\bigl(d^\floq{\ell_*}_{\Delta m,g}(\beta)\bigr)^2\,.
\eeq
One particularly interesting aspect of Bohr resonances is the disappearance of any $\alpha$-dependence from the Landau-Zener parameter $Z$ and from the backreaction $B$. This is in contrast with the steep power-laws found for hyperfine and fine resonances, and it means that the character of Bohr resonances is much more \emph{universal}.

\section{Resonant history of the cloud}
\label{sec:history}

In this section we draw a consistent picture of the co-evolution of the cloud and the binary, using the tools developed in Sections~\ref{sec:resonance-pheno} and \ref{sec:types-of-resonances}. Assuming a well-motivated initial state of the cloud (generally $\ket{211}$ or $\ket{322}$), an astrophysically relevant range for $\alpha$ (see equations \eqref{eqn:alpha-211} and \eqref{eqn:alpha-322}), and small $q$, the plethora of phenomena described in Section~\ref{sec:resonance-pheno} only occur in recognizable and relatively simple patterns. These constitute the ``realistic'' cases, which we systematically explore in this section, with the goal of understanding the state of the system by the time it becomes observable: for example, when it enters the LISA band. First, we discuss the generic behavior of the different types of resonances in Section \ref{sec:generalB}; then, in Section \ref{sec:evolution-211} and \ref{sec:evolution-322}, we study explicitly the history for a cloud initialized in the state $\ket{211}$ or $\ket{322}$.

\subsection{General behavior}
\label{sec:generalB}

The initial state $\ket{a}=\ket{n_a\ell_am_a}$ of the cloud, populated by superradiance, generally has $m_a=\ell_a=n_a-1$. Within the multiplet of states $\ket{n_a\ell_am}$ with $m\le m_a$, this is the one with highest energy, as can be readily seen from (\ref{eq:eigenenergy}). Hyperfine resonances, which occur the earliest in the inspiral, thus necessarily have $\Delta\epsilon<0$ and are of the floating type. To understand their behavior, it is important to keep in mind a few key points.
\begin{description}
\item[Adiabaticity.] The first question to answer is whether a given hyperfine resonance is adiabatic or not. We can apply the results of Section~\ref{sec:floating-adiabatic}. If $2\pi ZB>f(\varepsilon)^{3/2}$ then the resonance is adiabatic: the binary starts evolving as described in Section~\ref{sec:floating-evolution-e-beta} until the transition completes after a time $\Delta t\ped{float}$, or the resonance breaks due to any of the conditions derived in Section~\ref{sec:resonance-breaking}. Almost all hyperfine resonances turn out to be adiabatic in the entire parameter space, except in a narrow interval of almost counter-rotating inclinations, say $\pi-\delta_1<\beta\le\pi$, where $\delta_1$ is the size of the interval. This is because, on floating orbits, the resonance condition $\Omega_0^\floq{g}=\Delta\epsilon/g$ forces $g$ to be negative; on the other hand, $\Delta m=m-m_a<0$, and from (\ref{eqn:dgg}) we see that for $\beta\to\pi$ the parameter $Z$ goes to zero as a (high) power of $\cos(\beta/2)$. The explicit determination of the angle $\delta_1$ as function of the parameters will be performed in Sections~\ref{sec:evolution-211} and~\ref{sec:evolution-322}.

\item[Cloud's decay and $\Gamma$-breaking.] After saturation of the dominant superradiant mode $\ket{n_a\ell_am_a}$, all states of the multiplet $\ket{n_a\ell_am}$ with $m\ne m_a$ have $\Im(\omega)<0$, meaning that they decay back in the BH with an $e$-folding time $t\ped{decay}\equiv\abs{2\Im(\omega_{n_a\ell_am})}^{-1}$. It is thus necessary to compare $\Delta t\ped{float}$ and $t\ped{decay}$. One of the most important results of this work is the following: for intermediate or extreme-mass ratios, and typical values of $M\ped{c}$ and $\alpha$, the decay timescale $t\ped{decay}$ is many orders of magnitude smaller than floating timescale, $\Delta t\ped{float}$. It is not easy to prove this statement in full generality, due to the complicated dependence of $t\ped{decay}$ on the parameters. Nevertheless, for small $\alpha$ and $\tilde a$, using the Detweiler approximation \cite{PhysRevD.22.2323} and the results from Section~\ref{sec:hyperfine}, we have
\beq
\frac{t\ped{decay}}{\Delta t\ped{float}}\propto\frac{M}{M\ped{c}}q^2\alpha^{10-4\ell_a}\tilde a^{4/3}\,,
\label{eqn:t-decay-t-float}
\eeq
where $\tilde a\propto\alpha$ at the superradiant threshold. For $\alpha\to0$ and small enough values of $\ell_a$, this ratio becomes very small. In fact, for small $q$, any possible value of $\alpha$ results in $t\ped{decay}\ll\Delta t\ped{float}$. A more detailed comparison is given in Figure~\ref{fig:tfloat_tdecay}, where $t\ped{decay}$ is computed numerically through Leaver's \cite{Leaver:1985ax,Cardoso:2005vk,Dolan:2007mj,Berti:2009kk} and Chebyshev's \cite{Baumann:2019eav} methods for various values of $\alpha$, and the spin $\tilde a$ is set to correspond to the boundary of the BH superradiant region.

This result has a dramatic consequence: hyperfine transitions are never able to change the state of the cloud. Instead, the portion that is transferred to state $\ket{b}$ decays immediately back into the BH.\footnote{\label{fn:bh-parameters}As a consequence, the mass and spin of the BH change. Our framework is not able to capture this effect, which we accordingly ignore in this work.} The analysis of Section~\ref{sec:resonance-breaking} then applies, and the resonance $\Gamma$-breaks when the fraction of the cloud remaining in state $\ket{a}$ falls below the threshold determined in \eqref{eqn:gamma-breaking}. In a relatively large portion of parameter space, generally around counter-rotating orbits, that formula returns $\abs{c_a}^2>1$, meaning that the resonance $\Gamma$-breaks immediately. The outcome is effectively similar to a non-adiabatic resonance, that never even starts the floating phase. Similar to before, we will define an angular interval $\pi-\chi_1<\beta\le\pi$, within which the resonance is not effective. The $\varepsilon$-breaking and $Z$-breaking are instead less relevant for realistic parameters.

\begin{figure} 
\centering
\includegraphics{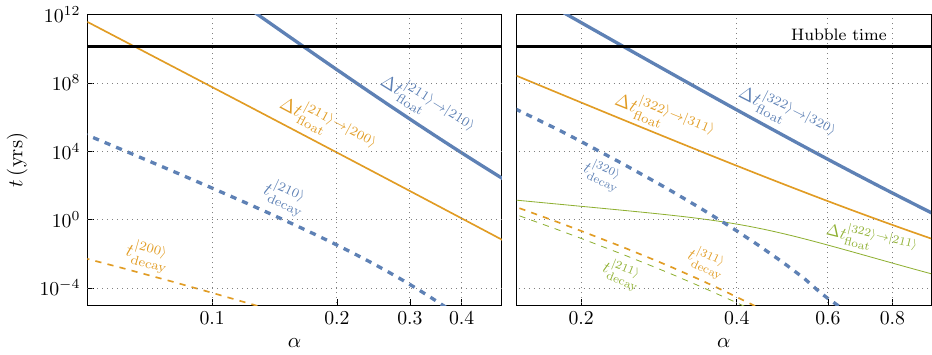}
\caption{Floating timescale $\Delta t\ped{float}$ (solid lines), compared to the decay timescale $t\ped{decay}$ (dashed lines) of the final state, for some selected resonances. We use benchmark parameters and determine the decay rate independently through Leaver's continued fraction method \cite{Leaver:1985ax,Cardoso:2005vk,Dolan:2007mj,Berti:2009kk} and the Chebyshev method in \cite{Baumann:2019eav}. Two resonances for a $\ket{211}$ initial state are shown (\emph{left panel}), namely $\ket{211}\to\ket{210}$ [{\color{Mathematica1}blue}] and $\ket{211}\to\ket{200}$ [{\color{Mathematica2}orange}]. Similarly, three resonances for a $\ket{322}$ initial state are shown (\emph{right panel}), namely $\ket{322}\to\ket{320}$ [{\color{Mathematica1}blue}], $\ket{322}\to\ket{311}$ [{\color{Mathematica2}orange}] and $\ket{322}\to\ket{211}$ [{\color{Mathematica3}green}]. The thick, normal and thin lines indicate hyperfine, fine or Bohr resonances respectively. Note the Bohr resonance falling outside of the ionization regime for large $\alpha$, changing the scaling of $\Delta t\ped{float}$ from $\alpha^{-3}$ to $\alpha^{-8}$, as predicted by \eqref{eqn:B_bohr} and~\eqref{eqn:P_ion}.}
\label{fig:tfloat_tdecay}
\end{figure}

\item[The strongest resonance.] As shown in Section~\ref{sec:eccentric-inclined-resonances}, on eccentric and inclined orbits a resonance between two given states is excited at many different orbital frequencies, depending on the value\footnote{As briefly mentioned in Section~\ref{sec:eccentric-inclined-resonances}, two separate indices, say $g_\varepsilon$ and $g_\beta$ are necessary when both eccentricity and inclination are not zero. However, this technicality is not crucial in understanding the history of the system.} of $\abs{g}=1,2,3,\ldots$ The strength of the coupling also depends on $\varepsilon$ and $\beta$. Keeping track of so many different resonances would be very complicated. However, the hierarchy $t\ped{decay}\ll\Delta t\ped{float}$ implies that as soon as an adiabatic floating resonance is encountered (and does not break early), the cloud is destroyed. This means that studying the ``strongest'' resonance (the one that destroys the cloud in the largest portion of parameter space) actually suffices to determine the fate of the cloud.

Up to moderate values of the eccentricity, the coupling $\eta^\floq{g}$ that remains nonzero in the limit of circular orbit is much larger than all the others. We can then approximate the ``strongest resonance'' by ignoring eccentricity altogether. Regarding inclined orbits instead, we observe that higher values of $g$ require contributions from higher values of the multipole index $\ell_*$: at the separations of hyperfine resonances, the lowest value of $\ell_*$ (typically the quadrupole $\ell_*=2$) produces the strongest coupling. Given two states, we will then study the resonance with the smallest value of $\abs{g}$.
\end{description}

Applying the previous considerations to each possible hyperfine resonance, we are able to determine whether the cloud is destroyed in the process or survives to later stages of the inspiral. However, the binary might be able to ``skip'' hyperfine resonances for other reasons. This is because some of them are placed at extremely large binary separations: typically $R_*/M\gtrsim\mathcal O(10^3)$ for a $\ket{211}$ initial state, and $R_*/M\gtrsim\mathcal O(10^4-10^5)$ for $\ket{322}$. These distances are large enough that not only other kinds of astrophysical interactions may play a role, but their presence is in some cases necessary, in order to bring the binary close enough for the merger to happen within a Hubble time. Quantitatively, for a quasi-circular inspiral, the initial separation as function of the time-to-merger $t_0$ is given by
\beq
\frac{R_*}M=\SI{2.3e+4}{}\,\biggl(\frac{t_0}{\SI{e+10}{yrs}}\biggr)^{1/4}\biggl(\frac{M}{10^4M_\odot}\biggr)^{-1/4}\Bigl(\frac{q}{10^{-3}}\Bigr)^{1/4}\,.
\eeq
In other words: if we want the binary to merge within a Hubble time, we might be forced to assume that it ``starts'' its evolution too close for hyperfine resonances to be encountered, especially for a cloud initialized in the $\ket{322}$ state. This can be achieved by a variety of formation mechanisms, including dynamical capture \cite{Amaro-Seoane:2012lgq,LISA:2022yao} and \emph{in-situ} formation \cite{LISA:2022yao,Stone:2016wzz,Bartos:2016dgn,McKernan_2018,Levin:2006uc}.

\vskip 4pt
If the system is able to skip through hyperfine resonances because they are either all non-adiabatic, or they $\Gamma$-break early, or the binary is formed at small enough separations, then the cloud can be present when fine resonances are encountered. Their phenomenology is largely similar to hyperfine ones, as they too are all of the floating type. We defer the discussion of some state-dependent aspects to Sections~\ref{sec:evolution-211} and~\ref{sec:evolution-322}. For the purpose of the present general discussion, it suffices to say that, once again, the cloud can survive this stage if $\pi-\delta_2<\beta\le\pi$ (for some angle $\delta_2$ to be determined), if the resonance $\Gamma$-breaks early in an interval $\pi-\chi_2<\beta\le\pi$, or if the binary is formed $\emph{in situ}$ at very small radii.

\vskip 4pt
Finally, if the cloud makes it to this point, it becomes potentially observable: the ``Bohr region'' can be in the LISA band and is rich of signatures of the cloud. These come in the form of ionization and Bohr resonances, the vast majority of which are sinking and non-adiabatic. State-dependent details will be discussed in Sections~\ref{sec:evolution-211} and~\ref{sec:evolution-322} and a summary of the observational signatures will be given in Section~\ref{sec:observational-signatures}. A diagrammatic representation of the three stages of the resonant history is shown in Figure~\ref{fig:history_211}.

\vskip 4pt
As a concluding remark, we note that the results derived here and in Section~\ref{sec:resonance-pheno} are specific to resonances involving two states only. We have explicitly checked that this is the case for the resonances discussed in the next sections, so we apply the results of Section~\ref{sec:resonance-pheno} without further modification.

\subsection{Evolution from a $\ket{211}$ initial state}
\label{sec:evolution-211}

The $\ket{211}$ state is the fastest-growing superradiant mode and represents therefore a natural assumption for the initial state of the cloud. The requirements that the superradiant amplification takes place, and does so on timescales no longer than a Gyr, set a constraint on $\alpha$:
\beq
\label{eqn:alpha-211}
0.02\biggl(\frac{M}{10^4M_\odot}\biggr)^{1/9}\lesssim\alpha<0.5\,.
\eeq
Once grown, the cloud will decay in GWs with a rate roughly proportional to $M\ped{c}^2\alpha^{14}$, assuming the scalar field is real. The resulting decay of $M\ped{c}$ is polynomial, rather than exponential in time; as such, we will not impose a further sharp bound on $\alpha$, and treat $M\ped{c}/M$ as an additional free parameter.

\vskip 4pt
There are two possible hyperfine resonances, with the states $\ket{210}$ and $\ket{21\,\minus1}$. Following the line of reasoning laid down in Section~\ref{sec:generalB}, we ignore the fact that the same resonances can be triggered at multiple points if the orbit is eccentric. Both resonances are then mediated by $g=-2$ and they are positioned at
\begin{align}
\ket{211}\overset{g=-2}{\longrightarrow}\ket{210} & \qquad \frac{R_0}M=\SI{8.3e+3}{}\,\Bigl(\frac\alpha{0.2}\Bigr)^{-4}\Bigl(\frac{\tilde a}{0.5}\Bigr)^{-2/3}\,,\\[8pt]
\ket{211}\overset{g=-2}{\longrightarrow}\ket{21\,\minus1} & \qquad \frac{R_0}M=\SI{5.2e+3}{}\,\Bigl(\frac\alpha{0.2}\Bigr)^{-4}\Bigl(\frac{\tilde a}{0.5}\Bigr)^{-2/3}\,,
\end{align}
where the value of the spin should be set equal to the threshold of superradiant instability of $\ket{211}$, that is, $\tilde a\approx4\alpha/(1+4\alpha^2)$. Both resonances become non-adiabatic in an interval $\pi-\delta_1<\beta\le\pi$, with the strongest constraint on $\delta_1$ given by $\ket{211}\to\ket{210}$. The value of $\delta_1$ is determined from \eqref{eqn:2piZB-epsilon0}: this means setting $2\pi ZB=f(\varepsilon_0)^{3/2}$, where $\varepsilon_0$ is the eccentricity at the onset of the resonance, and solving for $\beta$ as function of the parameters. Making use of the relations (\ref{eqn:hyperfine-B-tfloat-exponent}), (\ref{eqn:hyperfine-g=-2-Z}) and (\ref{eqn:dgg}), and evaluating numerically the overlap $\eta^\floq{2}$ between the two states, we find
\beq
\delta_1=\SI{7.5}{\degree}\,\biggl(\frac{M\ped{c}/M}{10^{-2}}\biggr)^{-1/6}\biggl(\frac{q}{10^{-3}}\biggr)^{1/12} \biggl(\frac{\alpha}{0.2}\biggr)^{4/3} \biggl(\frac{\tilde{a}}{0.5}\biggr)^{1/36}f(\varepsilon_0)^{1/4}\,.
\label{eqn:211-delta1}
\eeq
Although $\ket{211}\to\ket{210}$ is also non-adiabatic in a neighbourhood of $\beta=0$, such a co-rotating binary would still encounter the adiabatic floating resonance $\ket{211}\to\ket{21\,\minus1}$ later, so that the only ``safe'' inclinations are in the neighborhood of counter-rotating determined in (\ref{eqn:211-delta1}).

\begin{figure} 
\centering
\includegraphics{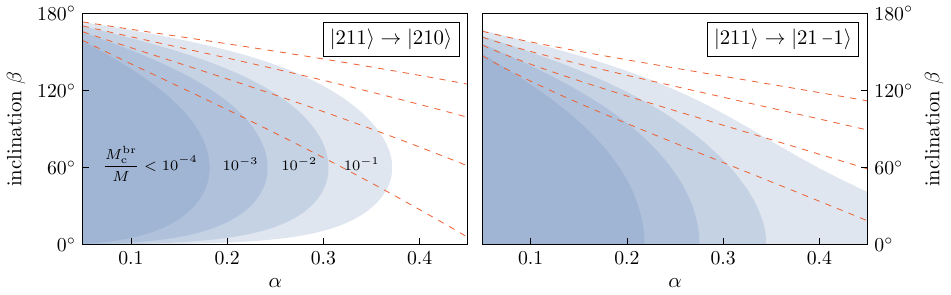}
\caption{Mass of the cloud $M\ped{c}\ap{br}$ at resonance $\Gamma$-breaking, as a function of $\alpha$ and $\beta$, for the two hyperfine resonances from the initial state $\ket{211}$. The mass of the cloud decreases during the resonance from its initial value $M\ped{c}$, and the resonance breaks when the value $M\ped{c}\ap{br}$ is reached. Values $M\ped{c}\ap{br}>M\ped{c}$ indicate that the resonance breaks immediately as it starts. The contours [{\color{Mathematica1}blue}] are calculated on circular orbits, as this gives a good approximation for the strongest constraint on $M\ped{c}\ap{br}$ even when overtones, due to orbital eccentricity, (i.e., higher values of $\abs{g}$ for the resonance between two given states) are taken into account. Due to the inaccuracy of the analytical approximations for the decay width $(\omega_{211})\ped{I}$, especially at large $\alpha$, we have determined the contours with Leaver's \cite{Leaver:1985ax,Cardoso:2005vk,Dolan:2007mj,Berti:2009kk} and Chebyshev \cite{Baumann:2019eav} methods. The dashed lines [{\color{Mathematica4}red}] are analytical approximations to the blue contours in the proximity of $\beta=\pi$, based on \eqref{eqn:211_chi_1}.}
\label{fig:deltas211}
\end{figure}

\vskip 4pt
Having determined when hyperfine resonances can be adiabatic, we now calculate where they break, using the results of Section~\ref{sec:resonance-breaking}. As anticipated in Section~\ref{sec:generalB}, the $\Gamma$-breaking is the most relevant mechanism of resonance breaking. To assess its impact, we observe that, because $B\propto M\ped{c}$, equation \eqref{eqn:gamma-breaking} can be written as a relation for the final mass of the cloud at resonance breaking, $M\ped{c}\ap{br}=M\ped{c}\abs{c_a}^2$, which can be computed as a function of $\alpha$ and $\beta$. If $M\ped{c}\ap{br}>M\ped{c}$ is found, then the resonance breaks immediately as it starts, as if it was non-adiabatic. The value of $M\ped{c}\ap{br}$ as function of $\alpha$ and $\beta$ is shown in in Figure~\ref{fig:deltas211}. Note that, in principle, the resonance always $\Gamma$-breaks before the cloud is completely destroyed, but its observational impact becomes negligible when $M\ped{c}\ap{br}$ is too small.

\vskip 4pt
The combined constraints due to the $\Gamma$-breaking of $\ket{211}\to\ket{210}$ and $\ket{211}\to\ket{21\,\minus1}$ imply that the cloud survives in a neighborhood of $\beta=\pi$, say $\pi-\chi_1<\beta<\pi$, similar to what we found for the adiabaticity of the resonances. An analytical approximation of $\chi_1$ for $\ket{211}\to\ket{210}$ based on Detweiler's formula \cite{PhysRevD.22.2323} is
\beq
\chi_1\approx\SI{38}{\degree}\biggl(\frac{M\ped{c}\ap{br}/M}{10^{-2}}\biggr)^{-1/6}\biggl(\frac\alpha{0.2}\biggr)^{7/6}\biggl(\frac{\tilde a}{0.5}\biggr)^{-5/18}f(\varepsilon\ped{br})^{1/6}\,,
\label{eqn:211_chi_1}
\eeq
where $\varepsilon\ped{br}$ is the eccentricity at resonance breaking. Formula~\eqref{eqn:211_chi_1} significantly underestimates the result for large $\alpha$, as shown in Figure~\ref{fig:deltas211}. Because $\chi_1>\delta_1$, this angular interval overwrites \eqref{eqn:211-delta1} as the portion of parameter space where the cloud survives hyperfine resonances.

\vskip 4pt
Finally, we check whether hyperfine resonances can $\varepsilon$-break or $Z$-break. Both $\varepsilon$ and $Z$ can vary significantly during the float, so we use the relation \eqref{eqn:epsilon-breaking} as an order-of-magnitude estimate. For generic values of the inclination, both hyperfine resonances have
\beq
\sqrt ZB\sim10^6\biggl(\frac{M\ped{c}/M}{10^{-2}}\biggr)\biggl(\frac{q}{10^{-3}}\biggr)^{-1}\biggl(\frac\alpha{0.2}\biggr)^{-6}\biggl(\frac{\tilde a}{0.5}\biggr)^{-1/3}\,.
\eeq
The resonances $\varepsilon$-breaks if $f(\varepsilon)=\sqrt ZB$, which is only satisfied at very high eccentricities, not smaller than $0.95$ for typical parameters. Such extreme eccentricities are only reachable if the initial inclination is very close to $\beta=\pi$, as can be seen from Figure~\ref{fig:streamplot_inclination-eccentricity}. But, as proved in \eqref{eqn:211-delta1} and \eqref{eqn:211_chi_1}, near-counter-rotating binaries do not undergo floating orbits at all, due to the resonances being either non-adiabatic or $\Gamma$-breaking immediately. As for the $Z$-breaking, one can conservatively ignore the term $Z/Z_0$ in \eqref{eqn:Z-breaking}, falling back to the same relation as \eqref{eqn:epsilon-breaking}.

\vskip 4pt
We conclude that the survival of the cloud to later stages of the inspiral is exclusively determined by the $\Gamma$-breaking. If the binary is outside the regions colored in Figure~\ref{fig:deltas211}, and computed in \eqref{eqn:211_chi_1}, it encounters the only possible fine resonance:
\beq
\ket{211}\overset{g=-1}{\longrightarrow}\ket{200}\qquad\frac{R_0}M=\SI{3.4e+2}{}\,\Bigl(\frac\alpha{0.2}\Bigr)^{-10/3}\,,
\label{eqn:R_0-211-200}
\eeq
whose angular dependence is determined through (\ref{eqn:dgg}) as usual. This resonance, however, has anomalous behavior for two reasons:
\begin{enumerate}
\item it is entirely mediated by the dipole $\ell_*=1$;
\item depending on the value of $\alpha$, it may fall inside the ionization regime ($P\ped{ion}\gtrsim P_\slab{gw}$) despite not being a Bohr resonance.
\end{enumerate}
As a consequence, its Landau-Zener parameter $Z$ does not scale as a pure power-law in $\alpha$ (nor $M\ped{c}$), and must be computed numerically. The explicit result is reported in Appendix~\ref{sec:211_200}. Similar to hyperfine resonances, we can compute angular intervals $\delta_2$ and $\chi_2$ where the resonance is non-adiabatic and $\Gamma$-breaks, respectively. The extremely large decay width of $\ket{200}$ (as all states with $\ell=0$), however, makes $\chi_2$ as large as to correspond with the whole possible range of inclinations, from $\SI{0}{\degree}$ to $\SI{180}{\degree}$. Fine resonances are thus effectively never excited for a cloud in a $\ket{211}$ state.

\begin{figure} 
\centering
\includegraphics{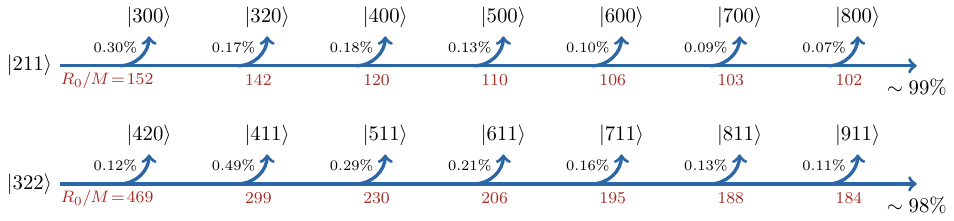}
\caption{Strongest sinking Bohr resonances on a counter-rotating orbit for a cloud in the $\ket{211}$ or $\ket{322}$ state. The percentages next to each resonance are the values of $\abs{c_b}^2$ for benchmark parameters, and they scale with $M\ped{c}$ and $q$ according to \eqref{eqn:scaling-cb2}, while the red numbers below are the resonant orbital separations $R_0$, in units of $M$.}
\label{fig:sinking-from-211-and-322}
\end{figure}

\vskip 4pt
Finally, if the binary arrives to the Bohr region with the cloud still intact, then it encounters the Bohr resonances, all of which are of the sinking type and fall inside the ionization regime (with the exception of $\ket{211}\to\ket{100}$). No extra circularization is provided by the hyperfine resonances, if they do not significantly destroy the cloud. Nevertheless, by the time the binary arrives to the Bohr regime, not only has it presumably evolved for a long time under the circularizing effect of GW radiation, but it also starts to ionize the cloud, further suppressing the eccentricity \cite{Tomaselli:2023ysb}. We will therefore assume that quasi-circular orbits are a good approximation by this point. The final population after each sinking resonance can be found using the approximation (\ref{eqn:sinkingpopulation}), which together with the scaling relations \eqref{eqn:B_bohr} and \eqref{eqn:Z_bohr}, implies
\beq
\abs{c_b}^2\approx3.7\,\biggl(\frac{Z}{B^2}\biggr)^{1/3}\propto\frac{M}{M\ped{c}}q^{4/3}\,.
\label{eqn:scaling-cb2}
\eeq
For the benchmark parameters, the values of $\abs{c_b}^2$ for the strongest sinking resonances (which are typically with states of the form $\ket{n00}$) are summarized in Figure~\ref{fig:sinking-from-211-and-322}, where we have assumed for simplicity a perfectly counter-rotating configuration ($\beta=\pi$). This is generally a good approximation, due to the relative smallness of the angle $\chi_1$. We see that all resonances are very non-adiabatic, in total transferring less than $1\%$ of the cloud to other states. Hence, ionization of $\ket{211}$ happens with minimal disturbance from Bohr resonances.

\vskip 4pt
The only floating Bohr resonance is $\ket{211}\to\ket{100}$. It is worth noting that this is also the only Bohr resonance falling outside the ionization regime (see Figure~\ref{fig:ionization-bohr-resonances}) and that recent numerical studies \cite{Brito:2023pyl} have shown that it has a resonance width much larger than all other resonances. This last observation means that the resonance might partially evade the analysis of the present paper, due to the nonlinear dependence of $P_\slab{gw}$ on $R_*$ playing an important role. In any case, we expect the extremely large decay width of $\ket{100}$ to $\Gamma$-break the resonance in most or all realistic cases, preventing the float from happening. 

\subsection{Evolution from a $\ket{322}$ initial state}
\label{sec:evolution-322}

The second-fastest growing mode is $\ket{322}$. In this case, the constraint on $\alpha$---imposing that the superradiance timescale is shorter than a Gyr---is
\beq
\label{eqn:alpha-322}
0.09\biggl(\frac{M}{10^4M_\odot}\biggr)^{1/13}\lesssim\alpha<1\,,
\eeq
while the rate of cloud decay in GWs is proportional to $M\ped{c}\alpha^{18}$.

\begin{figure} 
\centering
\includegraphics{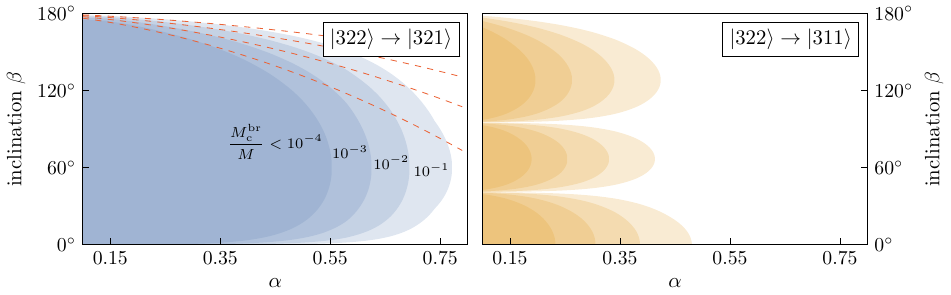}
\caption{Same as Figure~\ref{fig:deltas211}, for the strongest hyperfine (\emph{left panel}) and fine (\emph{right panel}) resonances from a $\ket{322}$ state. The analytical approximations of the contours are not shown in the latter case, as they quickly become inaccurate for moderate values of $\alpha$.}
\label{fig:deltas322}
\end{figure}

\vskip 4pt
Compared to Section~\ref{sec:evolution-211}, a larger number of hyperfine resonances are possible, with any state of the form $\ket{32m_b}$, with $-2\le m_b\le1$. All of these can happen with $g=-4$, in which case the hexadecapole $\ell_*=4$ is entirely responsible for the mixing of the states. However, the cases $m_b=0$ and $m_b=1$ can also resonate, at different separations, with $g=-2$: these are dominated by the quadrupole $\ell_*=2$ instead, which makes these resonances much stronger than the others. Their positions are
\begin{align}
\ket{322}\overset{g=-2}{\longrightarrow}\ket{321} & \qquad \frac{R_0}M=\SI{5.4e+4}{}\,\Bigl(\frac\alpha{0.2}\Bigr)^{-4}\Bigl(\frac{\tilde a}{0.5}\Bigr)^{-2/3}\,,\\[8pt]
\ket{322}\overset{g=-2}{\longrightarrow}\ket{320} & \qquad \frac{R_0}M=\SI{3.4e+4}{}\,\Bigl(\frac\alpha{0.2}\Bigr)^{-4}\Bigl(\frac{\tilde a}{0.5}\Bigr)^{-2/3}\,,
\end{align}
which should be evaluated at $\tilde a\approx2\alpha/(1+\alpha^2)$. The most stringent constraint on $\delta_1$ is given by $\ket{322}\to\ket{321}$ and equals
\beq
\delta_1=\SI{5.4}{\degree}\,\biggl(\frac{M\ped{c}/M}{10^{-2}}\biggr)^{-1/6}\biggl(\frac{q}{10^{-3}}\biggr)^{1/12} \biggl(\frac{\alpha}{0.2}\biggr)^{4/3} \biggl(\frac{\tilde{a}}{0.5}\biggr)^{1/36}f(\varepsilon_0)^{1/4}\,.
\label{eqn:322-delta1}
\eeq
The angle $\chi_1$, within which the same resonance $\Gamma$-breaks, is instead
\beq
\chi_1\approx\SI{4.8}{\degree}\biggl(\frac{M\ped{c}\ap{br}/M}{10^{-2}}\biggr)^{-1/6}\biggl(\frac\alpha{0.2}\biggr)^{11/6}\biggl(\frac{\tilde a}{0.5}\biggr)^{-5/18}f(\varepsilon\ped{br})^{1/6}\,,
\label{eqn:322_chi_1}
\eeq
also more accurately numerically computed and shown in Figure~\ref{fig:deltas322} (\emph{left panel}). Similar to the resonant history of $\ket{211}$, some resonances (such as $\ket{322}\to\ket{321}$) become weak around $\beta=0$, yet other resonances (such as $\ket{322}\to\ket{320}$) do not, thereby eliminating any possible ``safe interval'' around a co-rotating configuration.

\vskip 4pt
Differently from Section~\ref{sec:evolution-211}, there is no clear hierarchy between $\delta_1$ and $\chi_1$. Which one is largest depends not only on $\alpha$, but also on the chosen value of $M\ped{c}\ap{br}$. The angular interval that leads to the survival of the cloud in appreciable amounts is, however, generally dominated by the $\Gamma$-breaking, as even very light clouds, say $M\ped{c}\ap{br}/M<10^{-4}$, are able to give clear signatures in the Bohr region~\cite{Baumann:2021fkf}.

\vskip 4pt
As in the $\ket{211}$ case, the $\varepsilon$-breaking and $Z$-breaking prove to not be relevant for the resonant history: the value
\beq
\sqrt ZB\sim10^7\biggl(\frac{M\ped{c}/M}{10^{-2}}\biggr)\biggl(\frac{q}{10^{-3}}\biggr)^{-1}\biggl(\frac\alpha{0.2}\biggr)^{-6}\biggl(\frac{\tilde a}{0.5}\biggr)^{-1/3}\,.
\eeq
requires extremely high eccentricities ($\varepsilon\gtrsim0.98$) to give rise to a resonance breaking. The corresponding initial inclinations are extremely close to $\beta=\pi$ and would fall in the interval \eqref{eqn:322-delta1}, where the resonance is not adiabatic.

\vskip 4pt
A cloud in the $\ket{322}$ state can experience fine resonances with states with $\ell\ne0$. Their decay width is smaller than those of the states with $\ell=0$: as a consequence, fine resonances can destroy a significant portion of the cloud before they $\Gamma$-break. The fine resonance that gives the most stringent constraints on $\delta_2$ and $\chi_2$ is
\beq
\ket{322}\overset{g=-1}\longrightarrow\ket{311}\qquad\frac{R_0}M=\SI{2.3e+3}{}\,\Bigl(\frac\alpha{0.2}\Bigr)^{-10/3}\,.
\eeq
Analytical approximations for $\beta\approx\pi$ give\footnote{For $\alpha\gtrsim0.5$, this resonance may marginally fall inside the ionization regime. However, the value of $P\ped{ion}$ never becomes much larger than $P_\slab{gw}$. We therefore ignore this detail, which only slightly increases the value of $\delta_2$ compared to the one presented in (\ref{eqn:322-delta2}).}
\beq
\delta_2=\SI{3.2}{\degree}\,\biggl(\frac{M\ped{c}/M}{10^{-2}}\biggr)^{-1/4}\biggl(\frac{q}{10^{-3}}\biggr)^{1/8} \biggl(\frac{\alpha}{0.2}\biggr)^{31/24}f(\varepsilon_0)^{3/8}
\label{eqn:322-delta2}
\eeq
and
\beq
\chi_2\approx\SI{9}{\degree}\biggl(\frac{M\ped{c}\ap{br}/M}{10^{-2}}\biggr)^{-1/4}\biggl(\frac\alpha{0.2}\biggr)^{3/2}f(\varepsilon\ped{br})^{1/4}\,,
\label{eqn:322_chi_2}
\eeq
while a more accurate numerical determination of the mass of the cloud at resonance breaking is given in Figure~\ref{fig:deltas322} (\emph{right panel}). It is worth noting that the strength of the $\ket{322}\to\ket{311}$ resonance has a complicated $\beta$ dependence, due to the octupole $\ell_*=3\ne-g$ being the dominant term. Consequently, this resonance becomes weak not only around $\beta=\SI{180}{\degree}$, but also around $\beta=\SI{41}{\degree}$ and $\SI{95}{\degree}$ (as visible from Figure~\ref{fig:deltas322}). However, other fine resonances remain strong at these intermediate inclinations and so, once again, the cloud can only reach the Bohr region if the inclination is in a narrow interval around the counter-rotating configuration.

\vskip 4pt
In the Bohr region, the system encounters several sinking resonances, the strongest of which are with states of the form $\ket{n11}$. The final populations $\abs{c_b}^2$ are displayed in Figure~\ref{fig:sinking-from-211-and-322}. For benchmark parameters, about $2\%$ of the cloud is lost in the process. None of the floating resonances, with $n=1$ or $n=2$ states, becomes adiabatic within the interval of inclinations discussed above.

\vskip 4pt
Finally, in case the binary is formed at radii small enough to avoid constraints on the inclination coming from fine resonances, an interesting scenario opens up. The strongest floating Bohr resonance is $\ket{322}\overset{g=-1}\longrightarrow\ket{211}$, which becomes adiabatic, for benchmark parameters, for $\beta<\SI{155}{\degree}$.\footnote{Due to the weakness of the resonance compared to most the (hyper)fine ones, it is not possible to expand around $\beta=\pi$ and get a simple formula for the upper limit on the angle as function of the parameters. Nevertheless, a good approximation is given by the following cubic equation: $(\pi-\beta)^4+2.8(\pi-\beta)^6>0.056\times(10^5M\ped{c}q/M)^{1/2}$.} Among all possible scenarios we considered in Sections~\ref{sec:evolution-211} and \ref{sec:evolution-322}, this is the only case where the binary's evolution in the Bohr region features a new phenomenon, beyond ionization and non-adiabatic sinking resonances: namely, an adiabatic floating resonance. The companion's motion continues to ionize the cloud while this resonance takes place, potentially changing $M\ped{c}$ significantly before its end. This is also the only floating resonance with the actual potential to partially move the cloud to a different state, rather than merely destroying it: as can be seen in Figure~\ref{fig:tfloat_tdecay} (\emph{right panel}), the hierarchy $\Delta t\ped{float}\gg t\ped{decay}$ is not valid in the entire parameter space. Hence, depending on the parameters, when the resonance ends, the inspiral can either continue without the cloud, or with a cloud in a (decaying) $\ket{211}$ state and a reduced value of $M\ped{c}$. In the latter case, the discussion in Section~\ref{sec:evolution-211} applies from this point onwards.

\section{Observational signatures}
\label{sec:observational-signatures}

The dynamics of the cloud-binary system are intricate and depend on the parameters. In Section~\ref{sec:history} we determined when the cloud is entirely destroyed in the early inspiral, when it loses some of its mass upon resonance breaking, and when it remains intact until the binary enters the Bohr region. There are thus two main ways the cloud can leave an imprint on the GW waveform: (1)  modifications of the waveform due to interaction with the cloud, in case it is still present in the late stages of the inspiral (Section~\ref{sec:direct-signatures}); (2) permanent consequences on the binary parameters left by a cloud destroyed early in the inspiral (Section~\ref{sec:indirect-evidence}). A partially destroyed cloud, left by a broken resonance, may be able to combine both kinds of signatures.

\subsection{Direct signatures of the cloud}
\label{sec:direct-signatures}

As discussed extensively in Section~\ref{sec:history}, the requirement that the cloud survives the hyperfine and fine resonances forces either the inclination angle to be within $\mathcal{O}(10^\circ)$ of a counter-rotating configuration or the binary to form at radii too small to ever excite those resonances. Then, most phenomena producing direct observational evidence of the cloud happen when the binary reaches the Bohr region. Here, ionization takes over GW radiation as the primary mechanism of orbital energy loss. When $P\ped{ion}\gg P_\slab{gw}$, the evolution of the GW frequency $f_\slab{gw}$ approximately follows a universal shape \cite{Baumann:2022pkl},
\beq
f_\slab{gw}(t)=\frac{\alpha^3}M\,f\biggl(\frac{M\ped{c}q\alpha^3}{M^2}t\biggr)\,,
\label{eqn:universal-f-ionization}
\eeq
where the function $f$ can be explicitly determined from the shape of $P\ped{ion}$. This universal behavior of $f_\slab{gw}(t)$ constitutes a direct evidence of the presence of the cloud.

\vskip 4pt
On top of this, sinking resonances can cause non-negligible upward ``jumps'' of $f_\slab{gw}$ due to their backreaction\footnote{In our work, we only study the backreaction on the orbital parameters. When including the  backreaction on the geometry as well, the cloud's transitions could cause ``resonant'' features in the emitted GWs, see e.g.~Figure~1 of \cite{Duque:2023cac}.}, even if they are strongly non-adiabatic. For a Bohr resonance $\ket{n_a\ell_am_a}\to\ket{n_b\ell_bm_b}$, they are located at
\beq
f_\slab{gw}\ap{res}=\frac{\SI{26}{mHz}}g\biggl(\frac{10^4M_\odot}{M}\biggr)\biggl(\frac\alpha{0.2}\biggr)^3\biggl(\frac1{n_a^2}-\frac1{n_b^2}\biggr)\,,
\label{eqn:position-resonances}
\eeq
where $g=m_b-m_a$, and thus fall inside the LISA band for benchmark parameters.\footnote{Formula~\eqref{eqn:position-resonances}, with $n_b\to\infty$, also describes the position of the $g$-th ``kink'' of the function $f$ appearing in \eqref{eqn:universal-f-ionization}, corresponding to the $g$-th discontinuity of $P\ped{ion}$ (see Figure~\ref{fig:ionization-bohr-resonances}).}

\vskip 4pt
The amplitude of the jump can be computed explicitly from \eqref{eqn:dimensionless-omega-evolution} (assuming quasi-circular orbits):
\beq
\Delta f_\slab{gw}=\frac{\SI{0.61}{mHz}}{\Delta m^{1/3}}\,\biggl(\frac{10^4M_\odot}{M}\biggr)\biggl(\frac{M\ped{c}/M}{0.01}\biggr)\biggl(\frac{q}{10^{-3}}\biggr)^{-1}\biggl(\frac\alpha{0.2}\biggr)^3\biggl(\frac1{n_a^2}-\frac1{n_b^2}\biggr)^{4/3}\biggl(\frac{\abs{c_b}^2}{10^{-3}}\biggr)\,,
\label{eqn:jump-resonances}
\eeq
where the values of $\abs{c_b}^2$ and their dependence on the parameters are given in Figure~\ref{fig:sinking-from-211-and-322} and equation \eqref{eqn:scaling-cb2}. The increase in frequency comes with smaller, long-lived oscillations of the frequency, and with a slight increase of the eccentricity; both these effects have been shown in the right panels of Figure~\ref{fig:eccentric-backreacted-resonance} for example parameters. The dephasing introduced by a single sinking resonance on top of the one coming from ionization is $\Delta\Phi_\slab{gw}\approx\pi f_\slab{gw}\ap{res}\Delta f_\slab{gw}/\gamma$. This is of the order of thousands of radians, although the exact number can vary by a few orders of magnitude in different regions of the parameter space. Not only is this well above the expected LISA precision of $\Delta\Phi_\slab{gw}\sim2\pi$, but such a dephasing would happen in a very narrow frequency range, in contrast to most other environmental effects, including ionization. This unique behavior would aid parameter estimation by directly linking the cloud's parameters with $\Delta\Phi_\slab{gw}$ via \eqref{eqn:position-resonances} and \eqref{eqn:jump-resonances}, especially if multiple jumps are observed within one signal.

\vskip 4pt
As discussed in Sections~\ref{sec:evolution-211} and \ref{sec:evolution-322}, the only cases where a floating resonance can be observed in the Bohr region require a binary formation at very small radii, so that all early resonances are skipped without a strict requirement on the inclination angle. Resonances of the type $\ket{n_a\ell_am_a}\to\ket{100}$ happen very late in the inspiral (see Figure~\ref{fig:ionization-bohr-resonances}), where relativistic corrections are expected to be more important \cite{Brito:2023pyl}. The only other floating Bohr resonance encountered in Section~\ref{sec:history} is $\ket{322}\to\ket{211}$. This is an interesting case because it may not entirely destroy the cloud. The expected GW signal is a constant frequency $f_\slab{gw}$ given by equation \eqref{eqn:position-resonances}, for a total floating time of\footnote{This value assumes a quasi-circular co-rotating orbit. Moderate nonzero values of eccentricity or inclination introduce $\mathcal O(1)$ variations in $\Delta t\ped{float}$.}
\beq
\Delta t\ped{float}=\SI{5.8}{yrs}\,\biggl(\frac{M}{10^4M_\odot}\biggr)\biggl(\frac{q}{10^{-3}}\biggr)^{-2}\biggl(\frac\alpha{0.2}\biggr)^{-3}\,.
\label{eqn:deltatfloat-322-211}
\eeq
Although the cloud's mass is continuously reduced by ionization while the resonance takes place, the value given in \eqref{eqn:deltatfloat-322-211} remains independent of $M\ped{c}$ as long as it is large enough to guarantee $P\ped{ion}\gg P_\slab{gw}$. 

\subsection{Indirect signatures: impact on binary parameters}
\label{sec:indirect-evidence}

For sufficiently small orbital inclinations, as seen in Figures~\ref{fig:history_211}, \ref{fig:deltas211}, and \ref{fig:deltas322}, the cloud can be destroyed during one of the floating resonances in the early inspiral, to a level where it no longer affects the binary dynamics in an observable way. Then, by the time the system enters in band, its evolution is expected to follow the rules of vacuum general relativity. Nevertheless, the binary still carries the marks of the previously existing boson cloud, and of the resonance that destroyed it. These are due to the backreaction on the orbit from that floating resonance, and come in the form of a change in the eccentricity and tilt of the inclination angle.

\begin{figure}
\centering
\includegraphics{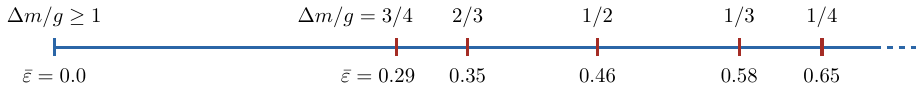}
\caption{Example values of the fixed point $\bar{\varepsilon}$ depending on $\Delta m/g$. Numbers can be found by solving equations \eqref{eqn:dimensionless-eccentricity-evolution} and \eqref{eqn:dimensionless-inclination-evolution} on a floating orbit.}
\label{fig:fixedpointsresonances}
\end{figure}

\vskip 4pt
While in Section~\ref{sec:history} we could simplify the analysis by studying only the strongest resonance, the impact on the orbital parameters strongly depends on which overtone (that is, which value of $g$) mediated the last adiabatic resonance encountered by the system.\footnote{If the system undergoes multiple floats, for example, because broken resonances leave a cloud massive enough to excite other adiabatic resonances, then the evolution of the eccentricity follows several nontrivial steps. Here, however, we focus on the last of those as it has the most direct observational consequences.} As shown in Figure \ref{fig:streamplot_inclination-eccentricity}, the orbital parameters follow specific sets of trajectories on the $(\varepsilon,\beta)$ plane, until the resonance breaks or completes. While floating orbits \emph{always} tilt the inclination angle towards a co-rotating configuration, the eccentricity is forced towards a fixed point, whose value depends on $\Delta m/g$. Some examples of the value of this fixed point are shown in Figure \ref{fig:fixedpointsresonances} for different values of $\Delta m / g$.

\begin{figure} 
\centering
\includegraphics{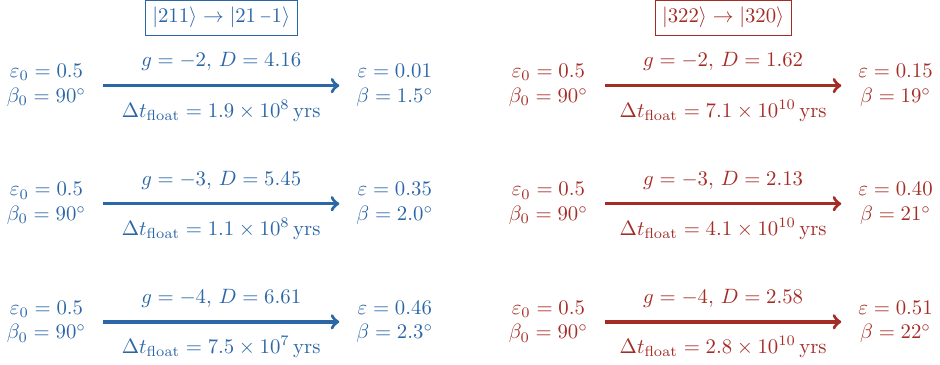}
\caption{Examples of backreaction on the eccentricity $\varepsilon$ and inclination $\beta$ during floating orbits that destroy the cloud entirely. We show the strongest resonance ($\Delta m/g=1$) and two overtones in each scenario, using the benchmark parameters $\alpha=0.2$, $q=10^{-3}$, $M\ped{c}=0.01M$, and $M=10^4M_{\odot}$. Each case is initialized with $\varepsilon_0=0.5$ and $\beta_0=\SI{90}{\degree}$ for illustrative purposes, but we observed that the final values of $\varepsilon$ and $\beta$ are very robust against the choice of different initial conditions. The final values of $\varepsilon$ and $\beta$, as well as $\Delta t\ped{float}$, are computed integrating numerically equations \eqref{eqn:dimensionless-dressed-schrodinger}, \eqref{eqn:dimensionless-omega-evolution}, \eqref{eqn:dimensionless-eccentricity-evolution}, and \eqref{eqn:dimensionless-inclination-evolution}. For benchmark parameters, the floating time of resonances from $\ket{322}$ exceeds the Hubble time; however, it strongly depends on the parameters, as derived in \eqref{eqn:hyperfine-B-tfloat-exponent}.}
\label{fig:examples-backreaction-orbit}
\end{figure}

\vskip 4pt
Assuming that the resonance does not break prematurely, the distance traveled by the binary in the $(\varepsilon,\beta)$ plane depends on the parameter $D$ alone, introduced in \eqref{eqn:D}. Very roughly, the system gets $e^D$ times closer to the eccentricity fixed point than it was before the resonance started. For our benchmark parameters, we find that most hyperfine and fine resonances have values of $D$ between 1 and 10, meaning that the fixed point is approached significantly. The value of $D$, however, strongly varies with the parameters, as we found in Section~\ref{sec:types-of-resonances}. In particular, it is inversely proportional to $q$, implying that only intermediate or extreme mass ratio binaries change significantly their orbital parameters during a floating resonance. Examples of variations of the parameters during a floating orbit are reported in Figure~\ref{fig:examples-backreaction-orbit} for the resonances $\ket{211}\to\ket{21\,\minus1}$ and $\ket{322}\to\ket{320}$.

\vskip 4pt
The eccentricity attractor points shown in Figure~\ref{fig:fixedpointsresonances} and Figure~\ref{fig:examples-backreaction-orbit} are a unique signature of the cloud-binary dynamics. Together with the suppression of the orbital inclination, this opens up the possibility of performing a statistical test of the parameters of a large number of binary inspirals, and comparing them with the ones predicted from a suitable model of their formation channels. This task is likely to be complicated by the existence of other astrophysical mechanisms or formation processes that can affect eccentricity and inclination, but diving into these details is beyond the scope of the present work. It should also be kept in mind that, after the completion of the resonance, GW emission will start circularizing the orbit once again, so that a ``backtracking'' of the eccentricity would be needed to test the existence of the eccentricity attractor points. For an in-depth analysis of the signatures of boson clouds on binary eccentricity and inclination, see the companion Letter \cite{Tomaselli:2024dbw}. In addition, reference \cite{Boskovic:2024fga} (which was completed at the same time as the present work) also explores the binary eccentricity in this context.

\vskip 4pt
Lastly, we note that the extremely long floating time associated with some hyperfine or fine resonances can stop many binaries from getting in band at all, consequently reducing the merger rate. For example, for our choice of benchmark parameters, the hyperfine resonances from the $\ket{322}$ state, shown in Figure~\ref{fig:examples-backreaction-orbit}, float for longer than the Hubble time.

\section{Conclusions}
\label{sec:conclusions}

In the context of gravitational-wave astronomy, binary black hole environments have long been proposed as a laboratory for fundamental physics. One such example are gravitational atoms, or clouds of ultralight bosons produced by superradiance around spinning black holes. Compared to other kinds of environment, the phenomenology of gravitational atoms is extremely rich. The two most striking types of interaction between the binary and the cloud are resonant phenomena \cite{Zhang:2018kib,Baumann:2018vus,Baumann:2019ztm} and friction effects \cite{Zhang:2019eid,Baumann:2021fkf,Baumann:2022pkl,Tomaselli:2023ysb}, both of which leave very distinct signatures on the emitted gravitational waveform. This complexity is a blessing for the potential detection and identification \cite{Cole:2022yzw} of such systems. However, it is a curse for the achievement of a complete characterization of their evolution.

\vskip 4pt
Previous studies have described the effects on the waveform as function of the state of the cloud and of the binary configuration at the time of observation. These are, however, the final products of a complex series of cloud-binary interactions that characterize former phases of the inspiral. Despite a number of relevant studies \cite{Takahashi:2021eso,Zhang:2018kib,Ding:2020bnl,Tong:2021whq,Du:2022trq,Tong:2022bbl,Fan:2023jjj,Takahashi:2023flk}, the combinations of cloud states and binary configurations compatible with this kind of evolution have not yet been determined.

\vskip 4pt
In this work, we finalize such a program by systematically studying the chronological sequence of resonances encountered during the binary inspiral. We do so in the most general possible set of assumptions: we allow for any value of the initial eccentricity, inclination and separation of the binary; at the same time, we keep the scaling with the cloud's parameters explicit, so to apply our results to the entire parameter space. Furthermore, we take into account the backreaction of the resonances on the orbit, and study how this impacts the behavior of the resonances themselves. This aspect, as well as the impact of inclination and eccentricity on resonances (and vice versa), have never been studied before, and each of these novel results turns out to play a crucial role in our analysis. Finally, we perform explicitly the exercise of ``following'' the evolution of the system from the initial states most likely to be populated by superradiance, $\ket{211}$ and $\ket{322}$, until the merger, and then summarize the gravitational wave signatures compatible with the scenarios studied.

\vskip 4pt
In principle, one might have expected the evolution of the system to be extremely complicated. The S-matrix approach developed in \cite{Baumann:2019ztm} suggests a tree of populated states branching more and more, every time a new resonance is encountered. In practice, however, we find that the hierarchy between the floating and decay timescales simplifies the picture dramatically.\footnote{As shown in Figure~\ref{fig:tfloat_tdecay} and equation \eqref{eqn:t-decay-t-float}, the separation of timescales is larger for smaller mass ratios. For equal-mass ratios $q\sim1$, the timescales could become comparable.} During every hyperfine or fine adiabatic resonance, the cloud loses mass until the resonance breaks, often to the point where it is no longer directly observable. The conclusion is then remarkably simple, and similar among the two cases studied explicitly. Only binaries close enough to a counter-rotating configuration, where these early resonances are very weak, are able to carry the cloud up to the point where it becomes observable through the effects of ionization and a large number of weak resonances. Our detailed study of the nonlinear behavior of resonances allows us to precisely quantify the angular interval of inclinations where this scenario is realized, see equations \eqref{eqn:211_chi_1}, \eqref{eqn:322_chi_1}, and \eqref{eqn:322_chi_2}, as well as Figures~\ref{fig:deltas211} and \ref{fig:deltas322}.

\vskip 4pt
The early disappearance of the cloud for generic orbital configurations may seem to suggest that the chances of detecting ultralight bosons using binary systems might be slim after all. It is certainly true that our work puts strict conditions for the direct observation of the cloud-binary interaction. To avoid all early resonances, either the initial separation must be small enough, or the orbital angular momentum must be approximately anti-aligned (within a tolerance interval whose size depends on the parameters) with the central BH's spin. The likelihood of either scenario depends on the astrophysical processes governing the formation of the binary system and is subject to large uncertainties---though see \cite{LISA:2022yao,Goodman:2003sf,Goodman:2002gv,Levin:2006uc} and \cite{Babak:2017tow,Amaro-Seoane:2012jcd} for the two cases, respectively. We note, however, that the event leading to the cloud destruction---an adiabatic floating resonance---necessarily exerts a strong backreaction on the binary's inclination and eccentricity, generally reducing both quantities significantly. This fact raises the possibility to reverse engineer the existence of a cloud from a statistical analysis of a large number of inspirals with well-measured orbital parameters.

\vskip 4pt
As anticipated in \cite{Tomaselli:2023ysb}, the present work answers the main remaining open questions left on the phenomenology of gravitational atoms in binaries, within a certain set of assumptions. We neglected a number of subleading effects. Some of them, like the accretion onto the secondary \cite{Baumann:2021fkf} and the cloud's self gravity \cite{Ferreira:2017pth,Hannuksela:2018izj} can be straightforwardly included in the binary's evolution, and do not change qualitatively the picture drawn here. Others, like the backreaction of the cloud's decay in GWs \cite{Cao:2023fyv}, the non-resonant overlap between growing and decaying states \cite{Tong:2022bbl,Fan:2023jjj}, and the change in the BH's mass and spin due to the absorption of the cloud during a resonance (see footnote~\ref{fn:bh-parameters}) can potentially introduce new relevant features, at large, small, and any separations, respectively. But perhaps most importantly, we stuck to a nonrelativistic analysis; steps forward towards a fully relativistic description have recently been taken in \cite{Brito:2023pyl,Cannizzaro:2023jle}. Future directions of studies in this field will likely focus on one or more of the above points.

\paragraph{Acknowledgements} We thank Enrico Cannizzaro for help with the numerical computation of the decay timescales. During the final stages of this work, we learned of the related work \cite{Boskovic:2024fga}: we are grateful to Rafael Porto for sharing their draft and coordinating the submission with us. TS is supported by the VILLUM Foundation (grant no.~VIL37766), the Danish Research Foundation (grant no.~DNRF162), and the European Union’s H2020 ERC Advanced Grant ``Black holes: gravitational engines of discovery'' grant agreement no.~Gravitas–101052587.

\newpage
\appendix
\addtocontents{toc}{\protect\vskip24pt}

\section{Hyperfine resonances and angular momentum}
\label{sec:hyperfine-angular-momenutm}

A nonzero black hole spin is responsible for the existence of the hyperfine energy splitting, as it breaks the spherical symmetry of the background spacetime. At the same time, we study the backreaction of resonances (hyperfine or not) on the orbit in the Newtonian approximation, assuming the conservation of the total angular momentum, which leads to equations \eqref{eqn:dimensionless-omega-evolution}, \eqref{eqn:dimensionless-eccentricity-evolution}, and \eqref{eqn:dimensionless-inclination-evolution}. This methodology might appear as fundamentally inconsistent, so let us inspect it more closely.

\vskip 4pt
The weak-field approximation of the Kerr metric, which is valid at large distances, reads
\beq
ds^2=-\biggl(1-\frac{2M}r\biggr)\dd t^2+\biggl(1+\frac{2M}r\biggr)\dd r^2+r^2(\dd\theta^2+\sin^2\theta\dd\phi^2)-\tilde aM\frac{4M}r\sin^2\theta\dd t\dd\phi\,.
\eeq
The last term is known to give rise to the Lense-Thirring precession, as the equation of motion of a scalar particle can be put in the form
\beq
\frac{\dd^2\vec r}{\dd t^2}=-\frac{M}{r^3}\vec r+4\frac{\dd\vec r}{\dd t}\times\vec B\,,
\eeq
where the \emph{gravitomagnetic field} $\vec B$ is related to the black hole spin as
\beq
\vec B=\vec\nabla\times\vec A\,,\qquad\vec A=-\frac{\vec J\times\vec r}{2r^3},\qquad\vec J=\tilde aM^2\hat z\,.
\eeq
The corresponding Hamiltonian is
\beq
H=\frac{(\vec p-4\mu\vec A)^2}{2\mu}-\frac{\mu M}r\approx\frac{\vec p^2}{2\mu}-\frac\alpha{r}+\frac{2\tilde aM^2}{r^3}L_z\,,
\label{eqn:hamiltonian-gravitomagnetism}
\eeq
where $\mu=\alpha/M$ is the mass of the particle. We can immediately check that the last term in \eqref{eqn:hamiltonian-gravitomagnetism} gives rise to the expected hyperfine splitting,
\beq
\braket{n\ell m|H|n\ell m}=2\tilde aM^2m\Braket{n\ell m|\frac1{r^3}|n\ell m}=2\tilde aM^2m\,\frac{(\mu \alpha)^3}{n^3\ell(\ell+1/2)(\ell+1)}\,,
\eeq
which perfectly matches the last term in \eqref{eq:eigenenergy}.

\vskip 4pt
The orbital angular momentum $\vec L=\vec r\times\vec p$ evolves as
\beq
\frac{\dd \vec L}{\dd t}=i[H,\vec L]=\frac{2}{r^3}\vec J\times\vec L,
\eeq
which is the expected Lense-Thirring precession. Applying this equation to the cloud-binary system gives rise to two additional terms on the right-hand sides of \eqref{eqn:Lz-balance} and \eqref{eqn:Lx-balance}, corresponding to the Lense-Thirring precession of the cloud (which vanishes in most cases, as $\vec S\ped{c}\parallel\vec J$ even during a transition, as we will see below) and of the binary. This precession is, however, \emph{parametrically small}. None of the other terms in \eqref{eqn:Lz-balance} and \eqref{eqn:Lx-balance} depend on the BH spin $\tilde a$, even in the case of hyperfine resonances, where the energy splitting is proportional to $\tilde a$. Not only for realistic parameters is this precession extremely slow, but it also does not disrupt the approach in the main text, as \eqref{eqn:Lx-balance} can be simply replaced by the analogous equation for the (precessing) equatorial projection of the angular momentum.

\vskip 4pt
Having justified the use of the conservation of total angular momentum, there is another potentially worrying aspect of the breaking of spherical symmetry, that has to do with the spin of the cloud when it is in a mixed state, for example during a transition. As long as the Hamiltonian is spherically symmetric, $\ket{n\ell m}$ are guaranteed to be eigenstates of the scalar field's orbital angular momentum $\vec L$. Its matrix elements are given by $L_z\ket{n\ell m}=m\ket{n\ell m}$ and, in the Condon–-Shortley convention, $L_\pm\ket{n\ell m}=\sqrt{\ell(\ell+1)-m(m\pm1)}\ket{n\ell,m\pm1}$, where $L_\pm=L_x\pm iL_y$. If the cloud is in a mixed state of the form $\ket{\psi}=c_a\ket{n_a\ell_am_a}+c_b\ket{n_b\ell_bm_b}$, then its $z$ component of the angular momentum is $m_a\abs{c_a}^2+m_b\abs{c_b}^2$, while the equatorial components vanish unless $\ell_a=\ell_b$ and $\abs{m_a-m_b}=1$.

\vskip 4pt
Remarkably, all the previous results still hold for the Hamiltonian \eqref{eqn:hamiltonian-gravitomagnetism}. That is because the perturbation $\sim L_z/r^3$ is diagonal on the basis $\ket{\ell m}$, only mixing states with different $n$. Even though the spacetime is not spherically symmetric, the angular structure of the eigenstates is unchanged. The equations in the main text then do not need any modification, except for the case of hyperfine transitions with $\abs{\Delta m}=1$. For a hyperfine transition with $m_b=m_a-1$, careful computation (in the Schrödinger, not dressed, frame) of the equatorial components of $\vec S\ped{c}$ shows that equation \eqref{eqn:Lx-balance} would need to be corrected with a term
\beq
\frac{\dd S_{\text{c},x}}{\dd\tau}\sim\frac{B}{g}\sqrt{\ell(\ell+1)-m_a(m_a-1)}\sqrt{Z}(\abs{c_b}^2-\abs{c_a}^2)\sin(C\tau/3)\,.
\eeq
This is a fast oscillating term that averages to zero on timescales much shorter than the evolution of the orbital parameters and the duration of the resonance. We thus ignore it in the main text.

\section{General resonance breaking}
\label{sec:breaKING}

The phenomenon of resonance breaking was discussed in Section~\ref{sec:resonance-breaking} in the simplified scenarios where only one of the following quantities is allowed to vary at a time: the eccentricity $\varepsilon$, Landau-Zener parameter $Z$, and cloud's mass $M\ped{c}$. We derive here the result in the general case. Taking the time derivative of \eqref{eqn:dimensionless-omega-evolution-Gamma}, we find
\beq
\begin{split}
\frac{\dd^2\omega}{\dd\tau^2}&=\frac{\dd f(\varepsilon)}{\dd\tau}+B\frac{\dd^2\abs{c_a}^2}{\dd\tau^2}=\frac{\dd f(\varepsilon)}{\dd\tau}+B\biggl(\frac{\dd^2c_a^*}{\dd\tau^2}c_a+2\frac{\dd c_a^*}{\dd\tau}\frac{\dd c_a}{\dd\tau}+c_a^*\frac{\dd^2c_a}{\dd\tau^2}\biggr)\\
&=\frac{\dd f(\varepsilon)}{\dd\tau}-2ZB(\abs{c_a}^2-\abs{c_b}^2)+\biggl(\frac{1}{2Z}\frac{\dd Z}{\dd\tau}-\Gamma\biggr)\biggl(\frac{\dd\omega}{\dd\tau}-f(\varepsilon)\biggr)+\omega\sqrt ZB(c_a^*c_b+c_ac_b^*)\,,
\end{split}
\label{eqn:breaKING-harmonic-oscillator}
\eeq
where the second line is obtained by repeated use of the Schrödinger equation \eqref{eqn:schrodinger-Gamma} together with \eqref{eqn:dimensionless-omega-evolution-Gamma}. Under the assumption that all coefficients appearing above evolve slowly during a floating orbit, equation \eqref{eqn:breaKING-harmonic-oscillator} has the structure of a damped harmonic oscillator, with solution
\beq
\omega=\frac{\frac{\dd f(\varepsilon)}{\dd\tau}-\frac{f(\varepsilon)}{2Z}\frac{\dd Z}{\dd\tau}+f(\varepsilon)\Gamma-2ZB(\abs{c_a}^2-\abs{c_b}^2)}{-\sqrt{Z}B(c_a^*c_b+c_ac_b^*)}+\text{damped oscillatory terms}\,.
\label{eqn:breaKING-harmonic-oscillator-solution}
\eeq
The resonance breaks whenever $c_b^*c_a+c_a^*c_b=0$. By direct application of the Schrödinger equation, we find
\beq
\sqrt Z\frac\dd{\dd\tau}(c_a^*c_b+c_ac_b^*)=-\omega\frac{\dd\abs{c_a}^2}{\dd\tau}-\Gamma\sqrt Z(c_a^*c_b+c_b^*c_a)\,.
\label{eqn:breaKING-dcacbcbcadt}
\eeq
By plugging in \eqref{eqn:breaKING-dcacbcbcadt} the non-oscillatory term of \eqref{eqn:breaKING-harmonic-oscillator-solution}, we arrive at an equation for the sole unknown $c_a^*c_b+c_b^*c_a$:
\beq
\frac{ZB}2\biggl(\frac\dd{\dd\tau}+2\Gamma\biggr)(c_a^*c_b+c_b^*c_a)^2=\biggl(\frac{\dd f(\varepsilon)}{\dd\tau}-\frac{f(\varepsilon)}{2Z}\frac{\dd Z}{\dd\tau}+f(\varepsilon)\Gamma-2ZB(\abs{c_a}^2-\abs{c_b}^2)\biggr)\frac{\dd\abs{c_a}^2}{\dd\tau}\,.
\label{eqn:breaKING-master}
\eeq
Remarkably, the evolution of the eccentricity, the variation of the Landau-Zener parameter and the decay of the cloud contribute additively to \eqref{eqn:breaKING-master}, each with its own term. In realistic cases, $\Gamma$ is large enough to force the population of state $\ket{b}$ to reach a saturation value $\abs{c_b}^2=f(\varepsilon)/(2\Gamma B)$, which is usually small enough to be neglected in \eqref{eqn:breaKING-master}. The point of resonance breaking, then, only involves the population left in the initial state, $\abs{c_a}^2$.

\section{Ionization at resonance}
\label{sec:ion-at-resonance}

The expressions for the ionization rate and power derived in \cite{Baumann:2021fkf} are valid under the assumption that the frequency $\Omega$ of the perturbation is away from any bound-to-bound state resonance. In this appendix, we relax this assumption by computing the new term contributing at resonance and showing that its effect is ultimately negligible. To allow for an easy match with the notations of \cite{Baumann:2021fkf}, here we denote the state initially populated by $\ket{b}$ and any other bound state by $\ket{a}$.

\vskip 4pt
Ignoring couplings between different continuum states (as justified in Appendix~A2 of \cite{Baumann:2021fkf}), the Hamiltonian of the gravitational atom reads,
\beq
\mathcal{H}=\sum_{b} \epsilon_{b}\ket{b}\!\bra{b} + \sum_{a \neq b} \eta_{ab}(t) \ket{a}\!\bra{b} + \sum_{K} \epsilon_{K}\ket{K}\!\bra{K}+\sum_{K, b}\left[\eta_{Kb}(t)\ket{K}\!\bra{b}+\text{h.c.}\right]\,,
\eeq
where $\ket{K}\equiv\ket{k;\ell m}$ is a continuum state multi-index, $\epsilon_K\approx k^2/(2\mu)$, and the couplings $\eta_{ab}(t)$ and $\eta_{Kb}(t)$ are the matrix elements of the perturbation \eqref{eqn:V_star}. As in \cite{Baumann:2021fkf}, by integrating out the continuum the Schrödinger equation can be recast in the following form\footnote{All quantities here depend on time, either through fast oscillatory terms, due to the evolving phase $\varphi_*$, or through the slow frequency chirp. For ease of notation, we will only explicitly write the time dependence of terms falling in the first class.}:
\beq
\label{eq:IonResoccupation}
i \frac{\dd c_{b}}{\dd t}=\mathcal{E}_{b}c_{b}(t)+\sum_{a \neq b}\left[\eta_{ba}(t) e^{i(\epsilon_b-\epsilon_a) t}+\mathcal{E}_{ba}(t)\right] c_{a}(t)\,,
\eeq
where we define the \emph{induced couplings},
\beq
\label{eq:inducedcouplingdef}
\mathcal{E}_{ba}(t)\equiv-i \int_{-\infty}^{t} \dd t^{\prime} \sum_{K} \eta^{*}_{K b}(t) \eta_{K a}(t^{\prime}) e^{-i(\epsilon_K-\epsilon_b)t+i(\epsilon_K-\epsilon_a)t'}\,.
\eeq
and the \emph{induced energies} $\mathcal E_b\equiv\mathcal E_{bb}$. The first term in \eqref{eq:IonResoccupation} controls the ionization of state $\ket{b}$, while the first term in the parenthesis is responsible for the $\ket{b}\to\ket{a}$ resonance. The last term, which is the focus of this appendix, is a coupling between $\ket{b}$ and $\ket{a}$ induced via the interaction with the continuum. Because $\mathcal{E}_{ba}(t)$ oscillates very rapidly unless $(m_b-m_a)\dot\varphi_*=\epsilon_b-\epsilon_a$, the parenthesis in \eqref{eq:IonResoccupation} can be neglected altogether whenever the system is not actively on resonance.

\vskip 4pt
Let us study what happens when this is the case instead. The same saddle-point approximation done in \cite{Baumann:2021fkf} can be applied to the case $a\ne b$, arriving to
\beq
\mathcal E_{ba}(t)=e^{i(\epsilon_b-\epsilon_a)t-i(m_b-m_a)\varphi_*(t)}\sum_{\ell,m}\biggl[-\frac{i\mu\,\eta_{Kb}^{*\floq{g_b}}\eta_{Ka}^\floq{g_a}}{2k_*^\floq{g_a}}\,\Theta\bigl((k_*^\floq{g_a})^2\bigr)\biggr]\,.
\eeq
Here, we defined $g_a=m-m_a$, evaluated $\ket{K}$ at $k_*^\floq{g_a}=\sqrt{2\mu((m-m_a)\dot\varphi_*+\varepsilon_a)}$, and expanded the bound-continuum coupling in its Floquet components, $\eta_{Ka}=\eta_{Ka}^\floq{g_a}e^{i(m-m_a)t}$ (and similarly for $a\leftrightarrow b$). To understand the effect of the induced coupling $\mathcal E_{ba}$, we can temporarily set $\eta_{ba}=0$ and write \eqref{eq:IonResoccupation} as
\beq
\frac{\dd\abs{c_b}^2}{\dd t}=\sum_{a}\sum_{\ell,m}\frac{\mu}{k_*^\floq{g_a}}\Theta\bigl((k_*^\floq{g_a})^2\bigr)\Re\Bigl[e^{i(\epsilon_b-\epsilon_a)t-i(m_b-m_a)\varphi_*(t)}\eta_{Kb}^{*\floq{g_b}}\eta_{Ka}^\floq{g_a}c_b^*(t)c_a(t)\Bigr]\,.
\label{eqn:ionization+mixedcoupling}
\eeq
Here, the term with $a=b$ reproduces the ionization term $\mathcal{E}_{b}c_{b}(t)$ in \eqref{eq:IonResoccupation}. Moreover, the evolution of state $\ket{a}$ is determined by the same formula, swapping $b\leftrightarrow a$. For $a\ne b$, however, this operation transforms the term in brackets into its complex conjugate, so its real part stays unchanged. We thus see that the induced coupling $\mathcal E_{ba}$ does \emph{not} contribute to a $\ket{b}\to\ket{a}$ transition alongside $\eta_{ba}$, as one might have expected from \eqref{eq:IonResoccupation} and as was speculated in \cite{Baumann:2021fkf}. Instead, both $\abs{c_b}^2$ and $\abs{c_a}^2$ experience an \emph{identical} depletion (in addition to ionization) or recombination, depending on the sign of the real part appearing in \eqref{eqn:ionization+mixedcoupling}; both cases are possible.

\vskip 4pt
We have validated the previous results by comparing them to an explicit numerical integration of the Schrödinger equation, with the continuum states modeled as a large set of discrete states, quadratically spaced in energy. By tuning the parameters to make the impact of the induced coupling clearly visible, we found that \eqref{eqn:ionization+mixedcoupling} gives, indeed, a very accurate description of the evolution of the populations around the resonance. In the main text, in particular for Bohr resonances, we are mainly concerned with the correction from the induced coupling to a naive approach where the contributions of ionization and the resonance are simply summed up. To determine its importance, we assume for simplicity that $\eta_{ab}=0$, $\abs{c_b}^2=1$ and $\abs{c_a}^2=0$ at $t=-\infty$ and employ a (further) saddle-point approximation in \eqref{eq:IonResoccupation} around the time $t_0$ such that $\dot\varphi_*=\Omega_0=(\epsilon_b-\epsilon_a)/(m_b-m_a)$. The population at $t=+\infty$ is then
\beq
\abs{c_a}^2=\frac{2\pi}{\abs{m_b-m_a}\gamma}\,\abs*{\sum_{\ell, m}\frac{\mu\,\eta_{Kb}^{*\floq{g_b}}\eta_{Ka}^\floq{g_a}}{2k_*^\floq{g_b}}\Theta\bigl((k_*^\floq{g_b})^2\bigr)}^2\,,
\label{eqn:final-saddle-point}
\eeq
where the couplings and $k_*^\floq{g_b}$ have to be evaluated at $\Omega=\Omega_0$. Similar to an argument already developed in \cite{Baumann:2021fkf}, this quantity $\abs{c_a}^2$ is $\mathcal O(q^3\alpha^4)$, and it has to compete with the $\eta^2/\gamma\sim\mathcal O(q\alpha^2)$ contributions due to direct coupling $\abs{\eta_{ba}}^2/\gamma$. Once again, we have validated \eqref{eqn:final-saddle-point} by comparing it to a direct numerical integration of the Schrödinger equation and evaluated it for a typical Bohr resonance, finding a final population of $\mathcal O(10^{-11})$. We conclude that simply adding the steady deoccupation introduced by ionization on top of the resonant transition studied in the main text is a good approximation for our purposes.

\section{$\ket{211}\to\ket{200}$ resonance}
\label{sec:211_200}

The strength of the fine resonance $\ket{211}\to\ket{200}$ has anomalous scaling with parameters, due to the dipole $\ell_*=1$ being entirely responsible for the coupling and the binary separation falling partially inside the region where ionization dominates over GW emission. We therefore determine the angle $\delta_2$, such that for $\pi-\delta_2<\beta\le\pi$ the resonance is non-adiabatic, as
\beq
\delta_2=\SI{6.7}{\degree}\,\biggl(\frac{M\ped{c}/M}{10^{-2}}\biggr)^{-1/4}\biggl(\frac{q}{10^{-3}}\biggr)^{1/8} \biggl(\frac{\alpha}{0.2}\biggr)^{7/8}f(\varepsilon_0)^{3/8}F(\alpha,M\ped{c})\,,
\label{eqn:211-delta2}
\eeq
where the formula holds for small $\delta_2$, and the function $F(\alpha,M\ped{c})$ is calculated numerically and shown in Figure~\ref{fig:211_200_F(alpha,Mc)}.

\begin{figure} 
\centering
\includegraphics{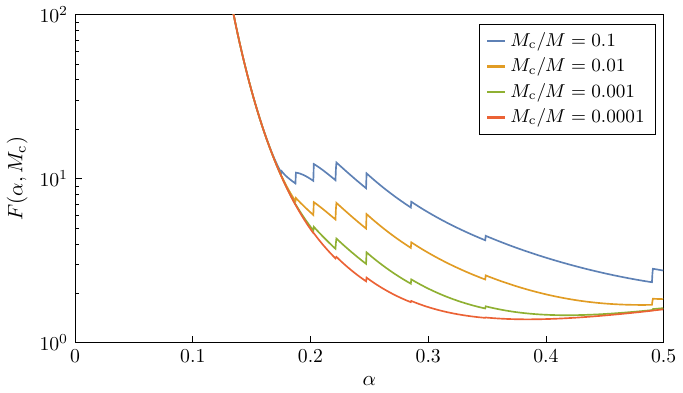}
\caption{Function $F(\alpha,M\ped{c})$ appearing in equation (\ref{eqn:211-delta2}), which defines the angular interval $\delta_2$ around a counter-rotating orbit where the resonance $\ket{211}\to\ket{200}$ is not adiabatic.}
\label{fig:211_200_F(alpha,Mc)}
\end{figure}

\section{Summary of key resonance variables}
\label{sec:variables}

\begin{center}
\begin{tabular}{clc}
\toprule
\textbf{Symbol} & \textbf{Meaning} & \textbf{Reference} \\
\midrule
$\varepsilon$ & Binary eccentricity & \\
$\beta$ & Binary inclination & \\
$g$ & Overtone number & \eqref{eqn:eta-circular} \\
$\gamma$ & Frequency chirp rate induced by GWs/ionization & \eqref{eqn:gamma_gws} \\
$\tau$ & Dimensionless time & \eqref{eqn:coefficients} \\
$\omega$ & Dimensionless frequency & \eqref{eqn:coefficients} \\
$Z$ & Landau-Zener parameter & \eqref{eqn:coefficients} \\
$B$ & Backreaction of a resonance & \eqref{eqn:BC} \\
$C$ & Inertia of $\varepsilon$ and $\beta$ w.r.t.~resonance backreaction & \eqref{eqn:BC} \\
$D$ & Distance parameter, $D=B/C$ & \eqref{eqn:D} \\
$\Gamma$ & Dimensionless decay width of the final state & \eqref{eqn:schrodinger-Gamma}\\
\bottomrule
\end{tabular}
\end{center}

\clearpage
\phantomsection

\addtocontents{toc}{\protect\vskip24pt}
\addcontentsline{toc}{section}{References}

\makeatletter

\interlinepenalty=10000

{\linespread{1.0542}
\bibliographystyle{utphys}
\bibliography{main}

\providecommand{\href}[2]{#2}\begingroup\raggedright\begin{thebibliography}{10}

\bibitem{Barausse:2014tra}
E.~Barausse, V.~Cardoso, and P.~Pani, ``{Can environmental effects spoil
  precision gravitational-wave astrophysics?}'',
  \href{http://dx.doi.org/10.1103/PhysRevD.89.104059}{{\em Phys. Rev. D}
  {\bfseries 89} no.~10, (2014) 104059},
  \href{http://arxiv.org/abs/1404.7149}{{\ttfamily arXiv:1404.7149 [gr-qc]}}.

\bibitem{CanevaSantoro:2023aol}
G.~Caneva~Santoro, S.~Roy, R.~Vicente, M.~Haney, O.~J. Piccinni, W.~Del~Pozzo,
  and M.~Martinez, ``{First Constraints on Compact Binary Environments from
  LIGO-Virgo Data}'',
  \href{http://dx.doi.org/10.1103/PhysRevLett.132.251401}{{\em Phys. Rev.
  Lett.} {\bfseries 132} no.~25, (2024) 251401},
  \href{http://arxiv.org/abs/2309.05061}{{\ttfamily arXiv:2309.05061 [gr-qc]}}.

\bibitem{LISA:2017pwj}
{\bfseries LISA} Collaboration, P.~Amaro-Seoane {\em et~al.}, ``{Laser
  Interferometer Space Antenna}'',
  \href{http://arxiv.org/abs/1702.00786}{{\ttfamily arXiv:1702.00786
  [astro-ph.IM]}}.

\bibitem{Baker:2019nia}
J.~Baker {\em et~al.}, ``{The Laser Interferometer Space Antenna: Unveiling the
  Millihertz Gravitational Wave Sky}'',
  \href{http://arxiv.org/abs/1907.06482}{{\ttfamily arXiv:1907.06482
  [astro-ph.IM]}}.

\bibitem{Maggiore:2019uih}
M.~Maggiore {\em et~al.}, ``{Science Case for the Einstein Telescope}'',
  \href{http://dx.doi.org/10.1088/1475-7516/2020/03/050}{{\em JCAP} {\bfseries
  03} (2020) 050}, \href{http://arxiv.org/abs/1912.02622}{{\ttfamily
  arXiv:1912.02622 [astro-ph.CO]}}.

\bibitem{Barausse:2007dy}
E.~Barausse and L.~Rezzolla, ``{The Influence of the hydrodynamic drag from an
  accretion torus on extreme mass-ratio inspirals}'',
  \href{http://dx.doi.org/10.1103/PhysRevD.77.104027}{{\em Phys. Rev. D}
  {\bfseries 77} (2008) 104027},
  \href{http://arxiv.org/abs/0711.4558}{{\ttfamily arXiv:0711.4558 [gr-qc]}}.

\bibitem{Speri:2022upm}
L.~Speri, A.~Antonelli, L.~Sberna, S.~Babak, E.~Barausse, J.~R. Gair, and M.~L.
  Katz, ``{Probing Accretion Physics with Gravitational Waves}'',
  \href{http://dx.doi.org/10.1103/PhysRevX.13.021035}{{\em Phys. Rev. X}
  {\bfseries 13} no.~2, (2023) 021035},
  \href{http://arxiv.org/abs/2207.10086}{{\ttfamily arXiv:2207.10086 [gr-qc]}}.

\bibitem{Eda:2013gg}
K.~Eda, Y.~Itoh, S.~Kuroyanagi, and J.~Silk, ``{New Probe of Dark-Matter
  Properties: Gravitational Waves from an Intermediate-Mass Black Hole Embedded
  in a Dark-Matter Minispike}'',
  \href{http://dx.doi.org/10.1103/PhysRevLett.110.221101}{{\em Phys. Rev.
  Lett.} {\bfseries 110} no.~22, (2013) 221101},
  \href{http://arxiv.org/abs/1301.5971}{{\ttfamily arXiv:1301.5971 [gr-qc]}}.

\bibitem{Kavanagh:2020cfn}
B.~J. Kavanagh, D.~A. Nichols, G.~Bertone, and D.~Gaggero, ``{Detecting dark
  matter around black holes with gravitational waves: Effects of dark-matter
  dynamics on the gravitational waveform}'',
  \href{http://dx.doi.org/10.1103/PhysRevD.102.083006}{{\em Phys. Rev. D}
  {\bfseries 102} no.~8, (2020) 083006},
  \href{http://arxiv.org/abs/2002.12811}{{\ttfamily arXiv:2002.12811 [gr-qc]}}.

\bibitem{Cole:2022yzw}
P.~S. Cole, G.~Bertone, A.~Coogan, D.~Gaggero, T.~Karydas, B.~J. Kavanagh,
  T.~F.~M. Spieksma, and G.~M. Tomaselli, ``{Distinguishing environmental
  effects on binary black hole gravitational waveforms}'',
  \href{http://dx.doi.org/10.1038/s41550-023-01990-2}{{\em Nature Astron.}
  {\bfseries 7} no.~8, (2023) 943--950},
  \href{http://arxiv.org/abs/2211.01362}{{\ttfamily arXiv:2211.01362 [gr-qc]}}.

\bibitem{Becker:2022wlo}
N.~Becker and L.~Sagunski, ``{Comparing accretion disks and dark matter spikes
  in intermediate mass ratio inspirals}'',
  \href{http://dx.doi.org/10.1103/PhysRevD.107.083003}{{\em Phys. Rev. D}
  {\bfseries 107} no.~8, (2023) 083003},
  \href{http://arxiv.org/abs/2211.05145}{{\ttfamily arXiv:2211.05145 [gr-qc]}}.

\bibitem{Karydas:2024fcn}
T.~K. Karydas, B.~J. Kavanagh, and G.~Bertone, ``{Sharpening the dark matter
  signature in gravitational waveforms I: Accretion and eccentricity
  evolution}'', \href{http://arxiv.org/abs/2402.13053}{{\ttfamily
  arXiv:2402.13053 [gr-qc]}}.

\bibitem{Kavanagh:2024lgq}
B.~J. Kavanagh, T.~K. Karydas, G.~Bertone, P.~Di~Cintio, and M.~Pasquato,
  ``{Sharpening the dark matter signature in gravitational waveforms II:
  Numerical simulations with the NbodyIMRI code}'',
  \href{http://arxiv.org/abs/2402.13762}{{\ttfamily arXiv:2402.13762 [gr-qc]}}.

\bibitem{Garg:2024oeu}
M.~Garg, A.~Derdzinski, S.~Tiwari, J.~Gair, and L.~Mayer, ``{Measuring
  eccentricity and gas-induced perturbation from gravitational waves of LISA
  massive black hole binaries}'',
  \href{http://arxiv.org/abs/2402.14058}{{\ttfamily arXiv:2402.14058
  [astro-ph.GA]}}.

\bibitem{Bamber:2022pbs}
J.~Bamber, J.~C. Aurrekoetxea, K.~Clough, and P.~G. Ferreira, ``{Black hole
  merger simulations in wave dark matter environments}'',
  \href{http://dx.doi.org/10.1103/PhysRevD.107.024035}{{\em Phys. Rev. D}
  {\bfseries 107} no.~2, (2023) 024035},
  \href{http://arxiv.org/abs/2210.09254}{{\ttfamily arXiv:2210.09254 [gr-qc]}}.

\bibitem{Aurrekoetxea:2023jwk}
J.~C. Aurrekoetxea, K.~Clough, J.~Bamber, and P.~G. Ferreira, ``{Effect of Wave
  Dark Matter on Equal Mass Black Hole Mergers}'',
  \href{http://dx.doi.org/10.1103/PhysRevLett.132.211401}{{\em Phys. Rev.
  Lett.} {\bfseries 132} no.~21, (2024) 211401},
  \href{http://arxiv.org/abs/2311.18156}{{\ttfamily arXiv:2311.18156 [gr-qc]}}.

\bibitem{Weinberg:1977ma}
S.~Weinberg, ``{A New Light Boson?}'',
  \href{http://dx.doi.org/10.1103/PhysRevLett.40.223}{{\em Phys. Rev. Lett.}
  {\bfseries 40} (1978) 223--226}.

\bibitem{Wilczek:1977pj}
F.~Wilczek, ``{Problem of Strong $P$ and $T$ Invariance in the Presence of
  Instantons}'', \href{http://dx.doi.org/10.1103/PhysRevLett.40.279}{{\em Phys.
  Rev. Lett.} {\bfseries 40} (1978) 279--282}.

\bibitem{Peccei:1977hh}
R.~D. Peccei and H.~R. Quinn, ``{CP Conservation in the Presence of
  Instantons}'', \href{http://dx.doi.org/10.1103/PhysRevLett.38.1440}{{\em
  Phys. Rev. Lett.} {\bfseries 38} (1977) 1440--1443}.

\bibitem{Arvanitaki:2009fg}
A.~Arvanitaki, S.~Dimopoulos, S.~Dubovsky, N.~Kaloper, and J.~March-Russell,
  ``{String Axiverse}'',
  \href{http://dx.doi.org/10.1103/PhysRevD.81.123530}{{\em Phys. Rev. D}
  {\bfseries 81} (2010) 123530},
  \href{http://arxiv.org/abs/0905.4720}{{\ttfamily arXiv:0905.4720 [hep-th]}}.

\bibitem{Svrcek:2006yi}
P.~Svrcek and E.~Witten, ``{Axions In String Theory}'',
  \href{http://dx.doi.org/10.1088/1126-6708/2006/06/051}{{\em JHEP} {\bfseries
  06} (2006) 051}, \href{http://arxiv.org/abs/hep-th/0605206}{{\ttfamily
  arXiv:hep-th/0605206}}.

\bibitem{Bergstrom:2009ib}
L.~Bergstrom, ``{Dark Matter Candidates}'',
  \href{http://dx.doi.org/10.1088/1367-2630/11/10/105006}{{\em New J. Phys.}
  {\bfseries 11} (2009) 105006},
  \href{http://arxiv.org/abs/0903.4849}{{\ttfamily arXiv:0903.4849 [hep-ph]}}.

\bibitem{Marsh:2015xka}
D.~J.~E. Marsh, ``{Axion Cosmology}'',
  \href{http://dx.doi.org/10.1016/j.physrep.2016.06.005}{{\em Phys. Rept.}
  {\bfseries 643} (2016) 1--79},
  \href{http://arxiv.org/abs/1510.07633}{{\ttfamily arXiv:1510.07633
  [astro-ph.CO]}}.

\bibitem{Hui:2016ltb}
L.~Hui, J.~P. Ostriker, S.~Tremaine, and E.~Witten, ``{Ultralight scalars as
  cosmological dark matter}'',
  \href{http://dx.doi.org/10.1103/PhysRevD.95.043541}{{\em Phys. Rev. D}
  {\bfseries 95} no.~4, (2017) 043541},
  \href{http://arxiv.org/abs/1610.08297}{{\ttfamily arXiv:1610.08297
  [astro-ph.CO]}}.

\bibitem{Ferreira:2020fam}
E.~G.~M. Ferreira, ``{Ultra-light dark matter}'',
  \href{http://dx.doi.org/10.1007/s00159-021-00135-6}{{\em Astron. Astrophys.
  Rev.} {\bfseries 29} no.~1, (2021) 7},
  \href{http://arxiv.org/abs/2005.03254}{{\ttfamily arXiv:2005.03254
  [astro-ph.CO]}}.

\bibitem{ZelDovich1971}
Y.~B. {Zel'Dovich}, ``{Generation of Waves by a Rotating Body}'', {\em Soviet
  Journal of Experimental and Theoretical Physics Letters} {\bfseries 14}
  (1971) 180.

\bibitem{ZelDovich1972}
Y.~B. {Zel'Dovich}, ``{Amplification of Cylindrical Electromagnetic Waves
  Reflected from a Rotating Body}'', {\em Soviet Journal of Experimental and
  Theoretical Physics} {\bfseries 35} (1972) 1085.

\bibitem{Starobinsky:1973aij}
A.~A. Starobinsky, ``{Amplification of waves reflected from a rotating ``black
  hole''\,}'', {\em Sov. Phys. JETP} {\bfseries 37} no.~1, (1973) 28--32.

\bibitem{Brito:2015oca}
R.~Brito, V.~Cardoso, and P.~Pani, ``{Superradiance}: {New Frontiers in Black
  Hole Physics}'', \href{http://dx.doi.org/10.1007/978-3-319-19000-6}{{\em
  Lect. Notes Phys.} {\bfseries 906} (2015) pp.1--237},
  \href{http://arxiv.org/abs/1501.06570}{{\ttfamily arXiv:1501.06570 [gr-qc]}}.

\bibitem{Zhang:2018kib}
J.~Zhang and H.~Yang, ``{Gravitational floating orbits around hairy black
  holes}'', \href{http://dx.doi.org/10.1103/PhysRevD.99.064018}{{\em Phys. Rev.
  D} {\bfseries 99} no.~6, (2019) 064018},
  \href{http://arxiv.org/abs/1808.02905}{{\ttfamily arXiv:1808.02905 [gr-qc]}}.

\bibitem{Baumann:2018vus}
D.~Baumann, H.~S. Chia, and R.~A. Porto, ``{Probing Ultralight Bosons with
  Binary Black Holes}'',
  \href{http://dx.doi.org/10.1103/PhysRevD.99.044001}{{\em Phys. Rev. D}
  {\bfseries 99} no.~4, (2019) 044001},
  \href{http://arxiv.org/abs/1804.03208}{{\ttfamily arXiv:1804.03208 [gr-qc]}}.

\bibitem{Zhang:2019eid}
J.~Zhang and H.~Yang, ``{Dynamic Signatures of Black Hole Binaries with
  Superradiant Clouds}'',
  \href{http://dx.doi.org/10.1103/PhysRevD.101.043020}{{\em Phys. Rev. D}
  {\bfseries 101} no.~4, (2020) 043020},
  \href{http://arxiv.org/abs/1907.13582}{{\ttfamily arXiv:1907.13582 [gr-qc]}}.

\bibitem{Baumann:2019ztm}
D.~Baumann, H.~S. Chia, R.~A. Porto, and J.~Stout, ``{Gravitational Collider
  Physics}'', \href{http://dx.doi.org/10.1103/PhysRevD.101.083019}{{\em Phys.
  Rev. D} {\bfseries 101} no.~8, (2020) 083019},
  \href{http://arxiv.org/abs/1912.04932}{{\ttfamily arXiv:1912.04932 [gr-qc]}}.

\bibitem{Baumann:2021fkf}
D.~Baumann, G.~Bertone, J.~Stout, and G.~M. Tomaselli, ``{Ionization of
  gravitational atoms}'',
  \href{http://dx.doi.org/10.1103/PhysRevD.105.115036}{{\em Phys. Rev. D}
  {\bfseries 105} no.~11, (2022) 115036},
  \href{http://arxiv.org/abs/2112.14777}{{\ttfamily arXiv:2112.14777 [gr-qc]}}.

\bibitem{Baumann:2022pkl}
D.~Baumann, G.~Bertone, J.~Stout, and G.~M. Tomaselli, ``{Sharp Signals of
  Boson Clouds in Black Hole Binary Inspirals}'',
  \href{http://dx.doi.org/10.1103/PhysRevLett.128.221102}{{\em Phys. Rev.
  Lett.} {\bfseries 128} no.~22, (2022) 221102},
  \href{http://arxiv.org/abs/2206.01212}{{\ttfamily arXiv:2206.01212 [gr-qc]}}.

\bibitem{Tomaselli:2023ysb}
G.~M. Tomaselli, T.~F.~M. Spieksma, and G.~Bertone, ``{Dynamical friction in
  gravitational atoms}'',
  \href{http://dx.doi.org/10.1088/1475-7516/2023/07/070}{{\em JCAP} {\bfseries
  07} (2023) 070}, \href{http://arxiv.org/abs/2305.15460}{{\ttfamily
  arXiv:2305.15460 [gr-qc]}}.

\bibitem{Brito:2023pyl}
R.~Brito and S.~Shah, ``{Extreme mass-ratio inspirals into black holes
  surrounded by scalar clouds}'',
  \href{http://dx.doi.org/10.1103/PhysRevD.108.084019}{{\em Phys. Rev. D}
  {\bfseries 108} no.~8, (2023) 084019},
  \href{http://arxiv.org/abs/2307.16093}{{\ttfamily arXiv:2307.16093 [gr-qc]}}.

\bibitem{Duque:2023cac}
F.~Duque, C.~F.~B. Macedo, R.~Vicente, and V.~Cardoso, ``{Axion Weak Leaks:
  extreme mass-ratio inspirals in ultra-light dark matter}'',
  \href{http://arxiv.org/abs/2312.06767}{{\ttfamily arXiv:2312.06767 [gr-qc]}}.

\bibitem{Takahashi:2021eso}
T.~Takahashi and T.~Tanaka, ``{Axion clouds may survive the perturbative tidal
  interaction over the early inspiral phase of black hole binaries}'',
  \href{http://dx.doi.org/10.1088/1475-7516/2021/10/031}{{\em JCAP} {\bfseries
  10} (2021) 031}, \href{http://arxiv.org/abs/2106.08836}{{\ttfamily
  arXiv:2106.08836 [gr-qc]}}.

\bibitem{Ding:2020bnl}
Q.~Ding, X.~Tong, and Y.~Wang, ``{Gravitational Collider Physics via
  Pulsar-Black Hole Binaries}'',
  \href{http://dx.doi.org/10.3847/1538-4357/abd803}{{\em Astrophys. J.}
  {\bfseries 908} no.~1, (2021) 78},
  \href{http://arxiv.org/abs/2009.11106}{{\ttfamily arXiv:2009.11106
  [astro-ph.HE]}}.

\bibitem{Tong:2021whq}
X.~Tong, Y.~Wang, and H.-Y. Zhu, ``{Gravitational Collider Physics via
  Pulsar\textendash{}Black Hole Binaries II: Fine and Hyperfine Structures Are
  Favored}'', \href{http://dx.doi.org/10.3847/1538-4357/ac36db}{{\em Astrophys.
  J.} {\bfseries 924} no.~2, (2022) 99},
  \href{http://arxiv.org/abs/2106.13484}{{\ttfamily arXiv:2106.13484
  [astro-ph.HE]}}.

\bibitem{Du:2022trq}
P.~Du, D.~Egana-Ugrinovic, R.~Essig, G.~Fragione, and R.~Perna, ``{Searching
  for ultra-light bosons and constraining black hole spin distributions with
  stellar tidal disruption events}'',
  \href{http://dx.doi.org/10.1038/s41467-022-32301-4}{{\em Nature Commun.}
  {\bfseries 13} no.~1, (2022) 4626},
  \href{http://arxiv.org/abs/2202.01215}{{\ttfamily arXiv:2202.01215
  [hep-ph]}}.

\bibitem{Takahashi:2023flk}
T.~Takahashi, H.~Omiya, and T.~Tanaka, ``{Evolution of binary systems
  accompanying axion clouds in extreme mass ratio inspirals}'',
  \href{http://dx.doi.org/10.1103/PhysRevD.107.103020}{{\em Phys. Rev. D}
  {\bfseries 107} no.~10, (2023) 103020},
  \href{http://arxiv.org/abs/2301.13213}{{\ttfamily arXiv:2301.13213 [gr-qc]}}.

\bibitem{Berti:2019wnn}
E.~Berti, R.~Brito, C.~F.~B. Macedo, G.~Raposo, and J.~L. Rosa, ``{Ultralight
  boson cloud depletion in binary systems}'',
  \href{http://dx.doi.org/10.1103/PhysRevD.99.104039}{{\em Phys. Rev. D}
  {\bfseries 99} no.~10, (2019) 104039},
  \href{http://arxiv.org/abs/1904.03131}{{\ttfamily arXiv:1904.03131 [gr-qc]}}.

\bibitem{Tomaselli:2024dbw}
G.~M. Tomaselli, T.~F.~M. Spieksma, and G.~Bertone, ``{The legacy of boson
  clouds on black hole binaries}'',
  \href{http://arxiv.org/abs/2407.12908}{{\ttfamily arXiv:2407.12908 [gr-qc]}}.

\bibitem{Hui:2022sri}
L.~Hui, Y.~T.~A. Law, L.~Santoni, G.~Sun, G.~M. Tomaselli, and E.~Trincherini,
  ``{Black hole superradiance with dark matter accretion}'',
  \href{http://dx.doi.org/10.1103/PhysRevD.107.104018}{{\em Phys. Rev. D}
  {\bfseries 107} no.~10, (2023) 104018},
  \href{http://arxiv.org/abs/2208.06408}{{\ttfamily arXiv:2208.06408 [gr-qc]}}.

\bibitem{Herdeiro:2021znw}
C.~A.~R. Herdeiro, E.~Radu, and N.~M. Santos, ``{A bound on energy extraction
  (and hairiness) from superradiance}'',
  \href{http://dx.doi.org/10.1016/j.physletb.2021.136835}{{\em Phys. Lett. B}
  {\bfseries 824} (2022) 136835},
  \href{http://arxiv.org/abs/2111.03667}{{\ttfamily arXiv:2111.03667 [gr-qc]}}.

\bibitem{East:2017ovw}
W.~E. East and F.~Pretorius, ``{Superradiant Instability and Backreaction of
  Massive Vector Fields around Kerr Black Holes}'',
  \href{http://dx.doi.org/10.1103/PhysRevLett.119.041101}{{\em Phys. Rev.
  Lett.} {\bfseries 119} no.~4, (2017) 041101},
  \href{http://arxiv.org/abs/1704.04791}{{\ttfamily arXiv:1704.04791 [gr-qc]}}.

\bibitem{Baumann:2019eav}
D.~Baumann, H.~S. Chia, J.~Stout, and L.~ter Haar, ``{The Spectra of
  Gravitational Atoms}'',
  \href{http://dx.doi.org/10.1088/1475-7516/2019/12/006}{{\em JCAP} {\bfseries
  12} (2019) 006}, \href{http://arxiv.org/abs/1908.10370}{{\ttfamily
  arXiv:1908.10370 [gr-qc]}}.

\bibitem{zener1932non}
C.~Zener, ``{Non-Adiabatic Crossing of Energy Levels}'', {\em Proceedings of
  the Royal Society of London} {\bfseries 137} no.~833, (1932) 696--702.

\bibitem{landau1932theorie}
L.~Landau, ``{Zur Theorie der Energie\"ubertragung}'', {\em Z. Sowjetunion}
  {\bfseries 2} (1932) 46--51.

\bibitem{wigner}
E.~P. Wigner, {\em Group Theory and Its Application to the Quantum Mechanics of
  Atomic Spectra}.
\newblock Academic Press, New York, 1959.

\bibitem{Peters:1963ux}
P.~C. Peters and J.~Mathews, ``{Gravitational radiation from point masses in a
  Keplerian orbit}'', \href{http://dx.doi.org/10.1103/PhysRev.131.435}{{\em
  Phys. Rev.} {\bfseries 131} (1963) 435--439}.

\bibitem{Peters:1964zz}
P.~C. Peters, ``{Gravitational Radiation and the Motion of Two Point Masses}'',
  \href{http://dx.doi.org/10.1103/PhysRev.136.B1224}{{\em Phys. Rev.}
  {\bfseries 136} (1964) B1224--B1232}.

\bibitem{PhysRevD.22.2323}
S.~Detweiler, ``Klein-gordon equation and rotating black holes'',
  \href{http://dx.doi.org/10.1103/PhysRevD.22.2323}{{\em Phys. Rev. D}
  {\bfseries 22} (Nov, 1980) 2323--2326}.

\bibitem{Leaver:1985ax}
E.~W. Leaver, ``{An Analytic representation for the quasi normal modes of Kerr
  black holes}'', \href{http://dx.doi.org/10.1098/rspa.1985.0119}{{\em Proc.
  Roy. Soc. Lond. A} {\bfseries 402} (1985) 285--298}.

\bibitem{Cardoso:2005vk}
V.~Cardoso and S.~Yoshida, ``{Superradiant instabilities of rotating black
  branes and strings}'',
  \href{http://dx.doi.org/10.1088/1126-6708/2005/07/009}{{\em JHEP} {\bfseries
  07} (2005) 009}, \href{http://arxiv.org/abs/hep-th/0502206}{{\ttfamily
  arXiv:hep-th/0502206}}.

\bibitem{Dolan:2007mj}
S.~R. Dolan, ``{Instability of the massive Klein-Gordon field on the Kerr
  spacetime}'', \href{http://dx.doi.org/10.1103/PhysRevD.76.084001}{{\em Phys.
  Rev. D} {\bfseries 76} (2007) 084001},
  \href{http://arxiv.org/abs/0705.2880}{{\ttfamily arXiv:0705.2880 [gr-qc]}}.

\bibitem{Berti:2009kk}
E.~Berti, V.~Cardoso, and A.~O. Starinets, ``{Quasinormal modes of black holes
  and black branes}'',
  \href{http://dx.doi.org/10.1088/0264-9381/26/16/163001}{{\em Class. Quant.
  Grav.} {\bfseries 26} (2009) 163001},
  \href{http://arxiv.org/abs/0905.2975}{{\ttfamily arXiv:0905.2975 [gr-qc]}}.

\bibitem{Amaro-Seoane:2012lgq}
P.~Amaro-Seoane, ``{Relativistic dynamics and extreme mass ratio inspirals}'',
  \href{http://dx.doi.org/10.1007/s41114-018-0013-8}{{\em Living Rev. Rel.}
  {\bfseries 21} (2018) 4}, \href{http://arxiv.org/abs/1205.5240}{{\ttfamily
  arXiv:1205.5240 [astro-ph.CO]}}.

\bibitem{LISA:2022yao}
{\bfseries LISA} Collaboration, P.~A. Seoane {\em et~al.}, ``{Astrophysics with
  the Laser Interferometer Space Antenna}'',
  \href{http://dx.doi.org/10.1007/s41114-022-00041-y}{{\em Living Rev. Rel.}
  {\bfseries 26} no.~1, (2023) 2},
  \href{http://arxiv.org/abs/2203.06016}{{\ttfamily arXiv:2203.06016 [gr-qc]}}.

\bibitem{Stone:2016wzz}
N.~C. Stone, B.~D. Metzger, and Z.~Haiman, ``{Assisted inspirals of stellar
  mass black holes embedded in AGN discs: solving the \textquoteleft{}final au
  problem\textquoteright{}}'',
  \href{http://dx.doi.org/10.1093/mnras/stw2260}{{\em Mon. Not. Roy. Astron.
  Soc.} {\bfseries 464} no.~1, (2017) 946--954},
  \href{http://arxiv.org/abs/1602.04226}{{\ttfamily arXiv:1602.04226
  [astro-ph.GA]}}.

\bibitem{Bartos:2016dgn}
I.~Bartos, B.~Kocsis, Z.~Haiman, and S.~M\'arka, ``{Rapid and Bright
  Stellar-mass Binary Black Hole Mergers in Active Galactic Nuclei}'',
  \href{http://dx.doi.org/10.3847/1538-4357/835/2/165}{{\em Astrophys. J.}
  {\bfseries 835} no.~2, (2017) 165},
  \href{http://arxiv.org/abs/1602.03831}{{\ttfamily arXiv:1602.03831
  [astro-ph.HE]}}.

\bibitem{McKernan_2018}
B.~McKernan, K.~E.~S. Ford, J.~Bellovary, N.~W.~C. Leigh, Z.~Haiman, B.~Kocsis,
  W.~Lyra, M.-M.~M. Low, B.~Metzger, M.~O’Dowd, S.~Endlich, and D.~J. Rosen,
  ``Constraining stellar-mass black hole mergers in agn disks detectable with
  ligo'', \href{http://dx.doi.org/10.3847/1538-4357/aadae5}{{\em The
  Astrophysical Journal} {\bfseries 866} no.~1, (Oct, 2018) 66}.

\bibitem{Levin:2006uc}
Y.~Levin, ``{Starbursts near supermassive black holes: young stars in the
  Galactic Center, and gravitational waves in LISA band}'',
  \href{http://dx.doi.org/10.1111/j.1365-2966.2006.11155.x}{{\em Mon. Not. Roy.
  Astron. Soc.} {\bfseries 374} (2007) 515--524},
  \href{http://arxiv.org/abs/astro-ph/0603583}{{\ttfamily
  arXiv:astro-ph/0603583}}.

\bibitem{Boskovic:2024fga}
M.~Bo\v{s}kovi\'c, M.~Koschnitzke, and R.~A. Porto, ``{Signatures of ultralight
  bosons in the orbital eccentricity of binary black holes}'',
  \href{http://arxiv.org/abs/2403.02415}{{\ttfamily arXiv:2403.02415 [gr-qc]}}.

\bibitem{Tong:2022bbl}
X.~Tong, Y.~Wang, and H.-Y. Zhu, ``{Termination of superradiance from a binary
  companion}'', \href{http://dx.doi.org/10.1103/PhysRevD.106.043002}{{\em Phys.
  Rev. D} {\bfseries 106} no.~4, (2022) 043002},
  \href{http://arxiv.org/abs/2205.10527}{{\ttfamily arXiv:2205.10527 [gr-qc]}}.

\bibitem{Fan:2023jjj}
K.~Fan, X.~Tong, Y.~Wang, and H.-Y. Zhu, ``{Modulating binary dynamics via the
  termination of black hole superradiance}'',
  \href{http://dx.doi.org/10.1103/PhysRevD.109.024059}{{\em Phys. Rev. D}
  {\bfseries 109} no.~2, (2024) 024059},
  \href{http://arxiv.org/abs/2311.17013}{{\ttfamily arXiv:2311.17013 [gr-qc]}}.

\bibitem{Goodman:2003sf}
J.~Goodman and J.~C. Tan, ``{Supermassive stars in quasar disks}'',
  \href{http://dx.doi.org/10.1086/386360}{{\em Astrophys. J.} {\bfseries 608}
  (2004) 108--118}, \href{http://arxiv.org/abs/astro-ph/0307361}{{\ttfamily
  arXiv:astro-ph/0307361}}.

\bibitem{Goodman:2002gv}
J.~Goodman, ``{Selfgravity and QSO disks}'',
  \href{http://dx.doi.org/10.1046/j.1365-8711.2003.06241.x}{{\em Mon. Not. Roy.
  Astron. Soc.} {\bfseries 339} (2003) 937},
  \href{http://arxiv.org/abs/astro-ph/0201001}{{\ttfamily
  arXiv:astro-ph/0201001}}.

\bibitem{Babak:2017tow}
S.~Babak, J.~Gair, A.~Sesana, E.~Barausse, C.~F. Sopuerta, C.~P.~L. Berry,
  E.~Berti, P.~Amaro-Seoane, A.~Petiteau, and A.~Klein, ``{Science with the
  space-based interferometer LISA. V: Extreme mass-ratio inspirals}'',
  \href{http://dx.doi.org/10.1103/PhysRevD.95.103012}{{\em Phys. Rev. D}
  {\bfseries 95} no.~10, (2017) 103012},
  \href{http://arxiv.org/abs/1703.09722}{{\ttfamily arXiv:1703.09722 [gr-qc]}}.

\bibitem{Amaro-Seoane:2012jcd}
P.~Amaro-Seoane, C.~F. Sopuerta, and M.~D. Freitag, ``{The role of the
  supermassive black hole spin in the estimation of the EMRI event rate}'',
  \href{http://dx.doi.org/10.1093/mnras/sts572}{{\em Mon. Not. Roy. Astron.
  Soc.} {\bfseries 429} no.~4, (2013) 3155--3165},
  \href{http://arxiv.org/abs/1205.4713}{{\ttfamily arXiv:1205.4713
  [astro-ph.CO]}}.

\bibitem{Ferreira:2017pth}
M.~C. Ferreira, C.~F.~B. Macedo, and V.~Cardoso, ``{Orbital fingerprints of
  ultralight scalar fields around black holes}'',
  \href{http://dx.doi.org/10.1103/PhysRevD.96.083017}{{\em Phys. Rev. D}
  {\bfseries 96} no.~8, (2017) 083017},
  \href{http://arxiv.org/abs/1710.00830}{{\ttfamily arXiv:1710.00830 [gr-qc]}}.

\bibitem{Hannuksela:2018izj}
O.~A. Hannuksela, K.~W.~K. Wong, R.~Brito, E.~Berti, and T.~G.~F. Li,
  ``{Probing the existence of ultralight bosons with a single
  gravitational-wave measurement}'',
  \href{http://dx.doi.org/10.1038/s41550-019-0712-4}{{\em Nature Astron.}
  {\bfseries 3} no.~5, (2019) 447--451},
  \href{http://arxiv.org/abs/1804.09659}{{\ttfamily arXiv:1804.09659
  [astro-ph.HE]}}.

\bibitem{Cao:2023fyv}
Y.~Cao and Y.~Tang, ``{Signatures of ultralight bosons in compact binary
  inspiral and outspiral}'',
  \href{http://dx.doi.org/10.1103/PhysRevD.108.123017}{{\em Phys. Rev. D}
  {\bfseries 108} no.~12, (2023) 123017},
  \href{http://arxiv.org/abs/2307.05181}{{\ttfamily arXiv:2307.05181 [gr-qc]}}.

\bibitem{Cannizzaro:2023jle}
E.~Cannizzaro, L.~Sberna, S.~R. Green, and S.~Hollands, ``{Relativistic
  Perturbation Theory for Black-Hole Boson Clouds}'',
  \href{http://dx.doi.org/10.1103/PhysRevLett.132.051401}{{\em Phys. Rev.
  Lett.} {\bfseries 132} no.~5, (2024) 051401},
  \href{http://arxiv.org/abs/2309.10021}{{\ttfamily arXiv:2309.10021 [gr-qc]}}.

\end{thebibliography}\endgroup
}
\makeatother

\end{document}